\documentclass[preprint,3p,fleqn]{elsarticle}

\let\originalleft\left
\let\originalright\right
\renewcommand{\left}{\mathopen{}\mathclose\bgroup\originalleft}
\renewcommand{\right}{\aftergroup\egroup\originalright}

\usepackage{amssymb}
\usepackage{multicol}

\journal{Computer Methods in Applied Mechanics and Engineering (CMAME)}

\usepackage{amsmath}
\usepackage{setspace}
\usepackage{amsfonts}
\usepackage{amsthm}
\usepackage{mathtools}
\usepackage{subfig}
\usepackage{lineno}
\usepackage{algorithm}
\usepackage{algpseudocode}
\usepackage{hyperref}
\usepackage{float}
\usepackage{array,multirow}
\usepackage{color}
\usepackage{tikz}
\usepackage{graphicx}
\usepackage[flushleft]{threeparttable}
\usepackage{blindtext}
\usepackage[colorinlistoftodos,prependcaption,textsize=tiny]{todonotes}
\usepackage{regexpatch}
\usepackage[export]{adjustbox}
\usepackage{booktabs}
\usepackage[normalem]{ulem}

\definecolor{lightgray}{gray}{0.80}

\usetikzlibrary{decorations.pathreplacing}
\usepackage[listings,theorems,skins,breakable]{tcolorbox}
\tcbset{skin=enhanced}
\newtcolorbox{lbracebox}[1][Word]{%
   frame hidden,enlarge left by=2cm,width=\linewidth-2cm,%
  overlay unbroken = {\draw [decorate,decoration={brace,amplitude=10pt},]%
                     (frame.south west)-- (frame.north west)
                    node [black,midway,left,xshift=-.6cm] {#1};},%
}

\usepackage{scalerel,amssymb}

\def\msquare{\mathord{\scalerel*{\Box}{gX}}}

\def\B{\color{black}}

\makeatletter
\xpatchcmd{\@todo}{\setkeys{todonotes}{#1}}{\setkeys{todonotes}{inline,#1}}{}{}
\makeatother

\theoremstyle{plain}
\newtheorem{theorem}{Theorem}[section]

\newtheorem{remark}[theorem]{Remark}

\theoremstyle{definition}

\newcommand{\mat}[1]{\mathsf{#1}} 
\newcommand{\vect}[1]{#1}
\newcommand{\List}[1]{\left(\right. #1 \left.\right) }
\newcommand{\eigfun}{\phi}

\newcommand{\eigval}{\lambda}
\newcommand{\eigvec}{\mat{v}_h}

\newcommand{\x}{\vect{x}}
\newcommand{\xp}{\vect{x}^\prime}
\newcommand{\px}{\hat{\x}}
\newcommand{\pxp}{\hat{\x}^\prime}
\newcommand{\pxk}[1]{\hat{\x}_{#1}}

\newcommand{\transposed}[1]{{#1}^\top}
\newcommand{\invtransposed}[1]{{#1}^{-\top}}
\newcommand{\dint}[1]{\,\mathrm{d}{#1}}
\newcommand{\ddint}[1]{\mathrm{d}{#1}}
\newcommand{\midx}[1]{\mathsf{#1}}

\newcommand{\bsspace}[1]{\mathcal{B}_{#1}}
\newcommand{\tbsspace}[1]{\tilde{\mathcal{B}}_{#1}}
\newcommand{\ubspline}[2]{B_{#1,#2}}
\newcommand{\bspline}[1]{B_{\midx{#1}}}
\newcommand{\geomap}[1]{F(#1)}
\newcommand{\geomapNOARG}[1]{F}

\newcommand{\detgeomap}[1]{\mathrm{det}\,\mathrm D\geomap{#1}}

\newcommand{\tbspline}[2]{\tilde{B}_{\midx{#1}}(#2)}
\newcommand{\tubspline}[3]{\tilde B_{#1,#2}(#3)}

\newcommand{\tinterpmat}[2]{\tilde{B}_{#1 #2}}
\newcommand{\invinterpmat}[2]{\tilde B^{-1}_{#1 #2}}

\newcommand{\matinterpmat}{\mat B}
\newcommand{\tmatinterpmat}{\mat{\tilde B}}

\newcommand{\nomatA}{A}
\newcommand{\nomatB}{Z}
\newcommand{\matA}{\mat \nomatA}
\newcommand{\matB}{\mat \nomatB}
\newcommand{\tmatA}{\mat{\tilde{\nomatA}}}

\newcommand{\tnomatA}{\tilde{\nomatA}}

\newcommand{\stochdom}{\Theta}
\newcommand{\w}{\theta}
\newcommand{\sigfield}{\Sigma}
\newcommand{\pmeasure}{\mathbb P}
\newcommand{\meanval}[1]{\mathbb{E}\left[ {#1} \right]}

\newcommand{\covfun}[2]{\Gamma(#1,{#2})}
\newcommand{\pcovfun}[2]{\hat{\Gamma}(#1,{#2})}

\newcommand{\randfield}{\alpha}
\newcommand{\drandfield}[1]{\tilde\alpha_{#1}}
\newcommand{\randvar}{\xi}

\newcommand{\dom}{\mathcal D}
\newcommand{\pdom}{{\hat{\mathcal{D}}}}

\newcommand{\physdim}{d}
\newcommand{\paradim}{\physdim}

\modulolinenumbers[5]

\begin{document}

\begin{frontmatter}



\title{A matrix-free isogeometric Galerkin method for Karhunen-Loève approximation of random fields using tensor product splines, tensor contraction and interpolation based quadrature}


\author[ibnm]{Michal L. Mika\corref{cor1}}
\ead{mika@ibnm.uni-hannover.de}

\author[ices]{Thomas J.R. Hughes}
\ead{hughes@ices.utexas.edu}

\author[ibnm]{Dominik Schillinger}
\ead{schillinger@ibnm.uni-hannover.de}

\author[ikm]{Peter Wriggers}
\ead{wriggers@ikm.uni-hannover.de}

\author[ibnm]{Ren\'e R. Hiemstra}
\ead{rene.hiemstra@ibnm.uni-hannover.de}

\cortext[cor1]{Corresponding author}
\address[ibnm]{Institut f{\"u}r Baumechanik und Numerische Mechanik, Leibniz Universit{\"a}t Hannover}
\address[ices]{Oden Institute for Computational Engineering and Sciences, The University of Texas at Austin}
\address[ikm]{Institut f{\"u}r Kontinuumsmechanik, Leibniz Universit{\"a}t Hannover}

\begin{abstract}
The Karhunen-Loève series expansion (KLE) decomposes a stochastic process into an infinite series of pairwise uncorrelated random variables and pairwise $L^2$-orthogonal functions. For any given truncation order of the infinite series the basis is optimal in the sense that the total mean squared error is minimized. The orthogonal basis functions are determined as the solution of an eigenvalue problem corresponding to the homogeneous Fredholm integral equation of the second kind, which is computationally challenging for several reasons. Firstly, a Galerkin discretization requires numerical integration over a $2d$ dimensional domain, where $d$, in this work, denotes the spatial dimension. Secondly, the main system matrix of the discretized weak-form is dense. Consequently, the computational complexity of classical finite element formation and assembly procedures as well as the memory requirements of direct solution techniques become quickly computationally intractable with increasing polynomial degree, number of elements and degrees of freedom. The objective of this work is to significantly reduce several of the computational bottlenecks associated with numerical solution of the KLE. We present a matrix-free solution strategy, which is embarrassingly parallel and scales favorably with problem size and polynomial degree. Our approach is based on (1) an interpolation based quadrature that minimizes the required number of quadrature points; (2) an inexpensive reformulation of the generalized eigenvalue problem into a standard eigenvalue problem; and (3) a matrix-free and parallel matrix-vector product for iterative eigenvalue solvers. Two higher-order three-dimensional $C^0$-conforming multipatch benchmarks illustrate exceptional computational performance combined with high accuracy and robustness.
\end{abstract}


\begin{highlights}
\item Interpolation based quadrature of the weak form of the integral eigenvalue problem that is optimal in terms of the number of evaluation points and with cost independent of polynomial degree;
\item Efficient formation of finite element arrays based on sum factorization of integrands defined on high-dimensional $C^0$-conforming multipatch domains;
\item Inexpensive reformulation of the generalized eigenvalue problem into an equivalent standard algebraic eigenvalue problem, which decreases computational cost significantly while improving conditioning;
\item Formulation of a matrix-free and parallel matrix-vector product for iterative eigenvalue solvers that scales quadratically with the number of degrees of freedom of the interpolation space.
\end{highlights}


\end{frontmatter}



\section{Introduction}
Most physical systems exhibit randomness, which, because of its lack of pattern or regularity, can not be explicitly captured by deterministic mathematical models. The randomness may be due to the nature of the phenomenon itself, called \textit{aleatoric} uncertainty, or due to a lack of knowledge about the system, referred to as \textit{epistemic} uncertainty. In the latter the uncertainty may be reduced by obtaining additional data about the system at hand. An example of an epistemic uncertainty encountered in engineering are the fluctuations of material properties throughout a body, which occur due to the inhomogeneity of the medium. Deterministic mechanical models typically feature empirically derived material parameters, such as material stiffness and yield stress,  that are assumed constant throughout the body. Their value is typically determined as a statistical volumetric average over a large set of laboratory specimens. This idealized model of reality may be insufficient in e.g. structural risk or reliability analysis and prediction, which is concerned with probabilities of violation of safety limits or performance measures, respectively \cite{melchers_structural_2017}. In this case the effects of uncertainty on the result of a computation need to be quantified.

Uncertainty in physical quantities that vary in space and or time may be adequately modeled by \emph{stochastic processes} or \emph{random fields} \cite{sudret_bruno_stochastic_2000}. This approach generalizes a deterministic system modeled by a partial differential equation to a stochastic system modeled by a stochastic partial differential equation or SPDE. Reliable predictions may be obtained by propagating uncertainties in input variables to those in the response. The main objective is to compute the response statistics, such as the mean and variance in the random solution field, or the probability that a set tolerance is exceeded. To compute these statistics it is necessary to discretize the SPDE, not only in space and time, but also in the stochastic dimensions. This can be a complicated task, not because of modeling randomness, but due to the \emph{curse of dimensionality}. Every random variable contributes one dimension to the problem. Hence, it is important to keep their total to a minimum.

\subsection{Discrete representation of random fields by the truncated Karhunen-Loève series expansion}
One of the relevant questions in stochastic analysis is how to represent random fields discretely, in a manner suitable
for use in numerical computation. The essential step is to break down the representation into a tractable
number of mutually independent random variables, whose combination preserves the stochastic variability
of the process \cite{eiermann_computational_2007, keese_review_2003}.
One representation that is of particular interest is the truncated Karhunen-Loève
series expansion or KLE \cite{karhunen_uber_1947, loeve_michel_functions_1948}.
The KLE decomposes a stochastic process into an infinite series of pairwise
uncorrelated random variables and pairwise $L^2$-orthogonal basis functions. Truncating the series expansion
after $M$ terms yields the best $M$-term linear approximation of the random field, in the sense that the total
mean squared error is minimized \cite{ghanem_stochastic_1991}.
The KLE is useful in practice when satisfactory accuracy is attained
with no more than 20-30 terms \cite{eiermann_computational_2007,stefanou_stochastic_2009}.

Computation of the truncated KLE requires the solution of a homogeneous Fredholm integral
eigenvalue problem (IEVP) of the second kind. In general this is only possible numerically. The most
popular numerical methods to solve IEVPs are the Nyström method, degenerate kernel methods and the
collocation and Galerkin method \cite{betz_numerical_2014, kress_linear_2014}. The Galerkin method is widely regarded as superior due to its
approximation properties and solid theoretical foundation. Specifically, it can be shown that
the eigenvalues converge monotonically towards the exact eigenvalues and, by construction, that
the modes preserve exactly the $L^2$-orthogonality property of the analytical mode-shapes \cite{eiermann_computational_2007}.

\subsection{Challenges in numerical solution of the KLE by means of the Galerkin method}
Efficient solution of the KLE using the Galerkin method is a computationally challenging task~\cite{eiermann_computational_2007}. The main challenges are the following:
\begin{itemize}
	\item[(i)] A Galerkin discretization requires numerical integration over a $2d$ dimensional domain, where $d$, in
this work, denotes the spatial dimension. The computational complexity of classical finite element formation and assembly
procedures scales as $\mathcal{O}\left(N_e^2 (p+1)^{3d} \right)$, where $N_e$ is the global number of elements, $p$ the polynomial degree and $d$ the spatial dimension.
	\item[(ii)] The main system matrix of the discretized weak-form is dense and requires $\mathcal O\left(8 N^2\right)$ bytes of memory in double precision arithmetic, where $N$ is the dimension of the trial space.
	\item[(iii)] Numerical solution requires one sparse backsolve $\mathcal{O}\left(N^2 \right)$ and one dense matrix-vector product $\mathcal{O}\left(N^2 \right)$ in each iteration of the eigenvalue solver, thus the solution time of the numerical eigenvalue solver scales $\mathcal{O}\left(N^2 \cdot N_{\text{iter}}\right)$, where $N_{\text{iter}}$ is the number of iterations required by the Lanczos solver.
\end{itemize}
Table \ref{tab:memoryrequirements} illustrates that explicit storage of the dense system matrix requires impracticable amounts of memory for problems involving more than $100K$ degrees of freedom. Hence, the computational complexity of classical finite element formation and assembly procedures as well as memory requirements of direct solution techniques become quickly computationally intractable with increasing polynomial degree, number of elements and degrees of freedom.

There has been a particular research effort devoted to alleviating the disadvantages of the Galerkin method. In~\cite{allaix_karhunen-loeve_2013, hackbusch_hierarchical_2005, khoromskij_application_2009} an approximation by (Kronecker product) hierarchical matrices is used to efficiently compute the dense matrices as well as to reduce the memory requirements. These matrices are sparse and allow for matrix multiplication, addition and inversion in $\mathcal O\left(N\log N\right)$ time (or for Kronecker product hierarchical matrices in $\mathcal O\left(N\right)$ time) where $N$ is the number of degrees of freedom. The generalized Fast Multipole Method, which also scales with $\mathcal O\left(N\log N\right)$, has been proposed in \cite{schwab_karhunenloeve_2006}. This method was shown to \textit{not} yield significant speed-ups for $p$ finite element methods and thus it is recommended for kernels of low regularity. Wavelet Galerkin-schemes \cite{phoon_implementation_2002} are also being used and can be coupled with compression techniques for boundary value problems \cite{dahmen_compression_2006}, but have the disadvantage, that the number of eigenmodes to be computed must be known in advance. The pivoted Cholesky decomposition \cite{harbrecht_efficient_2015} focuses on approximating the discretized random fields with sufficiently fast decaying eigenvalues. In this case a truncation of the pivoted Cholesky decomposition of the covariance operator allows for an estimation of the eigenvalues in the post-processing step in $\mathcal O\left(M^2 N\right)$ time, where $M$ is the truncation order of the Cholesky decomposition. One of the advantages of this method is the fact, that the number of eigenmodes required for a certain accuracy of the random field discretization can be estimated in advance.

\begin{table}[H]
\centering
\caption{Minimum memory required for storage of the main system matrix in the solution of the homogeneous Fredholm integral problem of the second kind assuming double-precision floating point arithmetic.}
\label{tab:memoryrequirements}
\resizebox{0.6\textwidth}{!}{%
\begin{tabular}{l | l | l | l | l}
    Number of degrees of freedom  	& $10^3$   	& $10^4$   	& $10^5$   	& $10^6$  \\ \hline
 	Matrix storage								& $8$ MB     & $800$ MB  & $80$ GB		& $8$ TB  \\
\end{tabular}%
}
\end{table}

\B
\subsection{Splines as a basis for random fields}
Splines are piecewise polynomials with increased smoothness across element boundaries compared to classical finite elements. Traditionally, splines have been primarily used as shape functions in computer aided design. More recently, with the introduction of isogeometric analysis \cite{hughes_isogeometric_2005}, splines have become more established as trial functions in finite element analysis. Although isogeometric analysis was originally introduced to improve the interoperability across several stages of the design to analysis process, it has proven its fidelity as an analysis technology. We refer to the monograph \cite{cottrell_isogeometric_2009} and references contained therein for an exposition of isogeometric analysis applied to deterministic problems in structural and fluid mechanics.

More recently, spline based isogeometric analysis has found its way into the stochastic community. Stochastic methods have been proposed to quantify uncertainty due to material randomness in linear elasticity \cite{li_spectral_2018, jahanbin_stochastic_2020}, static analysis of plates \cite{wang_stochastic_2019}, vibrational analysis of shells \cite{li_spectral_2019}, static and dynamic structural analysis of random composite structures \cite{ding_isogeometric_2019} and functionally graded plates \cite{hien_stochastic_2017, li_spectral_2018-1, li_spectral_2019-1}. In \cite{zhang_development_2018} a method is proposed to quantify the effect due to uncertainty in shape. Of these, the methods proposed in \cite{li_spectral_2018, li_spectral_2018-1, li_spectral_2019-1, li_spectral_2019}
use isogeometric analysis within a spectral stochastic finite element framework \cite{ghanem_stochastic_1991}, which is based on a KLE of random fields. The methods in \cite{ding_isogeometric_2019, hien_stochastic_2017, wang_stochastic_2019} use perturbation series of which \cite{wang_stochastic_2019} expands random fields in terms of the KLE. Standard polynomial chaos is used in \cite{zhang_development_2018}, while the methods in \cite{eckert_polynomial_2020}, \cite{jahanbin_stochastic_2020} and \cite{rahman_spline_2020}  discretize the stochastic dimensions in terms of splines. In particular, in \cite{eckert_polynomial_2020} tensor product B-splines are used to expand stochastic variables, \cite{jahanbin_stochastic_2020} proposes a spline-dimensional decomposition (SDD) and \cite{rahman_spline_2020} proposes a spline chaos expansion, thus extending generalized polynomial chaos \cite{xiu_wiener--askey_2002}.

To the best of our knowledge, \cite{arthur_solution_1973} is the first work in which splines have been used to approximate the truncated KLE. In his work the author applies a degenerate kernel approximation based on tensor product spline interpolation at the Greville abscissa. More recently, in the spirit of isogeometric analysis, non-uniform rational B-splines (NURBS) have been used to approximate the KLE using the Galerkin method~\cite{rahman_galerkin_2018} and the collocation method  \cite{jahanbin_isogeometric_2019}. These methods avoid the geometrical errors in the representation of CAD geometry typically made within the classical finite element method. The authors note that the use of splines in the geometry description as well as in discretization of the spatial and stochastic dimensions could enable a ``seamless uncertainty quantification pipeline''.

\begin{figure}[h]
    \centering
    \includegraphics[width=0.7\textwidth]{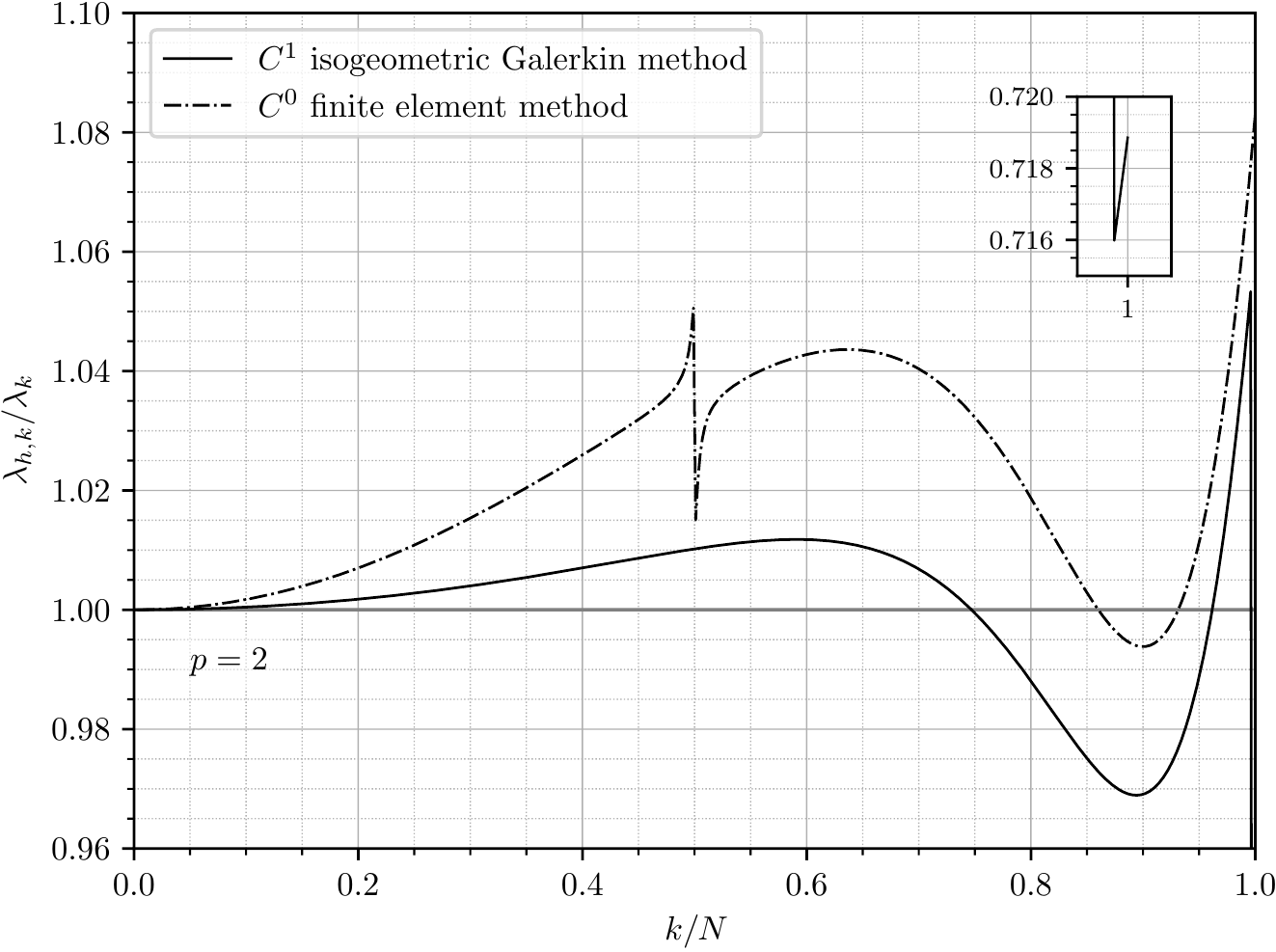}
    \caption{Normalized discrete eigenvalues corresponding to a univariate Fredholm integral eigenvalue problem with an exponential kernel (correlation length is one). Comparison of eigenvalues obtained by $C^1$ quadratic splines to $C^0$ quadratic piecewise polynomials. Both methods employ a standard Galerkin projection based on full Gauss quadrature. The reference solution used to normalize the results is computed by the approach described in Remark \ref{remark:spectral_reference_solution}}
    \label{fig:spectral_props}
\end{figure}

In the context of the present work we would like to highlight the superior spectral approximation properties of smooth splines as compared to classical $C^0$ finite element shape functions. Several studies \cite{cottrell_studies_2007, hughes_duality_2008, hughes_finite_2014, puzyrev_spectral_2018} have investigated the spectral approximation properties of splines in eigenvalue problems corresponding to second and fourth order differential operators and have demonstrated that splines have improved robustness and accuracy per degree of freedom across virtually the entire range of modes. The numerical results for the Fredholm integral eigenvalue problem are no different, as corroborated by the results shown in Figure \ref{fig:spectral_props}. It's precisely these properties that make splines appealing in the representation of random fields by means of the Karhunen-Loève expansion.

\subsection{Contributions}
We present a matrix-free isogeometric Galerkin method for Karhunen-Loève approximation of random fields by splines. Our solution methodology resolves several of the aforementioned computational bottlenecks associated with numerical solution of integral eigenvalue problems and enables solution of large-scale three-dimensional IEVPs on complex $C^0$-conforming multipatch domains. Below we summarize our main contributions.

\subsubsection*{Conversion to a standard eigenvalue problem}
We have chosen a specific trial space of rational spline functions whose Gramian matrix has a Kronecker product structure independent of the geometric mapping. This enables us to perform the backsolve, used to convert the IEVP to standard form, in $\mathcal{O}\left(N \cdot N^{1/d} \right)$ time by utilizing standard linear algebra techniques from \cite{golub_matrix_1996, saad_numerical_2011}.

\subsubsection*{Interpolation based quadrature}
We present an interpolation based quadrature technique designed and optimized specifically for the variational formulation of the Fredholm integral equation. The approach integrates a rich target space of functions with minimal number of quadrature points and outperforms existing competitive techniques in isogeometric analysis, such as quadrature by interpolation and table look-up \cite{mantzaflaris_integration_2015, pan_fast_2020} and weighted quadrature \cite{calabro_fast_2017, hiemstra_fast_2019, sangalli_matrix-free_2018}. The proposed interpolation based quadrature technique is inspired by a similar technique used within linear finite elements in \cite[Chapter 3.1.3]{keese_andreas_numerical_2004} and \cite{khoromskij_application_2009}. Instead, our approximation of the covariance function is based on higher order tensor product spline interpolation and resembles the kernel approximation made in~\cite{arthur_solution_1973}. Besides requiring as few quadrature points as possible, the interpolation based quadrature technique exposes Kronecker structure in the integral equations, reducing computational complexity significantly.

\subsubsection*{Matrix-free solution methodology}
We present a matrix-free solution methodology to avoid explicit storage of the dense system matrix associated with numerical computation of the KLE. The matrix-free solution methodology not only reduces the memory complexity from $\mathcal{O}(N^2)$ to $\mathcal{O}(N)$, but also significantly reduces the solution time. This is achieved by integrating the matrix-free solver with the proposed interpolation based quadrature technique. The latter exposes Kronecker structure in the resulting discrete integral equation, thereby reducing formation costs to $\mathcal{O}\left(N \cdot N^{1/d} \right)$ per iteration, leaving only the dense matrix-vector product that is associated with the Lanczos algorithm that remains $\mathcal{O}(N^2)$ per iteration. The integrative approach to quadrature and matrix-free solution techniques, exploiting Kronecker structure, is inspired by the matrix-free weighted quadrature method proposed recently in \cite{sangalli_matrix-free_2018}.

\subsubsection*{Open source implementation}
We provide an open-source Python implementation of the described techniques that is available for download at \url{https://github.com/m1ka05/tensiga}. All benchmarks have been obtained using this implementation.

\subsection{Outline}
In Section \ref{sec:background_and_notation} we briefly review the necessary mathematical and algorithmic background with regard to Kronecker products, B-splines and NURBS. In Section \ref{sec:discretization} we present the Karhunen-Lo\`eve series expansion of random fields and the weak formulation of the corresponding Fredholm integral eigenvalue problem of the second kind. In Section \ref{sec:efficient_mf_strategy} we introduce our methodology for numerical solution of the truncated KLE. This includes reformulation of the eigenvalue problem to standard form, interpolation based quadrature of the weak form of the Fredholm integral problem, and a matrix-free algorithm with low computational complexity and minimal memory requirements that is embarrassingly parallelizable. The computational complexity is described in more detail in Section \ref{sec:complexity}, where we compare our method with usual formation and assembly techniques used for standard Galerkin methods from the literature. Finally, in Section \ref{sec:examples}, we present a one-dimensional numerical study and several three-dimensional high-order numerical examples. A conclusion and an outlook with recommendations for future work are given in Section \ref{sec:conclusion}.

\section{Background and notation}
\label{sec:background_and_notation}
This section introduces some of the machinery that is used throughout the paper. The presented solution methodology for the Fredholm integral equation relies heavily on the properties of Kronecker products in combination with multidimensional tensor contraction \cite{golub_matrix_1996}. We briefly review the main properties used in this work and illustrate their use in the Kronecker matrix-vector product. The Kronecker structure of the involved matrices is a direct consequence of the chosen tensor product spline function spaces. We briefly introduce B-splines as a basis for polynomial splines and Non-Uniform Rational B-splines (NURBS) for smooth geometrical mappings. For additional details we refer the reader to standard reference books~\cite{cottrell_isogeometric_2009, piegl_nurbs_1995}.

\subsection{Evaluation of the computational cost of an algorithm} \label{sec:operation_counts}
The computational cost of the algorithms discussed in this work are evaluated in terms of \emph{floating point operations per second} (\emph{flops}). A single \emph{flop} represents the amount of work required to preform one floating point addition, subtraction, multiplication or division \cite{golub_matrix_1996}. Although the number of flops does not provide a complete assessment of the efficiency of an algorithm, it is widely used in the literature. Indeed, many other considerations such as cache-line efficiency and number of memory allocations can have a large impact on the performance of an algorithm. Typically, we are interested in the leading terms that dominate the computational cost of an algorithm and record the performance in terms of an order-of-magnitude estimate of the number of flops, written in Big-Oh notation as $\mathcal{O}(\cdot)$.

\subsection{Kronecker products and tensor contraction}
Let $\mat{A} \in \mathbb{R}^{m \times n}, \; \mat{B} \in \mathbb{R}^{p \times q}$ and $\mat{C} \in \mathbb{R}^{s \times t}$ denote real valued matrices. The Kronecker product $ \mat{A} \otimes \mat{B} \in \mathbb{R}^{m \cdot p \times n \cdot q}$ is a matrix defined as
\begin{align}
\mat{A} \otimes \mat{B} :=
	\begin{bmatrix}
		A_{11} \mat{B} 	&	\cdots 		& A_{1n} \mat{B} \\
				\vdots 			& 					& \vdots 				\\
		A_{m1} \mat{B} 	&	\cdots 		& A_{mn} \mat{B}
	\end{bmatrix}
\end{align}
Kronecker products satisfy the following properties
\begin{subequations}
\begin{align}
	\left(\mat{A} \otimes \mat{B}\right) \otimes \mat{C} &= \mat{A} \otimes \left(\mat{B} \otimes \mat{C}\right)  &(\text{associativity}) \label{eq:kron_assoc}\\
	\left( \mat{A} \otimes \mat{B} \right) \left( \mat{C} \otimes \mat{D} \right) &=
		\left(\mat{A}\mat{C} \right) \otimes  \left(\mat{B} \mat{D} \right) &(\text{mixed product property})											\label{eq:kron_mixedprod}\\
	\left( \mat{A} \otimes \mat{B} \right)^{-1} &=  \mat{A}^{-1} \otimes  \mat{B}^{-1}  &(\text{inverse of a Kronecker product})	\label{eq:kron_inv}	\\
	\left( \mat{A} \otimes \mat{B} \right)^{\top} &=  \mat{A}^{\top} \otimes  \mat{B}^{\top}  &(\text{transpose of a Kronecker product})	\label{eq:kron_transp}
\end{align}
\end{subequations}

Let $\mat{X} \in \mathbb{R}^{n_1 \times \cdots \times n_d}$, $\mat{Y} \in \mathbb{R}^{m_1 \times \cdots \times m_d}$ denote two $d$-dimensional arrays. Vectorization of $\mat{X}$ is a linear operation that maps $\mat{X}$ to a vector $\mathrm{vec}(\mat{X}) \in  \mathbb{R}^{n_1 \cdot \ldots \cdot n_d}$ with entries
\begin{align}
\mathrm{vec}(\mat{X})_i := X_{i_1 \ldots i_d}, \qquad \text{where }
i = i_1+ (i_2-1) n_1 + (i_3-1) n_1 \cdot n_2 + \ldots + (i_d-1) n_1 \cdot  \ldots \cdot n_{d-1}.
\end{align}
One recurring theme in this paper involving Kronecker matrices is efficient matrix vector multiplication. Let $\mat{D}_{k} \in \mathbb{R}^{m_k \times n_k}$ denote a set of $d$ matrices $\{ D_{i_k j_k}, \; k=1,\ldots,d \}$, $i_k = 1,\ldots,m_k$ and $j_k = 1,\ldots,n_k$. The matrix-vector product
\begin{subequations}
\begin{align}
	\mathrm{vec} \left(\mat{Y} \right) &= \Big(\mat{D}_d \otimes \cdots \otimes \mat{D}_1 \Big) \mathrm{vec} \left(\mat{X} \right)
	& \mathcal O (M \cdot N) \; \text{flops}
\end{align}
can be written as a tensor contraction instead
\begin{align}
	Y_{i_1 \cdots i_d} &= \sum_{j_1 \cdots j_d} D_{i_1 j_1} \cdots D_{i_d j_d} X_{j_1 \cdots j_d}
	& \mathcal O (\max \left(N \cdot  m_1,  \; n_d \cdot M \right)) \; \text{flops}
\end{align}
\end{subequations}
Here $N = n_1 \cdot \ldots \cdot n_d$ and $M = m_1 \cdot \ldots \cdot m_d$. The second approach scales nearly linearly with matrix size and significantly outperforms standard matrix vector multiplication which scales quadratically with the matrix size. In practice, highly optimized linear tensor algebra libraries can be used to perform the tensor contraction such as the tensor algebra compiler (TACO)~\cite{kjolstad_tensor_2017}. Our Python implementation uses Numpy's reshaping and matrix-matrix product routines, which call low-level BLAS routines. The implemented reshapes do not require any expensive and unnecessary data copies.

\subsection{B-splines}
Consider a $\paradim$-dimensional parametric domain $\pdom = [0,1]^\paradim \subset \mathbb R^\paradim$ with local coordinates $\px = (\px_1, \ldots, \px_\paradim)$. Let $\List{ \ubspline{i_k}{p_k}(\pxk{k}), \; i_k=1, \ldots , n_k}$ denote the univariate B-spline basis of polynomial degree $p_k$ and dimension~$n_k$, corresponding to the $k$th parametric coordinate $\pxk{k}$. We consider multivariate B-splines as tensor products of univariate B-splines
\begin{equation}
	\bspline{i} (\px) = \prod_{k=1}^\paradim \ubspline{i_k}{p_k}(\pxk{k}), \quad
	\midx{i} :=(i_1, \ldots, i_\paradim).
\end{equation}
Here $\midx i\in \mathcal I$ is a multi-index in the set $\mathcal I := \{ (i_1, \ldots, i_\paradim) : 1 \leq i_k \leq n_k \}$. The collection of all multivariate B-spline basis functions spans the space
\begin{equation}
	\bsspace{h} := \mathrm{span}\left\{ \bspline{i}(\px) \right\}_{\midx{i} \in \mathcal{I}}.
\end{equation}
It is important to note that splines allow for increased continuity between polynomial elements as compared to classical $C^0$-continuous finite element basis functions. This turns out to have significant impact on the spectral accuracy of the Galerkin method. This is evidenced by several studies \cite{cottrell_studies_2007,hughes_duality_2008,hughes_finite_2014,puzyrev_spectral_2018} and will be discussed in some detail in this work.

\subsection{Geometrical mapping} \label{sec:geomap}
Let $F : \pdom \rightarrow \dom$ map a point $\px$ from the parametric domain $\pdom$ to a point $\x$ in the physical domain $\dom$. We assume that the map $F$ and its inverse are smooth such that the Jacobian matrix $\left[\mathrm{D}\geomap{\px}\right]_{ij} := \frac{\partial F_i}{\partial \pxk{j}}$ and its inverse are well-defined. In this work $F$ is represented as a linear combination of Non-Uniform Rational B-splines (NURBS). NURBS are rational functions of B-splines that enable representation of common engineering shapes with conic sections, which cannot be represented by polynomial B-splines~\cite{piegl_nurbs_1995}. The discretization method presented in this work makes heavy use of tensor product properties of the involved function spaces. Since NURBS do not have a tensor product structure, we use them only to represent the geometry and do not consider them as a basis for the function spaces.

\section{Isogeometric Galerkin discretization of the Karhunen-Loève series expansion}
\label{sec:discretization}
The Karhunen-Lo\`eve series expansion (KLE) decomposes a stochastic process or field into an infinite linear combination of $L^2$-orthogonal functions and uncorrelated stochastic random variables. In this section we present the probability theory underlying the KLE and discuss its discretization by means of the Galerkin method.

\subsection{Karhunen-Loève expansion of random fields \label{sec:KLE}}
Consider a complete probability space $(\stochdom, \sigfield, \pmeasure)$. Here $\stochdom$ denotes a sample set of random events, $\sigfield$~is the $\sigma$-algebra of Borel subsets of $\stochdom$ and $\pmeasure$ is a probability measure $\pmeasure\, : \,\sigfield\rightarrow [0,1]$. A random field $\randfield(\cdot ,  \w) \; : \; \stochdom \mapsto L^2(\dom)$ on a bounded domain $\dom \in \mathbb{R}^d$ is a collection of deterministic functions of $\x \in \dom$, called realizations, that are indexed by events $\w \in \stochdom$. A subset of realizations $ \randfield(\cdot ,  \stochdom_s), \; \stochdom_s \in \sigfield$, has a probability of occurrence of $\pmeasure(\stochdom_s)$.

Let $\meanval{\cdot}$ denote the expectation operator corresponding to the probability measure $\pmeasure$. Assuming $\randfield \in L^2(\dom \times \stochdom)$ its first and second order moments exist and are given by
\begin{subequations}
\begin{align}
	\mu(\x) &:= \meanval{\randfield(\x, \w)}\quad \text{and} \label{eq:mean}	\\
\covfun{\x}{\xp} &:= \meanval{(\alpha(\x, \w) - \mu(\x))(\alpha(\xp, \w) - \mu(\xp))} \label{eq:kernel}.
\end{align}
\end{subequations}
Here $\mu$ is called the mean or expected value of $\randfield$ over all possible realizations, and $\Gamma : \dom \times \dom \rightarrow \mathbb R$ is called its \emph{covariance function} or \emph{kernel}. By definition, the kernel is bounded, symmetric and positive semi-definite~\cite{ghanem_stochastic_1991}. Because the kernel is square integrable, that is, $\Gamma \in L^2(\dom \times \dom)$, it is in fact a Hilbert-Schmidt kernel, see \cite{lang_real_2012}.

A random field is \emph{stationary} or \emph{homogeneous} if its statistical properties do not vary as a function of the position $\x \in \dom$. This implies that the covariance function can be written as a function of the difference $\x - \xp$. Furthermore, for \emph{isotropic} random fields the statistical properties are invariant under rotations, which means the covariance is a function of Euclidean distance $\| \x - \xp \|_2$.

\begin{remark} \label{remark:geodistance}
Although, the Euclidean distance is widely used in the literature its use is not always justified. In general, the geodesic distance, i.e. the shortest distance between points $\x$ and $\xp$ along all paths contained in $\dom$, is the true measure of distance. The Euclidean distance can vary significantly from the geodesic distance especially if the correlation length is relatively large and the domain is non-convex. The geodesic distance is, however, difficult and expensive to compute, which explains its non-use. In this work we also use the Euclidean distance measure and assume its choice is a reasonable one in the context of the applied numerical benchmark problems.
\end{remark}

\begin{figure}
	\centering
	\subfloat[exponential kernel]{\includegraphics[trim = 0cm  0cm 10cm 0cm,clip,width=0.49\textwidth]{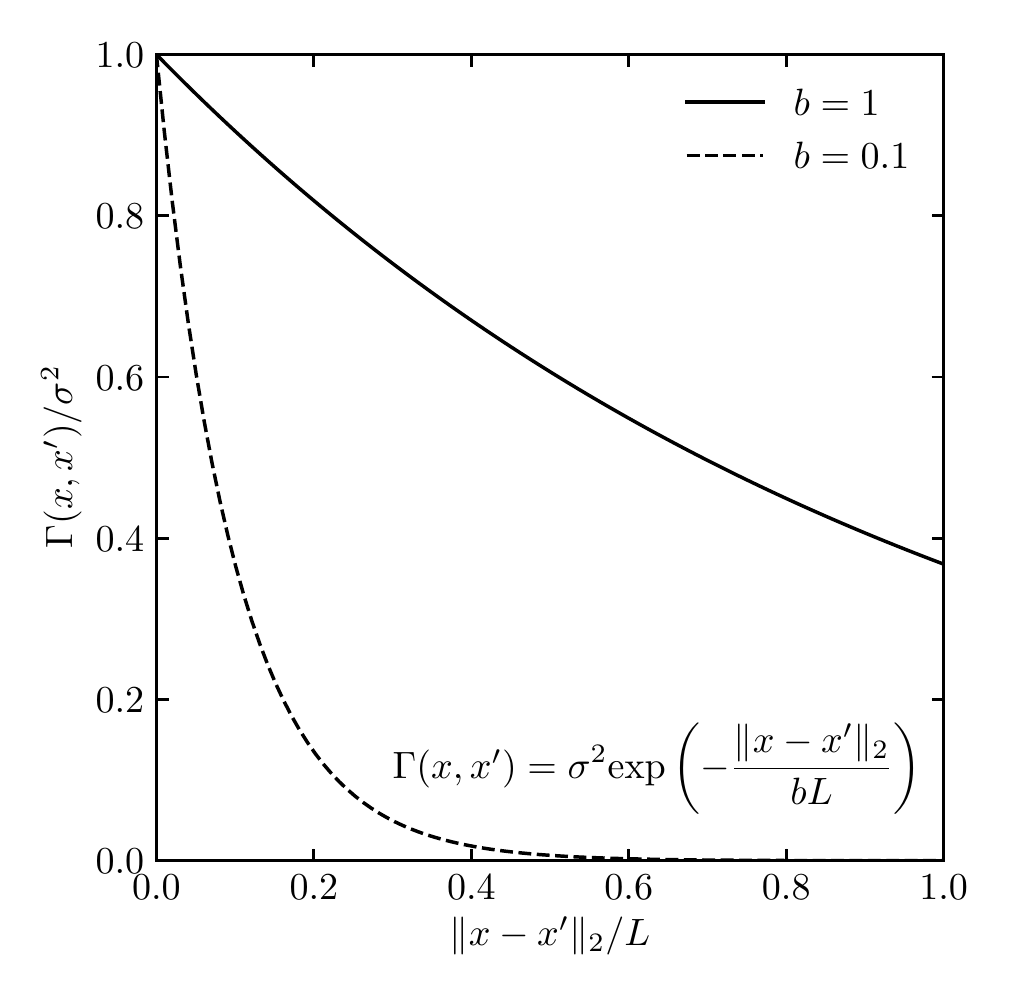}}
	\subfloat[squared exponential kernel]{\includegraphics[trim = 10cm  0cm 0cm 0cm,clip,width=0.49\textwidth]{images/covs.pdf}}
	\caption{The exponential and squared exponential (Gaussian) covariance functions for different correlation lengths with $b=\{0.1, 1.0\}$. Note the difference in the the continuity of both kernels at $\x=\xp$. The exponential kernel is $C^0$, while the square exponential kernel is $C^\infty$  at  $\x=\xp$.}
	\label{fig:kernels}
\end{figure}

Figure \ref{fig:kernels} shows two common examples of covariance functions that correspond to stationary isotropic random fields: the \emph{exponential} and the \emph{Gaussian} or \emph{squared exponential kernel}. Important parameters that influence the locality of these correlation functions are the variance $\sigma^2$ and correlation length $bL$. Here $L$ denotes a characteristic length and $b$ is a dimensionless factor.

The KLE of a random field $\randfield(\cdot, \w) $ requires the solution of an integral eigenvalue problem. Consider the linear operator
\begin{align}
	\label{eq:hilbert_schmidt}
	T \; :\; L^2(\dom) \mapsto L^2(\dom), \quad \left(T  \eigfun \right) (\x) := \int_{\dom} \covfun{\x}{\xp}\eigfun(\xp) \dint{\xp}.
\end{align}
The operator $T$ is compact. In fact, $T$ is a Hilbert-Schmidt operator, since the covariance function is a Hilbert-Schmidt kernel. Furthermore, since the covariance function is positive semi-definite and symmetric~\cite{ghanem_stochastic_1991}, $T$~is a self-adjoint positive semi-definite linear operator. The eigenfunctions $\{\eigfun_i \}_{i\in \mathbb N}$ of $T$ are defined by the homogeneous Fredholm integral eigenvalue problem of the second kind,
\begin{equation}
	\label{eq:fredholm}
	T\eigfun_i = \eigval_i \eigfun_i, \quad \eigfun_i \in L^2(\dom) \text{ for } i \in \mathbb{N}.
\end{equation}
The important properties of the eigenpairs are (1) the normalized eigenfunctions $\{\eigfun_i \}_{i\in \mathbb N}$ are orthonormal, that is, $(\eigfun_i, \eigfun_j)_{L^2(\dom)} = \delta_{ij}$, and thus form a basis for $L^2(\dom)$; and (2) the corresponding eigenvalues form a sequence $\eigval_1 \geq \eigval_2 \geq \ldots \geq 0$, which in general decays with increasing mode number.

Because $\Gamma$ in \eqref{eq:kernel} is symmetric and positive semi-definite, it possesses the spectral decomposition \cite{courant_methods_1989,mercer_functions_1909}%
\begin{equation}
	\label{eq:spectral_decomposition}
	\covfun{\x}{\xp} = \sum\limits_{i=1}^{\infty}
	\eigval_i \eigfun_i(\x)\eigfun_i(\xp).
\end{equation}

With these definitions, the KLE of a random field $\randfield \in L^2(\dom \times \stochdom)$ is defined by the following series~\cite{karhunen_uber_1947}%
\begin{align}
	\randfield(\x, \w) = \mu(\x) + \sum\limits_{i=1}^{\infty} \sqrt{\eigval_i} \eigfun_i(\x) \randvar_i(\w), \quad \text{where}\quad
	\randvar_i(\w) := \frac{1}{\sqrt{\eigval_i}} \int_{\dom} \left(\randfield(\x, \w) - \mu(\x) \right) \eigfun_{i}(\x)\dint{\x}.
\end{align}
While $\{ \eigfun_i \}_{i\in \mathbb N} $ are pairwise $L^2$-orthogonal on $\dom$, the $\{  \randvar_i \}_{i\in \mathbb N}$ are pairwise uncorrelated zero-mean random variables \cite{ghanem_stochastic_1991}. For this reason the KL expansion is sometimes said to be bi-orthogonal.

\subsection{Truncated Karhunen-Loève expansion}
In order to represent a random field in a discrete numerical computation it is necessary to discretize the continuous probability space. This can be achieved by truncating the KLE after $M$ terms and thus reducing the dimension of the stochastic space to $M$ uncorrelated random variables
\begin{equation}
	\drandfield{M}(\x, \w) = \mu(\x) + \sum\limits_{i=1}^{M} \sqrt{\eigval_i} \eigfun_i(\x) \randvar_i(\w).
\end{equation}
The mean of a random field is not affected by the discretization. The variance of the discretization on the other hand can be derived from the spectral decomposition in equation~\eqref{eq:spectral_decomposition}
\begin{equation}
	\meanval{(\drandfield{M}(\x, \w) - \mu(\x))^2} = \sum\limits_{i=1}^M \eigval_i \eigfun_i^2(\x).
\end{equation}
The variance of the discretized random field converges uniformly in $\dom$ and in $L^2(\stochdom,\sigfield, \pmeasure)$ towards the true variance \cite{rahman_galerkin_2018}
\begin{equation}
	\lim_{M\to\infty} \sum\limits_{i=1}^{M} \eigval_i \eigfun^2_i(\x) = \covfun{\x}{\x}.
\end{equation}
Furthermore, it can be shown that the KLE is optimal with respect to the global mean-squared error among all series expansions of truncation order $M$ \cite{ghanem_stochastic_1991}.

\subsection{Variational formulation}
The variational formulation or weak form of the integral eigenvalue problem introduced in equation \eqref{eq:fredholm} states

\textit{Find $\{\eigval, \eigfun \} \in \mathbb{R}_0^{+} \times L^2(\dom)$ such that}
\begin{equation}
	\int_{\dom} \left(\int_{\dom^\prime} \covfun{\x}{\xp} \eigfun(\xp) \dint{\xp} - \eigval \eigfun(\x)\right) \psi(\x)\dint{\x} = 0 \quad\forall \psi \in L^2(\dom).
	\label{eq:weakform}
\end{equation}
Confining the solution to the finite-dimensional subspace $\mathcal S_h \subset L^2(\dom)$ we obtain the discrete variational formulation

\textit{Find $\{\eigval, \eigfun \} \in \mathbb{R}_0^{+} \times \mathcal S_h$ such that}
\begin{equation}
	\int_{\dom} \left(\int_{\dom^\prime} \covfun{\x}{\xp} \eigfun_h(\xp)\dint{\xp} - \eigval_h \eigfun_h(\x)\right) \psi_h(\x)\dint{\x} = 0 \quad\forall \psi_h \in \mathcal S_h.
\end{equation}
This is the Galerkin method for the homogeneous Fredholm integral eigenvalue problem of the second kind~\cite{atkinson_numerical_1997,rahman_galerkin_2018}. Within the trial space under consideration, the Galerkin method produces the best $L^2$ approximation of the analytical modes. The resulting discrete modes preserve exactly the $L^2$ orthogonality property of the analytical mode-shapes. Furthermore, it can be shown that a variational treatment using the Galerkin method leads to eigenvalues that converge monotonically, under mesh refinement, towards the true eigenvalues~\cite{ghanem_stochastic_1991}.

\subsection{Choice of the trial space}
The choice of the trial space $\mathcal S_h$ provides some freedom in the design of the Galerkin method. The recently proposed isogeometric Galerkin method for the KLE of random fields uses NURBS for the test and trial spaces \cite{rahman_galerkin_2018}. This choice is motivated by the fact that the geometrical mapping is defined using NURBS and it is natural to remain within the isoparametric paradigm. This method shares the same technical challenges as all classical Galerkin methods applied to this class of problems~\cite{allaix_karhunen-loeve_2013, rahman_galerkin_2018}: the formation and assembly costs, which have a time complexity of $\mathcal O\left(N_e^2\cdot p^{3d}\right)$, as well as the storage requirements, which have space complexity of $\mathcal O\left(N^2\right)$, become quickly intractable with increasing number of elements $N_e$, polynomial degree $p$, dimension $d$ and number of degrees of freedom~$N$. A practical Galerkin method must address these difficulties in the design of the method.

We abandon the isoparametric concept and choose a different space to represent the finite-dimensional solution. Our choice offers multiple computational advantages without sacrificing higher-order accuracy and robustness. We define the trial space for the Galerkin method as
\begin{equation}
	\mathcal S_h := \mathrm{span} \left\{\frac{\bspline{i}(\px)} {\sqrt{ \detgeomap{\px}   }} \right\}_{\midx{i} \in \mathcal I}.
\label{eq:basis}
\end{equation}
Because the geometrical mapping $F$ is smooth and invertible the Jacobian determinant is never singular, that is, $\detgeomap{\px} > 0$ for all $\px \in \pdom$. Importantly, the functions are linearly independent due to linear independence of B-splines. In general, however, these basis functions will not form a partition of unity. Instead, the characterizing property is that products of these functions are integral preserving, that is, they transform as volume forms
\begin{equation*}
	\int_{\dom} \frac{\bspline{i}(\px)}{\sqrt{ \detgeomap{\px}}} \frac{\bspline{j}(\px)}{\sqrt{ \detgeomap{\px}}} \dint x = \int_{\dom} 	\frac{\bspline{i}(\px) \bspline{j}(\px)}{\detgeomap{\px}} \dint x = \int_{\pdom} \bspline{i}(\px) \bspline{j}(\px) \dint \px.
\end{equation*}

\subsection{Matrix formulation}
After substituting the desired subspace for the test and trial functions and performing minor algebraic manipulations, the discretized Galerkin method results in a generalized algebraic eigenvalue problem
\begin{equation}
	\matA \eigvec = \eigval_h \matB \eigvec,
\end{equation}
where the system matrices are formed by evaluating
\begin{align}
	\matA_{\midx i\midx j}
		&= \int_\pdom \int_{\pdom^\prime} \covfun{\x(\px)}{\x(\pxp))} \frac{\bspline{i}(\px)}{\sqrt{ \detgeomap{\px} }} \frac{\bspline{j}(\pxp)}{\sqrt{ \detgeomap{\pxp} }} \detgeomap{\px} \,\detgeomap{\pxp} \dint{\pxp}\ddint{\px} 	\nonumber 	\\ \label{eq:GalA_mat}
		&= \int_\pdom \int_{\pdom^\prime} \covfun{\x(\px)}{\x(\pxp))} \bspline{i}(\px) \bspline{j}(\pxp) \sqrt{\detgeomap{\px} \detgeomap{\pxp}} \dint{\pxp}\ddint{\xp}
\end{align}
and
\begin{align}
	\matB_{\midx{i}\midx{j}}
		&= \int_\pdom \frac{\bspline{i}(\px)} {\sqrt{ \detgeomap{\px} }} \frac{\bspline{j}(\px)} {\sqrt{ \detgeomap{\px} }}\, \detgeomap{\px} \dint{\px}	\nonumber \\
		&= \int_\pdom \bspline{i}(\px) \bspline{j}(\px) \dint{\px}.
\end{align}
As a result of the chosen solution space, the mass matrix $\matB$ has a Kronecker structure and can be decomposed into $k=1,\ldots,\paradim$ univariate mass matrices
\begin{equation}
	\matB_k := \nomatB_{i_k j_k} = \int_0^1 \ubspline{i_k}{p_k}(\pxk{k}) \ubspline{j_k}{p_k}(\pxk{k}) \dint{\pxk{k}}, \qquad i_k,j_k = 1,\ldots,n_k.
\label{eq:univariate_mass_matrix}
\end{equation}
The system mass matrix $\matB$ can be then written as
\begin{equation}
	\matB = \matB_{\paradim} \otimes \cdots \otimes \matB_{1}.
\label{eq:tensor_structure_B}
\end{equation}
Instead of computing and storing the matrix $ \matB$, we precompute and store the matrices $\matB_{k}, \; k=1, \ldots ,\paradim$. Furthermore, it is the Kronecker structure that allows us to inexpensively reformulate the generalized eigenvalue problem to a standard algebraic eigenvalue problem.

\begin{remark} \label{remark:massmatrix}
In practice $\matB_k$ in \eqref{eq:univariate_mass_matrix} is computed exactly up to machine precision using Gauss-Legendre numerical quadrature with $p+1$ quadrature points per element, where $p$ is the polynomial degree in component direction $k$. For alternative ways of computing integrals of piecewise polynomial products see \cite{vermeulen_integrating_1992}. Because the domain of integration is one-dimensional the formation and assembly costs of $\mathcal O(n_{e} p^3)$ as well as the storage costs of $\mathcal O(pn)$ bytes are negligible compared to the total solver costs. Here $n_{e}$ is the number of univariate elements and $n$ is the univariate number of degrees of freedom in component direction $k$.
\end{remark}

\subsection{Discretization of multipatch geometries}
\label{sec:multipatch}

Single patch domains can only represent simple geometric models. In general multipatch domains need to be considered. Because the integral operator in \eqref{eq:hilbert_schmidt} does not involve derivatives, it does not require any smoothness from the finite element spaces. Hence, the techniques presented in this paper are valid for multi-patch domains. The examples in this paper involve $C^0$-conforming multi-patch domains. Alternatively, non-conforming $C^{-1}$ discretizations are also possible with minor change.

\section{Efficient matrix-free solution strategy}
\label{sec:efficient_mf_strategy}
There are two major challenges when applying the Galerkin method to  discretize the homogeneous Fredholm integral eigenvalue problem in \eqref{eq:fredholm}. Firstly, the variational formulation requires integration over a $2d$-dimensional domain to evaluate the matrix entries in $\matA$. This leads to formation and assembly costs with complexity $\mathcal O(N_e^2 (p+1)^{3d})$, where $N_e$ is the global number of elements, $p$ the polynomial degree and $d$ the spatial dimension. Secondly, because the matrix is dense, $\matA$ requires insurmountable memory storage for any practical problem of interest. Several techniques have been presented in the literature in order to deal with these challenges, for example by approximation with low-rank matrices like the hierarchical matrices \cite{allaix_karhunen-loeve_2013,hackbusch_hierarchical_2005,khoromskij_application_2009} or by using Fast Multipole Methods \cite{schwab_karhunenloeve_2006}. In this work we present a combination of four techniques to deal with the aforementioned challenges:
\begin{enumerate}
\item Reformulation of the generalized eigenvalue problem into an equivalent standard eigenvalue problem;
\item Interpolation based quadrature for variational formulations of integral equations;
\item Efficient formation of finite element arrays based on Kronecker matrix-vector product;
\item Formulation of a matrix-free and parallel matrix-vector product for the Lanczos algorithm.
\end{enumerate}\B
The reformulation into a standard algebraic eigenvalue problem significantly reduces the computational cost and simultaneously improves conditioning. By exploiting the Kronecker structure of the right-hand-side mass matrix we can perform this reformulation with negligible overhead. The proposed non-standard quadrature technique that we call \emph{interpolation based integration} is tailored for variational formulations of integral equations. The technique is optimal in the sense that few quadrature points are required while integrating a rich space of tensor product functions on the $2d$-dimensional domain $\dom \times \dom$. Importantly, the technique lends itself to multidimensional tensor contraction due to the Kronecker structure of the involved matrices. This significantly speeds up the evaluation of integrals over high-dimensional domains and scales favorably with polynomial degree. Finally, all techniques are combined within a matrix-free evaluation scheme that is embarrassingly parallel and requires minimal memory storage. The formation and assembly costs of our approach are negligible compared to the remaining solver costs of the Lanczos eigenvalue solver, which is $\mathcal O( \tilde{N}^2 \cdot N_{\text{iter}} / N_{\text{thread}})$. Here $\tilde{N}$ is the global number of degrees of freedom of the interpolation space, $N_{\text{iter}}$ is the number of iterations of the eigensolver and $N_{\text{thread}}$ is the number of simultaneous processes. In the following we discuss each of the proposed techniques in more detail.

\subsection{Reformulation into a standard algebraic eigenvalue problem}
Let us consider a Cholesky factorization of the mass matrix $\matB = \mat L \transposed{\mat L}$ and define a linear transformation of the eigenvectors $\eigvec^\prime := \transposed{\mat L}\eigvec$. The generalized eigenvalue problem can then be rewritten (see \cite[Chapter 9.2.2]{saad_numerical_2011}) as a standard eigenvalue problem with unchanged eigenvalues corresponding to new eigenvectors $\eigvec^\prime$
\begin{equation}
	\matA^\prime \eigvec^\prime= \lambda_h \eigvec^\prime, \quad \text{where}\quad \matA^\prime :=\mat L^{-1} \matA \invtransposed{\mat L}.
\end{equation}
It is expected that the new system matrix $\matA^\prime$ has improved conditioning compared to $\matA$. The kernel  is positive-definite and symmetric. In practice, it often quickly tends to zero for increasing distance $\lVert\x-\xp\rVert_2$. In the limiting case, where $\covfun{\x}{\xp}\rightarrow \delta(\x,\xp)$, the system matrix $\matA\rightarrow \matB$ and hence the preconditioner would be ideal.

Although improved conditioning is beneficial, the main reason for the chosen transformation is efficiency. Solution of a standard algebraic eigenvalue problem is much less expensive than solution of a generalized eigenvalue problem. The transformation itself is inexpensive. Using the Kronecker structure of $\matB$ and the properties \eqref{eq:kron_mixedprod} and \eqref{eq:kron_transp} we may write
\begin{align}
	\matB 	&= \matB_d \otimes \ldots \otimes \matB_1  \nonumber \\
			 	&= \mat L_{d} \transposed{\mat L_{d}} \otimes \cdots \otimes
					\mat L_{1} \transposed{\mat L_{1}} \nonumber \\
				&= \left(\mat L_{d} \otimes \cdots \otimes \mat L_1\right) \transposed{\left(\mat L_{d} \otimes \cdots\otimes \mat L_{1}\right)} = \mat{L} \transposed{\mat L}.
\label{eq:cholesky}
\end{align}
Here $\mat L_k \mat {L_k}^\top, \; k=1, \ldots d$, denote the Cholesky factorizations corresponding to the univariate mass matrices~$\matB_k $. Hence, instead of performing the Cholesky factorization for the complete system matrix $\matB \in \mathbb{R}^{N \times N}$, which is the standard procedure in most solvers for generalized algebraic eigenvalue problems, we merely need the Cholesky factorizations for $\matB_k \in \mathbb{R}^{n_k \times n_k}, \; k = 1,\ldots,d$.

The factorization is precomputed once before using it in the eigenvalue solver. The associated computational cost is reduced from $\mathcal{O}\left(N^3\right)$ to $\mathcal{O}\left(n^3\right)$ flops,  where $n = \max(n_1, ..., n_d)$ and $N = n_1 \cdot  ... \cdot n_d$. Subsequently, the cost of applying the factorization in a single iteration of the eigenvalue solver is reduced from $\mathcal{O}\left(N^2\right)$ to $\mathcal{O}\left(n \cdot N\right)$. Besides a reduction in computational cost, this approach significantly reduces the required memory storage. \B

\subsection{Interpolation based quadrature for integral equations} \label{sec:GalerkinImprv}
One of the most straightforward ways of improving efficiency is to design quadrature rules that require fewer evaluation points. This is especially true for the variational formulation of the Fredholm integral equation which requires numerical integration over a $2d$-dimensional domain. In practice, accurate and efficient quadrature rules are designed as follows. First, one chooses a space of functions $\mathbb{T}$, called the target space for numerical quadrature, whose elements should be exactly integrated by the new quadrature rule. If this space is in some sense rich enough then the error due to the quadrature can be bounded by the discretization error, which is needed to show optimal rates of convergence of the numerical method, see~\cite{Hiemstra:2017}. The next objective is to find a quadrature rule that requires as few points as possible to integrate all functions in $\mathbb{T}$.

We present a non-standard quadrature technique that generalizes the approach to quadrature presented for linear finite elements in \cite{keese_andreas_numerical_2004, khoromskij_application_2009} to higher order splines. \B Our approach is tailored toward evaluating integrals found in variational formulations of integral equations and achieves a very low number of evaluation points while integrating exactly a rich space of functions. The target space $\mathbb{T}$ is chosen such that the quadrature scheme exactly evaluates the integral
\begin{align}
	\int_{\pdom}  \int_{\pdom^\prime} \tilde{G}(\px, \pxp) \bspline{i}(\px) \bspline{j}(\pxp) \dint{\pxp} \ddint{\px}, \qquad
	\tilde{G} \in \tbsspace{h}(\pdom) \otimes \tbsspace{h}(\pdom^\prime).
	\label{eq:target_space_quadrature}
\end{align}
using $\tilde{N}^2$ points. Here $\tbsspace{h}$ is another $d$-dimensional spline space that can be chosen independently of $\bsspace{h}$ and $\tilde{N}$ is its dimension. In practice this space can be chosen to fit well with the integrand in the variational formulation of the integral equation. This provides additional flexibility to the quadrature scheme.

Because $ \tilde{G} \in \tbsspace{h}(\pdom) \otimes \tbsspace{h}(\pdom^\prime) $ is a real-valued $2d$-variate spline function it can be expanded in terms of B-spline basis functions and real-valued coefficients $\{\tilde{G}_{\midx k\midx l}\}_{\midx k,\midx l \in \tilde{\mathcal I}}$ as
\begin{equation*}
	\tilde{G}(\px, \pxp) := \sum\limits_{\midx k,\midx l\in\tilde{\mathcal I}}  \tilde{G}_{\midx k\midx l} \tbspline{\midx{k}}{\px} \tbspline{\midx{l}}{\pxp}
	\quad
	\text{with}\quad \midx{k}, \midx{l}\in\tilde {\mathcal I} := \{ (i_1, \ldots, i_\paradim) : 1 \leq i_k \leq \tilde{n}_k \}.
\end{equation*}
Comparing the multidimensional integrand in \eqref{eq:GalA_mat} with the one in \eqref{eq:target_space_quadrature} we may conclude that the degrees of freedom $\{\tilde{G}_{\midx k\midx l}\}_{\midx k,\midx l \in \tilde{\mathcal I}}$ should be chosen such that $\tilde{G}$ is a good approximation of the function $G \; : \; \pdom \times \pdom \mapsto \mathbb{R}^+$ defined as
\begin{equation}
	G(\px,\pxp) := \pcovfun{\px}{\pxp} \B\sqrt{\detgeomap{\px}\detgeomap{\pxp}}.
\label{eq:integrand}
\end{equation}
Here $\pcovfun{\px}{\pxp}$ is the pull-back of the kernel $\covfun{\x}{\xp}$ from the physical to the parametric space using the geometrical mapping $F$.

\begin{remark}
\label{remark_on_smoothness}
To maintain optimal accuracy in numerical quadrature, locally, the smoothness of the interpolation space should not exceed smoothness of the integrand in \eqref{eq:integrand}. Indeed, Figure \ref{fig:l2error_exp_kref} shows that error convergence due to quadrature of an exponential kernel, which features reduced regularity at $x=x'$, is suboptimal. Similarly, the reduced regularity due to the Jacobian determinant term needs to be taken into account when choosing the interpolation space. Being set in the framework of splines and isogeometric analysis, our method provides sufficient flexibility in enforcing required smoothness.
\end{remark}

The approximation $\tilde{G}$ can be estimated in different ways. We follow a similar approach to the degenerate kernel approximation in \cite{arthur_solution_1973} and choose to collocate $G$ at the Greville abscissa~\cite{de_boor_practical_1978}. This approach is both simple and combines high order accuracy with a minimal number of evaluation points. We note that related ideas based on quasi-interpolation have been presented in \cite{calabro_efficient_2018, falini_adaptive_2019, giannelli_study_2019} for formation and assembly of boundary integral equations and in \cite{mantzaflaris_integration_2015, pan_fast_2020} for matrix assembly in Galerkin discretization of PDEs.

Let  $\tilde \matinterpmat = \tinterpmat{\midx i}{\midx j} := \tbspline{j}{\px_{\midx i}}$ denote the $d$-variate spline collocation matrix evaluated at the Greville abscissa $\px_{\midx i} \in \pdom, \; \midx i \in \tilde{\mathcal{I}}$. The interpolation problem states

\textit{Find $\{\tilde{G}_{\midx k\midx l}\}_{\midx k,\midx l \in \tilde{\mathcal I}} $ such that}%
\begin{equation}
	\sum\limits_{\midx k,\midx l\in\tilde{\mathcal I}}  \tilde{G}_{\midx k\midx l} \tbspline{\midx{k}}{\px_{\midx i}} \tbspline{\midx{l}}{\pxp_\midx j}
	= G(\px_{\midx i}, \pxp_{\midx j}) \quad \forall \, \midx{i}, \midx{j} \in \tilde{\mathcal I}.
\end{equation}
This is equivalent to the matrix problem $\mat G = \tmatinterpmat \mat{\tilde G} \tmatinterpmat^\top$. Hence, the matrix of coefficients can be computed as $ \mat{\tilde{ G}} =  \tmatinterpmat^{-1} \mat{G} \tmatinterpmat^{-\top}$. The computational cost of the interpolation can be significantly reduced from $\mathcal O(\tilde{N}^3)$ to  $\mathcal O(\tilde{n} \cdot \tilde{N})$ flops, where $\tilde{n} = \max{(\tilde{n}_1, ..., \tilde{n}_d)}$, \B by exploiting the Kronecker structure of $\tmatinterpmat$. We decompose $\tmatinterpmat$ into $\paradim$ univariate collocation matrices $\tmatinterpmat_k$, $k=1,\ldots,\paradim$, and use property \eqref{eq:kron_inv} to write its inverse as
\begin{equation}
	\tmatinterpmat^{-1} = \tmatinterpmat^{-1}_\paradim \otimes \cdots \otimes \tmatinterpmat^{-1}_1 \quad \text{with} \quad
	\tmatinterpmat_k :=  \tinterpmat{i_k}{j_k} =  \tubspline{j_k}{\tilde{p}_k}{\pxk{i_k}}.
\label{eq:interpolation_matrix}
\end{equation}
In practice we compute $d$ $\mat{L} \mat{U}$ factorizations, each corresponding to a univariate matrix $\tmatinterpmat_k$, to apply the inverse of $\tmatinterpmat$ to a vector. Note, that this approach is similar to the approach we took in \eqref{eq:cholesky} for the Cholesky factorization of $\matB$.

\subsection{Matrix formation}
By substituting $G$ in \eqref{eq:GalA_mat} with $\tilde{G}$ we can approximate matrix $\matA$ by a matrix $\tmatA$ with entries
\begin{align*}
\label{eq:approxAsysmat}
\tnomatA_{\midx i\midx j}
	&:= \int_{\pdom}  \int_{\pdom^\prime} \tilde{G}(\px, \pxp) \bspline{i}(\px) \bspline{j}(\pxp) \dint{\pxp} \ddint{\px}  \\
	&= \sum\limits_{\midx k,\midx l\in \tilde {\mathcal I}} \tilde{G}_{\midx k\midx l} \int_\pdom  \int_{\pdom^\prime}
				\tbspline{\midx{k}}{\px} \tbspline{\midx{l}}{\pxp} \bspline{i}(\px) \bspline{j}(\pxp) \dint{\pxp} \ddint{\px}    \\
	&= \sum\limits_{\midx k,\midx l\in \tilde {\mathcal I}} \tilde{G}_{\midx k\midx l}
				\int_\pdom  \tbspline{\midx{k}}{\px} \bspline{i}(\px) \dint{\px}
				\int_{\pdom^\prime} \tbspline{\midx{l}}{\pxp} \bspline{j}(\pxp) \dint{\pxp}.
\end{align*}
Hence, using the tensor product structure of the interpolation space we have separated the $2d$ dimensional integral into a product of two $d$ dimensional integrals.  In matrix notation we may write $\tmatA = \mat{M}^\top \mat{\tilde{G}} \mat{M}$, or
\begin{equation}
	\tnomatA_{\midx i\midx j}  = \sum\limits_{\midx k,\midx l\in \tilde {\mathcal I}} \tilde{G}_{\midx k\midx l} M_{\midx k \midx i} M_{\midx l\midx j}.
\label{eq:ApproxGalA_mat}
\end{equation}
Here $\mat{M} := M_{\midx i \midx j}$ is a mass matrix
\begin{equation}
	M_{\midx i \midx j} = \int_\pdom \tbspline{i}{\px} \bspline{j}(\px) \dint{\px}.
\end{equation}
Similarly, as we did for matrix $\matB$ in \eqref{eq:tensor_structure_B}, we can exploit the Kronecker structure and decompose $\mat M$ into~$\paradim$ univariate mass matrices $\mat M_k := M_{i_k j_k}$
\begin{equation}
	\mat M = \mat M_\paradim \otimes \cdots \otimes \mat M_1 \quad \text{with} \quad
	M_{i_k j_k} = \int_0^1 \tubspline{i_k}{\tilde{p}_k}{\pxk{k}} \ubspline{j_k}{p_k}(\pxk{k}) \dint{\pxk{k}}.
\label{eq:univariate_mass_matrix2}
\end{equation}
As in the case of \B $\matB_k$, $k=1,...,d$, these univariate matrices are computed up to machine precision as discussed in  Remark \ref{remark:massmatrix}. The approximation error $\matA - \tmatA$ is entirely due to the interpolation error $G - \tilde{G}$. Hence, accurate approximation of $G$ should result in an accurate approximation of $\matA$.

\subsection{Matrix-free solution strategy}
The interpolation based quadrature technique introduced in the previous section involves computation of the matrix of coefficients $\mat{\tilde{G}} := \tilde{G}_{\midx k \midx l}$. This matrix is dense and has $\tilde{N}^{2}$ entries. Consequently, storage of $\mat{\tilde{G}}$ is just as inconvenient as storing $\tmatA$ and becomes quickly intractable with problem size. In this section we propose a matrix-free evaluation of the matrix-vector product $\mat{v}^\prime \mapsto \tmatA \mat{v}^\prime$ that does not require explicit access to matrix $ \mat{\tilde{G}} $ or $\tmatA$.

\subsubsection{Basic setup}
We have the following standard algebraic eigenvalue problem
\begin{subequations}
\begin{equation}
	\tmatA^\prime \mat{v}^\prime = \lambda_h \mat{v}^\prime.
\end{equation}
Here, the system matrix $\tmatA^\prime$ can be written as
\begin{equation}
	\tmatA^\prime = \mat{L}^{-1}  \mat{M}^\top  \tmatinterpmat^{-1} \mat{J} \mat{\Gamma} \mat{J} \tmatinterpmat^{-\top} \mat{M} \mat{L}^{-\top},
\end{equation}
where $\mat{J}$ is a diagonal matrix with diagonal entries given by the square roots of Jacobian determinants evaluated at the Greville abscissa, and $\matinterpmat$, $\mat{M}$ and $\mat{L}$ are all Kronecker product matrices. Consequently, a matrix-vector product with any of these matrices can be performed close to linear time complexity. The matrix vector product $\mat{v}^\prime \mapsto \tmatA^\prime \mat{v}^\prime$ can be subdivided into the following operations
\begin{align}
	\mat{\Gamma} &:=\pcovfun{\px_{ \midx{k}}}{\pxp_{ \midx{l}}}						& (\text{Evaluation of the kernel at the Greville abscissa}) \\
	\mat{G} &= \mat{J} \mat{\Gamma} \mat{J}												& (\text{Scaling of the kernel}) 					\\
	\mat{\tilde{G}} &= \tmatinterpmat^{-1} \mat{G} \tmatinterpmat^{-\top}							& (\text{Interpolation of the scaled kernel})	\\
	\tmatA &= \mat{M}^{\top} \mat{\tilde{G}} \mat{M}						& (\text{Evaluation of the integrals})								\\
	\tmatA^\prime &= \mat{L}^{-1} \tmatA \mat{L}^{-\top}	& (\text{Application of the preconditioner})
\end{align}
\end{subequations}
In the following subsection we present a matrix-free matrix-vector product that incorporates each of the above steps. Except for the diagonal matrix $\mat{J}$, none of the above matrices are stored explicitly. Only the corresponding univariate matrices are stored and used in the Kronecker products, while the entries of $\mat{\Gamma}$ are computed on the fly.

\subsubsection{Matrix-free algorithm}
Let ${\hat\Gamma}_{k_1\ldots k_d l_1\ldots l_d}:=\pcovfun{\px_{1,k_1},\ldots,\px_{d,k_d}}{\pxp_{1,l_1},\ldots,\pxp_{d,l_d}} \in \mathbb R^{\tilde{n}_1\times \cdots \times \tilde{n}_d \times \tilde{n}_1\times \cdots \times \tilde{n}_d}$ denote the function values of the kernel evaluated at the tensor product grid of the Greville abscissa in $\pdom \times \pdom$. The proposed evaluation order of the matrix-free matrix-vector product is summarized in Algorithm \ref{alg:mf_eval}.

\begin{algorithm}[H]
\textbf{Input}:
	$v_{i_1\ldots i_d} \in \mathbb{R}^{n_1 \times \cdots \times n_d}$,
	$J_{l_1\ldots l_d} \in \mathbb{R}^{\tilde{n}_1 \times \cdots \times \tilde{n}_d}$,
	$\tilde B_{i_k j_k} \in \mathbb{R}^{\tilde{n}_k \times \tilde{n}_k}$ and
	$M_{l_k j_k} \in \mathbb{R}^{\tilde{n}_k \times n_k}$\\
\textbf{Output}: $v^\prime_{i_1\dots i_d} \in \mathbb{R}^{n_1\times \cdots \times n_d}$
\begin{algorithmic}[1]
	\State $V_{j_1\ldots j_d} 		\leftarrow L^{-1}_{ i_1  j_1}\cdots L^{-1}_{ i_d  j_d} v_{ i_1\ldots i_d }$	\Comment{Preconditioning from right}
	\State $X_{k_1\ldots k_d} 	\leftarrow M_{k_1 j_1}\cdots M_{k_d j_d} V_{j_1\ldots j_d}$
	\State $Y_{l_1\ldots l_d} 	\leftarrow \invinterpmat{ k_1}{l_1 }\cdots \invinterpmat{ k_d}{l_d } X_{k_1\ldots k_d}$	\Comment{Use $\mat L \mat U$-factorization of $\mat{\tilde B}_k, \; k=1,...,d$}
	\State $Y^\prime_{l_1\ldots l_d} \leftarrow J_{l_1\ldots l_d} \odot Y_{l_1\ldots l_d}$
	\State $Z^\prime_{k_1\ldots k_d} \leftarrow {\hat \Gamma}_{ k_1\ldots k_d l_1\ldots l_d} Y^\prime_{l_1\ldots l_d}$		\Comment{Evaluate in parallel without forming $\mat{\Gamma}$}
	\State $Z_{k_1\ldots k_d} \leftarrow J_{k_1\ldots k_d} \odot Z^\prime_{k_1\ldots k_d}$
	\State $Y_{j_1\ldots j_d} \leftarrow \invinterpmat{ j_1 }{  k_1 }\cdots \invinterpmat{ j_d }{ k_d } Z_{k_1\ldots k_d}$ 	\Comment{Use $\mat L \mat U$-factorization of $\mat{\tilde B}_k, \; k=1,...,d$}
	\State $V_{l_1\ldots l_d} \leftarrow M_{j_1 l_1}\cdots M_{j_d l_d} Y_{j_1\ldots j_d}$
	\State $v_{i_1\ldots i_d}^\prime \leftarrow L^{-1}_{i_1 l_1}\cdots L^{-1}_{i_d l_d} V_{l_1\ldots l_d}$
  \Comment{Preconditioning from left}
\caption{Matrix-free evaluation of the matrix-vector product $\mat{v}^\prime \mapsto \mat{\tilde{A}}^\prime \mat{v}^\prime $}\label{alg:mf_eval}
\end{algorithmic}
\end{algorithm}

The matrix-free matrix vector product $\mat{v}^\prime \mapsto \tmatA^\prime \mat{v}^\prime$ is evaluated in nine separate stages. Stage one applies back-substitution of the upper triangular matrix $ \mat{L}^{\top}$ and exploits its Kronecker structure. Stage two applies a matrix vector product with matrix $\mat{M}$ and again exploits its Kronecker structure. Stage three applies back-substitution using the factorization of the interpolation matrix $\mat{\tilde{B}}$. Again, Kronecker structure is essential to reduce both the space and time complexity of the back-substitution. In stage four the coefficient vector is element-wise multiplied by the square root of the Jacobian determinant evaluated at the Greville abscissa. Here, element-wise multiplication is denoted by the $\odot$ symbol. Stage five dominates the computational cost of Algorithm \ref{alg:mf_eval}. This stage represents a dense matrix-vector product. To perform this step without explicitly forming matrix $\mat{\Gamma}$ we compute its entries on the fly, \B one row at a time\B. We compute products of the coefficient vector with several rows of $\mat{\Gamma}$ in parallel. Stages six to nine are equivalent to stages four to one, due to the symmetry of the operator.

Due to the iterative solution process, the matrix-vector product needs to be evaluated at each iteration. The number of iterations is dependent on the number of required eigenmodes, the conditioning of the algebraic eigenvalue problem and the efficiency of the eigensolver. In this work we use the standard implicitly restarted Lanczos method  \cite{golub_matrix_1996}.

\section{Computational complexity analysis}
\label{sec:complexity}
The goal of a time-complexity analysis is to obtain an estimate of the computational cost that scales linearly with time. This cost is expressed in terms of certain parameters that depend on the problem size, the dimension and the polynomial degree. For this purpose, let us introduce the following notation:

\begin{table}[H]
	\begin{tabular}{ll}
		$n$ 							& number of degrees of freedom of the trial space in one component direction;		\\
		$\tilde{n}$					& number of degrees of freedom of the interpolation space in one component direction;		\\
		$N:=n^d$ 					& total number of degrees of freedom of the trial space;												\\
		$\tilde N:=\tilde n^d$ & total number of degrees of freedom of the interpolation space;										\\
		$N_e$ 						& total number of spatial elements in the trial space;													\\
		$N_q$ 						& number of quadrature points in a standard quadrature loop.										\\
		$N_{\text{iter}}$		& number of iterations of the matrix-free algorithm;													\\
		$N_{\text{thread}}$	& number of simultaneous shared memory processes in the matrix-vector product.
	\end{tabular}
\end{table}

\subsection{Standard finite element procedures}
In the following we present the computational complexity of standard finite element procedures for higher-order finite elements. We use the tensor product structure of the high-dimensional space $\pdom \times \pdom$ to minimize the involved computations. The general setting for this analysis is (1) $\pdom \times \pdom$ has $N_e^2$ elements; and (2) we assume a quadrature rule $Q(f) := \sum_{k=1}^{N_q} w_k f(x_k)$, with $1 \leq N_q  \leq (p+1)^d$, to integrate the products on every $d$-dimensional element $\msquare^d$ in $\pdom$.

The leading term in formation and assembly is determined by the cost of forming the element matrices. Consider the following element matrix
\begin{align*}
	\nomatA^e_{\midx i\midx j} &= \int_{\msquare^d} \bspline{\midx i}(\px) \int_{\msquare^d} \covfun{\px}{\pxp} \bspline{\midx j}(\pxp) \dint{\pxp} \ddint{\px} \\ 		&\approx  \sum_{k=1}^{N_q}  w_k \bspline{\midx i}(\px_k) \sum_{l=1}^{N_q}  w_l \covfun{\px_k}{\pxp_l} \bspline{\midx j}(\pxp_l) \\
		&= \sum_{k=1}^{N_q}  C_{\midx ik}  {D}_{k\midx j} \quad \text{with}\quad {D}_{k\midx j} = \sum_{l=1}^{N_q}  w_l  \covfun{\px_k}{\px_l} \bspline{\midx j}(\px_l)
\end{align*}
with $\midx i,\midx j \in\mathcal I$. We see that $\nomatA^e_{\midx i\midx j}$ can be formed by the matrix product of matrices $\mat{C} \in \mathbb{R}^{(p+1)^d \times  N_q}$ and $\mat{D} \in \mathbb{R}^{N_q \times (p+1)^d}$. This matrix product costs $\mathcal O\left(N_q (p+1)^{2d}\right)$. The formation of $C_{\midx ik}$ is negligible. The formation of $D_{k\midx j}$ on the other hand is $\mathcal O\left(N^2_q (p+1)^{d}\right)$. Since $N_q \leq (p+1)^d$ the leading term is $N_q (p+1)^{2d}$. Hence, the total cost of forming one element matrix is $\mathcal O\left(N_q (p+1)^{2d}\right)$. In total we have to integrate over all $N_e^2$ multidimensional elements of $\pdom \times \pdom$. With that, the total cost of forming $\matA$ is $\mathcal O\left(N_e^2 N_q (p+1)^{2d}\right)$. Using a Gauss-Legendre quadrature rule with $(p+1)$ quadrature points in every coordinate direction gives in total $N_q = (p+1)^d$ quadrature points, and we can expect a leading cost proportional to $\mathcal O\left(N_e^2 (p+1)^{3d}\right)$.

\subsection{Finite element procedures employing sum factorization}
Estimates presented in the previous subsection hold for classical \emph{hp} finite element procedures employing a standard quadrature loop. Next we discuss the complexity of finite element procedures that employ sum factorization instead of a standard quadrature loop. Sum factorization significantly speeds up the element array formation by exploiting the tensorial structure of both the finite element basis and the used quadrature rules \cite{antolin_efficient_2015, bressan_sum_2019, orszag_spectral_1980,tino_eibner_fast_2005,vos_h_2010}. It's worth noting that, due to the structure of the integral operator, the sum factorization technique looks somewhat different than is standard in the \emph{hp} finite element method.

The setting for this analysis is (1)~$\pdom \times \pdom$ has $N_e^2$ rectangular elements; (2) we use a tensor product basis of polynomial degree $p$ on every element; and (3) we use a tensor product of univariate Gauss-Legendre quadrature rules $Q(f) := \sum_{k=1}^{p+1} \omega_k f(x_k)$ to integrate the products on every $d$-dimensional element $\msquare^d$ in~$\pdom$. Consider the element matrix
\begin{align*}
	\nomatA^e_{\midx i\midx j}
			&= \int_{\msquare^d} \bspline{\midx i}(\x) \int_{\msquare^d} \covfun{\x}{\xp} \bspline{\midx j}(\xp) \dint{\xp}\ddint{\x} \\ 															&\approx  \sum_{k_1=1}^{p+1} \ubspline{i_1}{p}(\x_{1,k_1}) \sum_{k_2=1}^{p+1} \ubspline{i_2}{p}(\x_{2,k_2})  \cdots \sum_{k_d=1}^{p+1} \ubspline{i_d}{p}(\x_{d,k_d})  \\
			& \quad \quad \sum_{l_1=1}^{p+1} \ubspline{j_1}{p}(\xp_{1,l_1}) \sum_{l_2=1}^{p+1} \ubspline{j_2}{p}(\xp_{2,l_2}) \cdots \sum_{l_d=1}^{p+1} \covfun{x_{1,k_1},\ldots,x_{d,k_d}}{x^\prime_{1,l_1},\ldots,x^\prime_{d,l_d}} \ubspline{j_d}{p}(x^\prime_{d,l_d})
\end{align*}
	with $\midx i,\midx j \in \mathcal I$. Sum factorization is essentially tensor contraction. The kernel evaluated at the grid of quadrature points is a tensor $\Gamma_{k_1...k_d l_1 ... l_d} \in \mathbb{R}^{(p+1) \times \cdots \times (p+1)}$ and is contracted with matrices $\ubspline{j_z}{p}(x^\prime_{z, l_z}) \in \mathbb{R}^{(p+1) \times (p+1)}$, for $z=1,\ldots,d$, and subsequently with matrices $\ubspline{i_z}{p}(x_{z, k_z}) \in \mathbb{R}^{(p+1) \times (p+1)}$ for $z=1,\ldots,d$. The cost of every contraction is $\mathcal O\left((p+1)^{2d+1}\right)$  flops. In total there are $2d$ such tensor contractions. Hence, the element matrix formation cost for $\matA^e$ is $\mathcal O\left(2d (p+1)^{2d+1}\right)$ flops. With $N^2_e$ elements, the leading cost of forming~$\matA$ is $\mathcal O\left(2d N_e^2 (p+1)^{2d+1}\right)$ flops.

\subsection{Proposed strategy using interpolation based quadrature}
\label{sec:ibq-complexity}
In order to analyze the computational complexity of the proposed solution strategy we must address each stage of the matrix-free matrix-vector product introduced in Algorithm~\ref{alg:mf_eval}. Let us consider the complexity in one iteration of the matrix-free algorithm.

\begin{table}[H]
	\begin{tabular}{ll}
	Stage 1		& has a cost $\mathcal O\left(d n^{d+1}\right)$. \\
	Stage 2		& has a cost depending on the chosen projection space,\\
					& $\quad\text{for } n>\tilde n$ the cost is $\mathcal O\left(d p n^{d}\right)$, \\
					& $\quad\text{for } n=\tilde n$ the cost is $\mathcal O\left(d p n^{d}\right)$, \\
					& $\quad\text{for } n<\tilde n$ the cost is $\mathcal O\left(d p \tilde n^{d}\right)$. \\
	Stage 3		& has a cost $\mathcal O\left(d\tilde n^{d+1}\right)$.\\
	Stage 4		& has a cost $\mathcal O\left(\tilde n^{d}\right)$.\\
	Stage 5		& has a cost $\mathcal O\left(\tilde N^2/N_{\mathrm{thread}}\right)$.
	\end{tabular}
\end{table}
The remaining steps $6,7,8$ and $9$ are equivalent to steps $4,3,2$ and $1$. In step 2 we assume sparse matrix algebra. In dense algebra, $p$ can be replaced by $n$. In steps 1~and~3 we assume that the Cholesky and $\mat L \mat U$ factorizations of the univariate matrices of size $n \times n$~and~$\tilde{n} \times \tilde{n}$, respectively, have been precomputed and are available. Subsequently, the solver costs are $\mathcal O\left(n^2\right)$ and $\mathcal O\left(\tilde{n}^2\right)$ flops, respectively, for each application of the factorization. The cost of a single iteration is typically dominated by step 5, which does not depend on the polynomial degree $p$. Fortunately, this step is embarrassingly parallel. Hence, the time complexity of the matrix free algorithm is $\mathcal O\left(\tilde N^2 N_{\mathrm{iter}} / N_{\mathrm{thread}}\right)$ flops.

\subsection{Storage comparison}
Both matrix-free and non-matrix-free methods need to store the resulting eigenmodes. Storage of the results takes roughly $L \cdot 8N$ bytes, where $L$ is the number of eigenmodes that need to be computed. Additionally, the standard approach that stores the dense left-hand-side system matrix $\matA \in \mathbb R^{N \times N}$ requires storage of $N \times N$ floating point numbers. Using double precision floating point arithmetic the storage requirements are $8 N^2$ bytes. The additional storage of the matrix-free approach scales linearly with problem size, with an asymptotic leading term of roughly $2 \cdot 8 N$ bytes, again using double precision floating point arithmetic. If step 4 of the matrix-free algorithm is performed using shared memory parallelism then one can expect this to increase to $(1+N_{\text{thread}}) \cdot 8 N$ where $N_{\text{thread}} $ is the number of simultaneous processes. Consequently, the storage cost of the matrix-free approach is typically governed by storage of the results and is thus optimal. Table \ref{tab:storage1} and \ref{tab:storage2} summarize the leading terms in storage of the two alternatives.
\begin{table}[h!]
	\centering \caption{The leading terms in storage costs of a method that explicitly stores the main system matrix. Here $L$ refers to the number of eigenmodes that need to be stored. The storage cost is dominated by storage of the system matrix.}
	\resizebox{0.65\textwidth}{!}{
		\begin{tabular}{l | r | r | r | r}
			Number of degrees of freedom  	& $10^3$   				& $10^4$   				& $10^5$   					& $10^6$  				\\ \hline
			Results storage								& $L \cdot 8$ KB   	& $L \cdot 80$ KB  & $L \cdot 800$ KB    & $L \cdot 8$ MB  	\\
			Matrix storage									& $8$ MB    			& $800$ MB   			& $80$ GB      			& $8$ TB
		\end{tabular}}
	\label{tab:storage1}
\end{table}

\begin{table}[h!]
	\centering \caption{The leading terms in storage costs of a matrix-free approach. Here $L$ refers to the number of eigenmodes that need to be stored. The storage cost of the matrix-free method is typically dominated by the storage of the results.}
	\resizebox{0.65\textwidth}{!}{
		\begin{tabular}{l | r | r | r | r}
			Number of degrees of freedom  	& $10^3$   				& $10^4$   				& $10^5$   					& $10^6$  				\\ \hline
			Results storage								& $L \cdot 8$ KB    & $L \cdot 80$ KB  & $L \cdot 800$ KB   	& $L \cdot 8$ MB  	\\
			Matrix-free approach  					& $2 \cdot 8$ KB    & $2 \cdot 80$ KB  & $2 \cdot 800$ KB  	& $2 \cdot 8$ MB
		\end{tabular}}
	\label{tab:storage2}
\end{table}

\section{Numerical results} \label{sec:examples}
In this section we present numerical results that illustrate the accuracy, robustness and computational efficiency of the proposed matrix-free isogeometric Galerkin method. An spectral study of the accuracy is performed in the case of one dimension. This study case gives insight into the accuracy attained by interpolation based quadrature of the covariance function and its affect on approximating the eigenvalue spectra. Next, several three-dimensional $C^0$-conforming multipatch benchmarks illustrate the computational performance attained for a range of polynomial degrees and different refinement strategies of the interpolation as well as the solutions space. In the first two three-dimensional examples, the error in the discrete linear operator introduced by the interpolation based quadrature is showcased in the $2$- and the Frobenius-matrix-norm. In the last two three-dimensional examples we study the effect of solution space refinement. All computations in the benchmark cases are performed entirely in a \textit{single} process, without taking advantage of parallel execution, neither in the kernel evaluation itself nor in the linear algebra packages behind the implementation. The machine used in these study cases is a laptop equipped with Intel(R) Core(TM) i7-9750H CPU @ 2.60GHz and 2x16 GB of non-ECC DDR4 2666MHz RAM. We further provide a plot showcasing the scalability of the method for one of the examples. The Python implementation in this work relies heavily on the packages Numpy \cite{van_der_walt_numpy_2011} and Scipy~\cite{scipy_10_contributors_scipy_2020} for the linear algebra and solver functionalities. In order to achieve high performance, crucial parts of the code base are just-in-time compiled using the LLVM-based Python-compiler Numba \cite{lam_numba_2015, lattner_llvm_2004}.

\subsection{One-dimensional case study}
Consider a one-dimensional random field defined on the domain $\dom = [0,1] \subset \mathbb R$. We investigate
\begin{itemize}
	\item[(i)]  the relative $L^2(\dom)$ interpolation error of the kernel, $G - \tilde G$, with respect to uniform \emph{h}-refinement enforcing $C^{p-1}$ continuity across element boundaries. We consider the cases where $G$ is the Gaussian and exponential kernel with $b = 0.1$, a characteristic domain length $L = 1$ and a variance $\sigma^2 = 1$.
	\item[(ii)] the normalized spectra corresponding to an exponential kernel with $b=1$.
\end{itemize}

\begin{remark} Normalized spectra corresponding a Gaussian kernel cannot be reliably computed across the full range of eigenvalues because the smallest eigenvalues quickly approach zero up to machine precision.
\end{remark}

\subsubsection*{Kernel approximation}
\begin{figure}%
    \centering
    \subfloat[Gaussian kernel \label{fig:l2error_gau_kref}]{\includegraphics[width=0.75\textwidth]{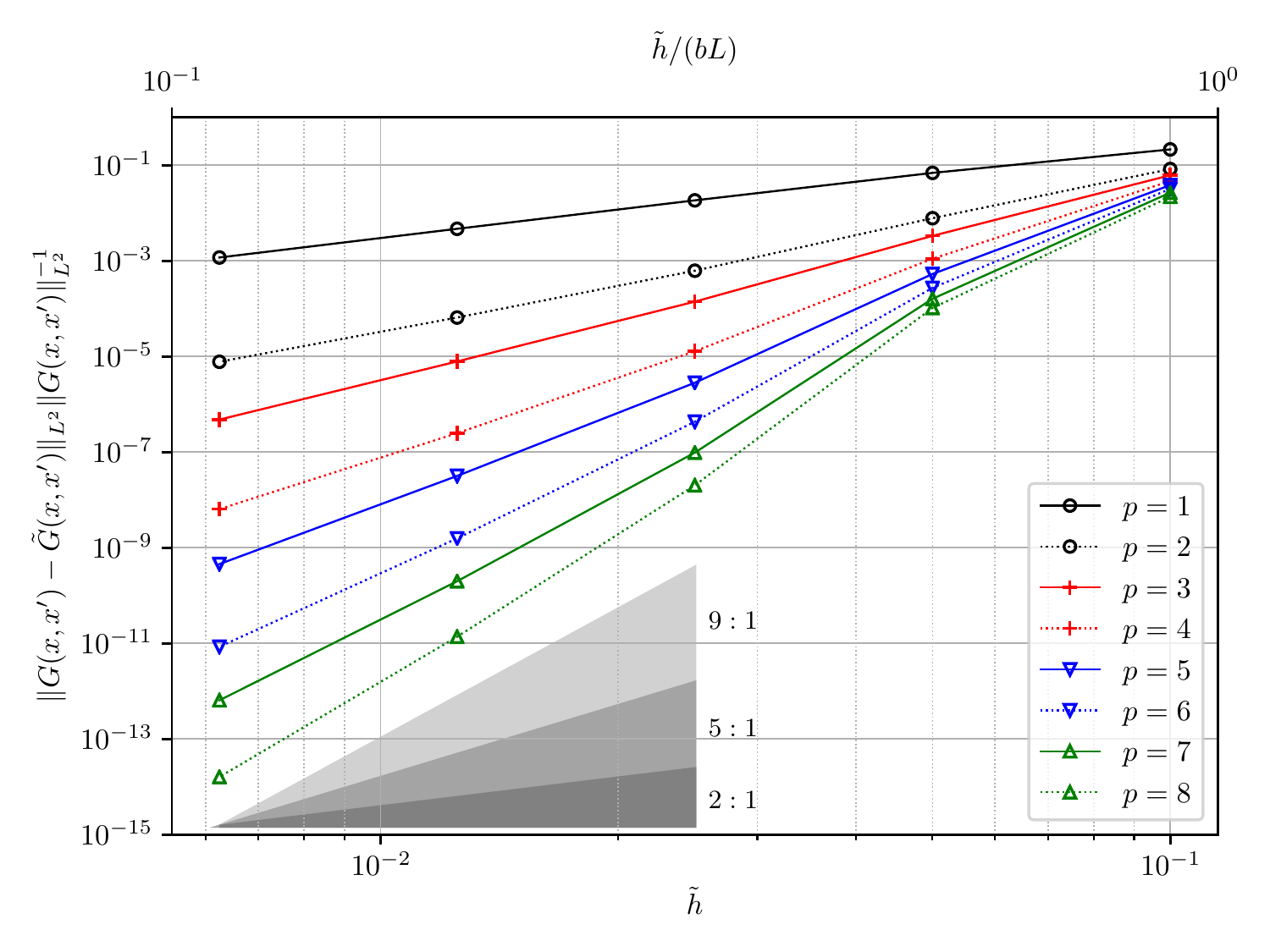}}	\\
    \subfloat[Exponential kernel \label{fig:l2error_exp_kref}]{\includegraphics[width=0.75\textwidth]{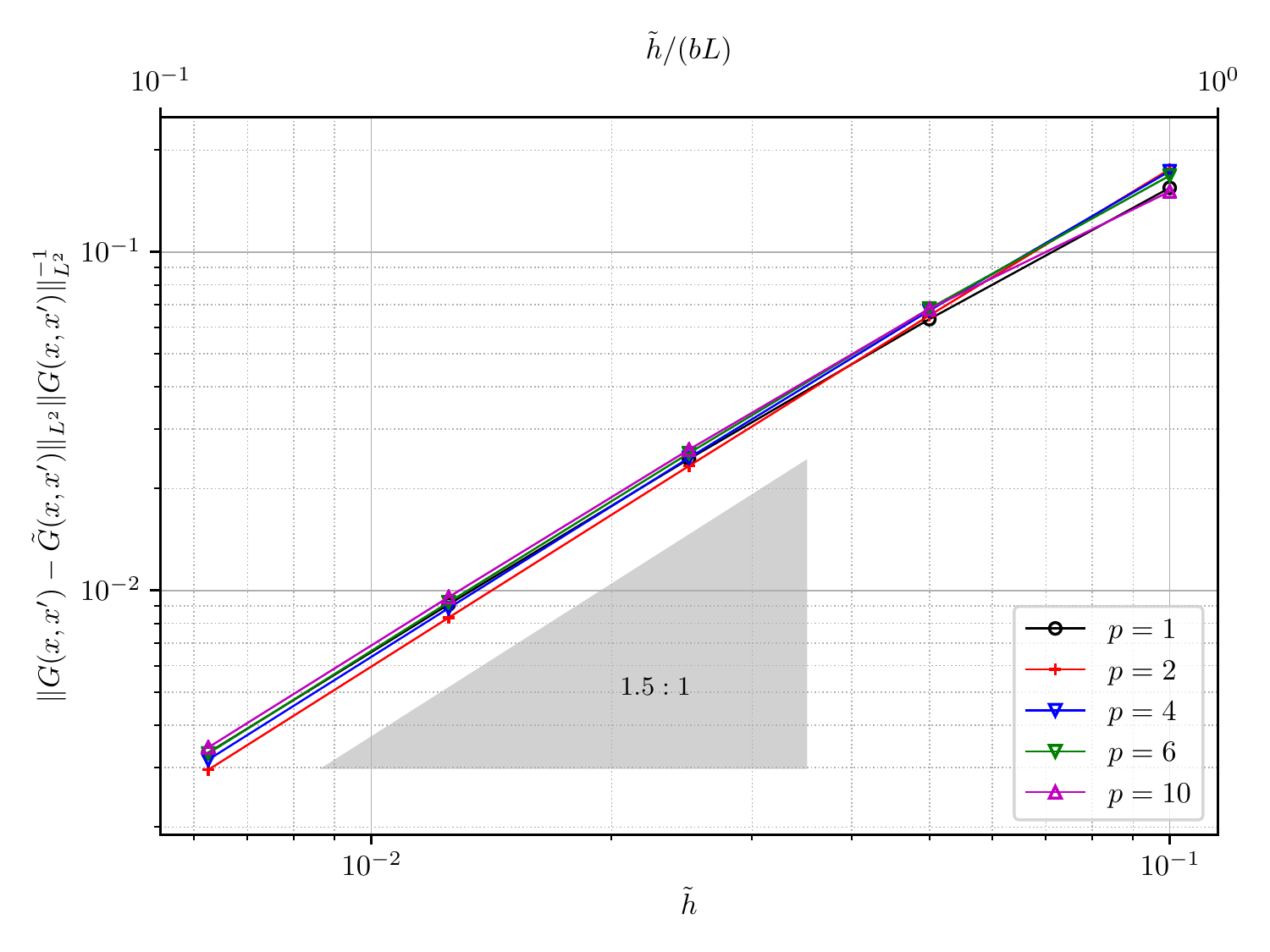}}
	\caption{Normalized $L^2$ interpolation error in the one-dimensional study case for multiple polynomial degrees. The error is given with respect to the mesh size of the interpolation space (bottom axis), as well as the mesh size of the interpolation space normalized by the correlation length (top axis). The convergence rates are approximately $\mathcal{O} (\tilde{h}^{p+1} )$ in (a) and $\mathcal{O} ( \tilde{h}^{3/2} )$ in (b).} %
\end{figure}

Figure \ref{fig:l2error_gau_kref} shows the convergence towards the Gaussian kernel for polynomial degrees 1 through 8. It seems that the even degrees $p=\{2,4,6,8\}$ perform relatively better than the preceding odd degrees $p=\{1,3,5,7\}$. It is evident that higher-order interpolation of a smooth kernel leads to a higher-order convergence rate in the approximation. The smooth interpolation space is not as suitable for approximation of kernels that have low regularity. Approximation of the exponential kernel in Figure \ref{fig:l2error_exp_kref}, which is $C^0$ along $\x=\xp$, shows that higher-order continuity of the basis does not lead to an increased convergence rate. This behavior is in agreement with convergence estimates for spline approximation of arbitrary smoothness \cite{sande_explicit_2020}.

\subsubsection*{Spectral approximation}
\begin{figure}
    \centering
    \includegraphics[width=0.75\textwidth]{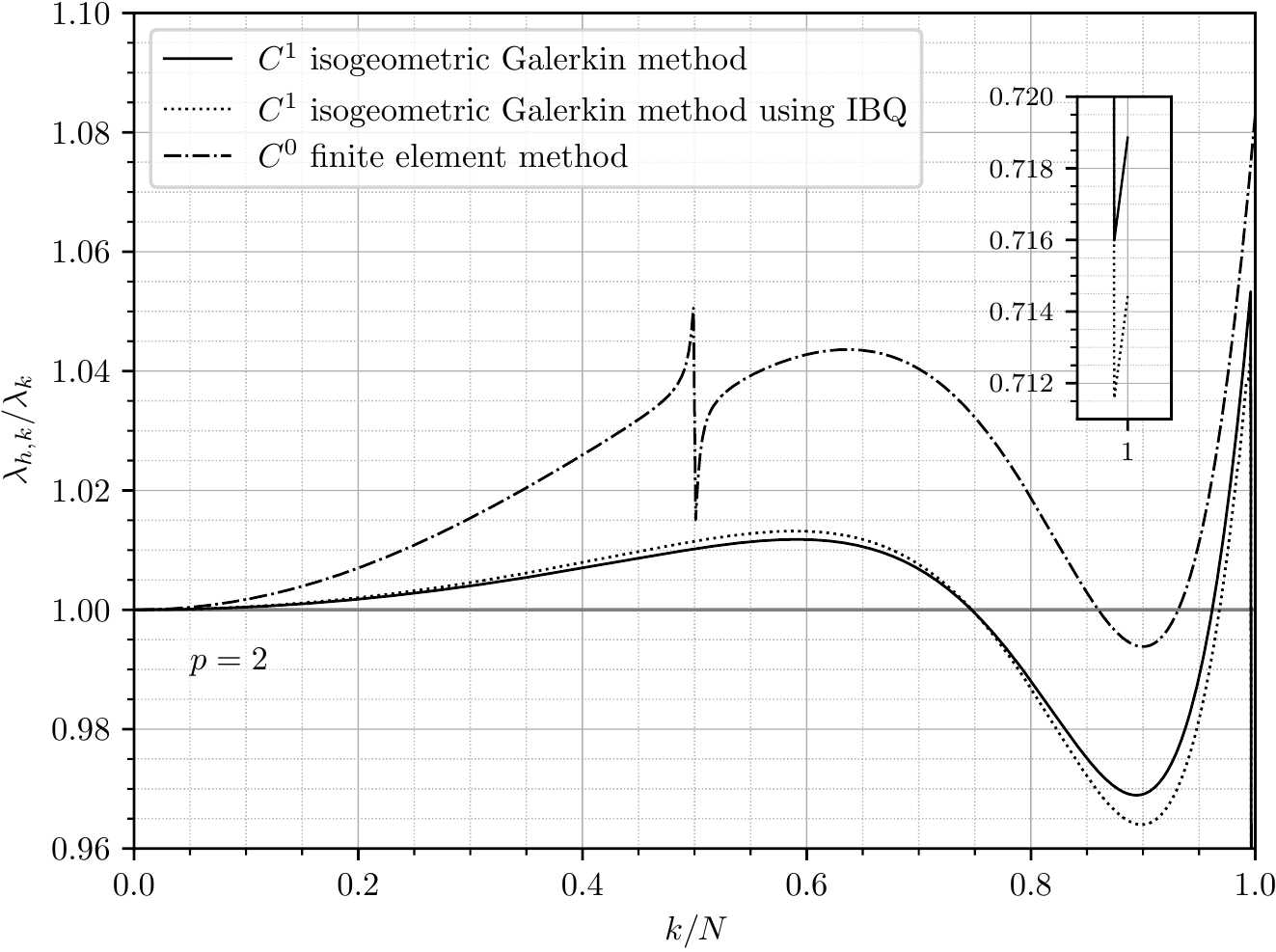} %
    \caption{Ratio of the approximated eigenvalues to the reference eigenvalues over a full spectrum of 501 eigenmodes in  the one-dimensional study case with an exponential kernel and a correlation length equal to the domain length.}%
    \label{fig:spectrum_ibq}
\end{figure}

Although the proposed method, in its current form, is best suited for smooth kernels like the Gaussian kernel, excellent approximation of the eigenvalues corresponding to non-smooth kernels is still possible. Figure \ref{fig:spectrum_ibq} depicts the full spectrum corresponding to the exponential kernel with $b=1$. The proposed Galerkin method using interpolation based quadrature (IBQ) is compared to the isogeometric Galerkin method proposed in \cite{rahman_galerkin_2018} and a classical $C^0$ finite element solution in the case of polynomial degree $p=2$. The interpolation space is set to $\tilde h = 0.005\cdot bL$. The proposed method (IBQ) exhibits the same advantageous characteristics as the standard isogeometric Galerkin method \cite{rahman_galerkin_2018} and exhibits no branching phenomena as in the case of the $C^0$-continuous finite element approximation. Due to their increased continuity across element boundaries, splines achieve a higher accuracy per degree of freedom and an increased robustness as compared with classical $C^0$ finite element methods. These results are in agreement with several other studies that have investigated spectral approximations corresponding to eigenvalue problems in structural mechanics \cite{cottrell_studies_2007,hughes_duality_2008,hughes_finite_2014,puzyrev_spectral_2018}.

\begin{remark} \label{remark:spectral_reference_solution}
The reference solution is computed using a standard isogeometric Galerkin method \cite{rahman_galerkin_2018} with fifty thousand degrees of freedom. The first twenty eigenvalues have been validated up to machine accuracy by the analytical approach described in \cite[Ch. 2.3.3, page 28-35]{ghanem_stochastic_1991}. The analytical computation of these eigenvalues involves solving for roots of a complex equation and is for that reason avoided beyond the first twenty eigenvalues.
\end{remark}

\subsection{Random field with exponential kernel in a three-dimensional half-open cylindrical domain}
In the first three-dimensional example we investigate a random field defined in a half-open cylindrical domain as shown in Figure \ref{fig:ex1_geo}. We consider Gaussian and exponential kernels with a correlation length $bL$ equal to the half of the characteristic length~$L$. The variance of the random field is $\sigma^2 = 1$. We note that the example with an exponential kernel is also studied in \cite{rahman_galerkin_2018}.

\subsubsection*{Example 1--1}

\begin{figure}
    \centering
    \includegraphics[width=0.4\textwidth]{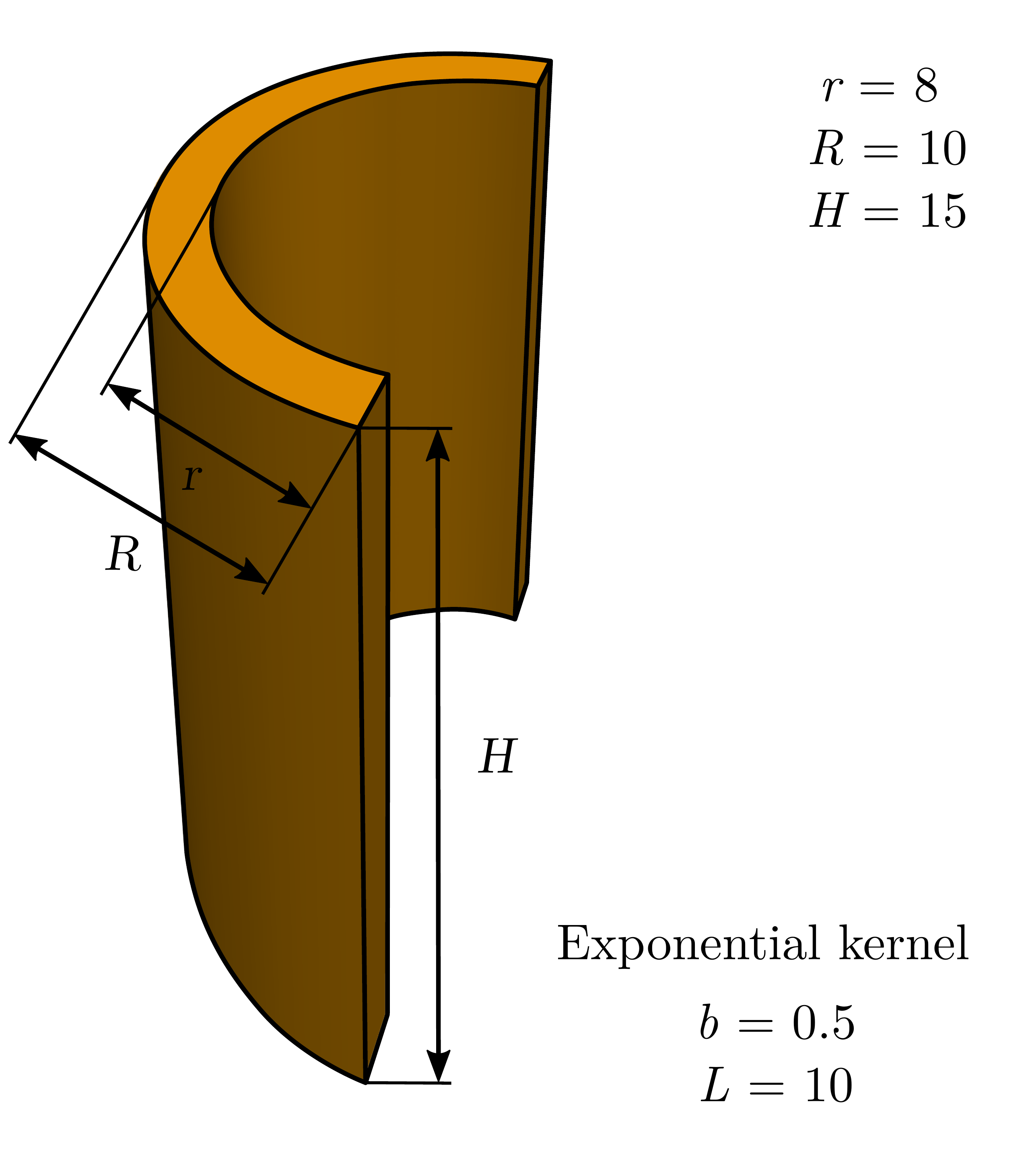} %
    \caption{Half-open cylindrical geometry in the first three-dimensional benchmark. The geometry is modeled using polynomial degrees $p = \{2, 1, 1\}$ and knot vectors $\Xi_1 = (0,0,0, 0.5,0.5, 1,1,1), \; \Xi_2 = (0,0,1,1), \; \Xi_3 = (0,0,1,1)$. This case is also studied in \cite{rahman_galerkin_2018}.}%
    \vspace{5em}
    \label{fig:ex1_geo}%
	\begin{minipage}{0.18\textwidth} 
		\centering
		\includegraphics[width=\linewidth]{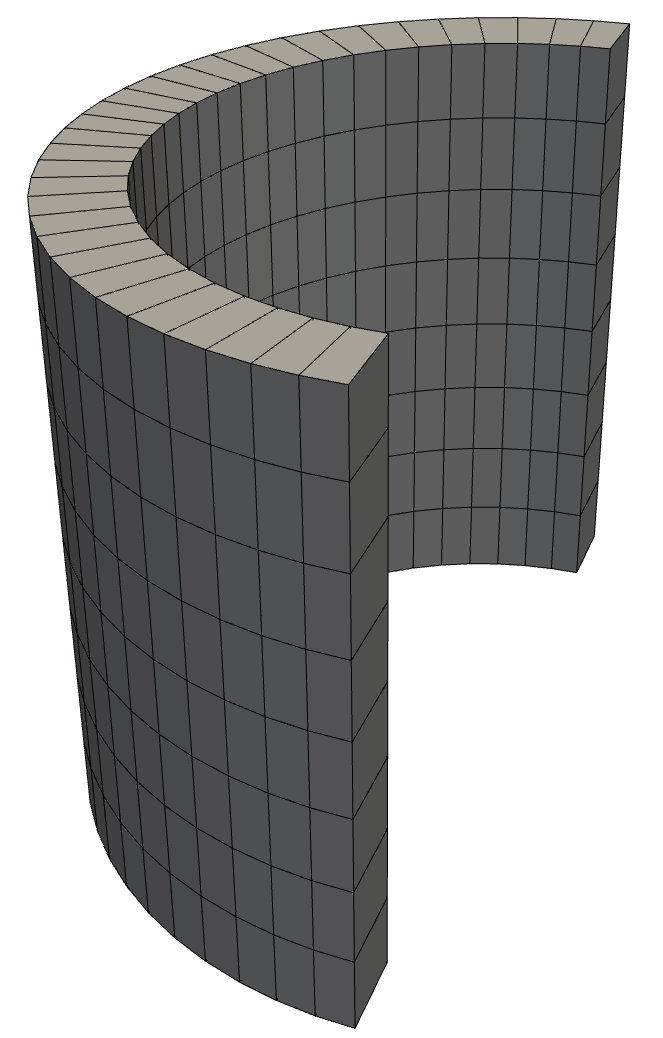}
		{\centering \scriptsize Case 1}\\
		{\centering \scriptsize $32\times 1 \times 8$ elements \par}
	\end{minipage}
	\begin{minipage}{0.18\textwidth} 
		\centering
		\includegraphics[width=\linewidth]{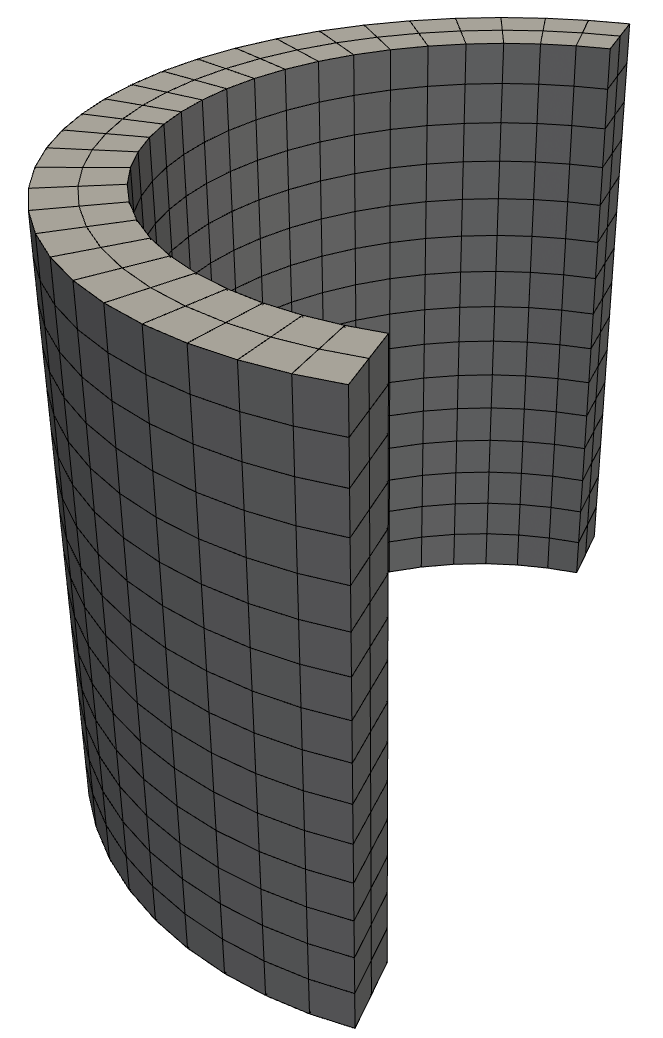}
		{\centering \scriptsize Case 2}\\
		{\centering \scriptsize $28\times 2 \times 15$ elements \par}
	\end{minipage}
	\begin{minipage}{0.18\textwidth} 
		\centering
		\includegraphics[width=\linewidth]{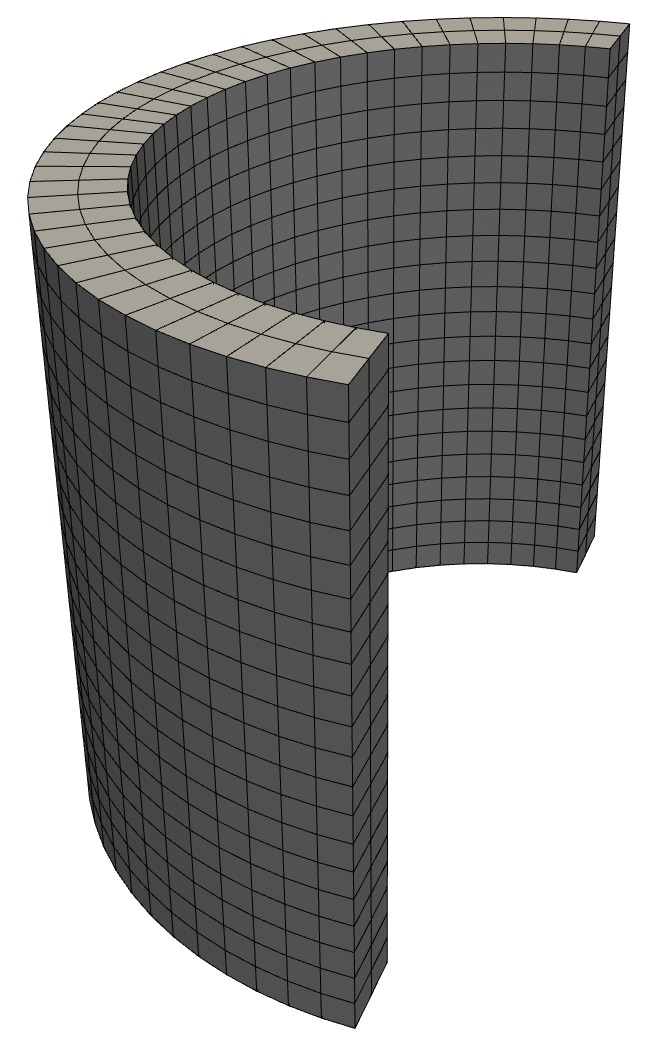}
		{\centering \scriptsize Case 3}\\
		{\centering \scriptsize $38\times 2 \times 21$ elements \par}
	\end{minipage}
	\begin{minipage}{0.18\textwidth} 
		\centering
		\includegraphics[width=\linewidth]{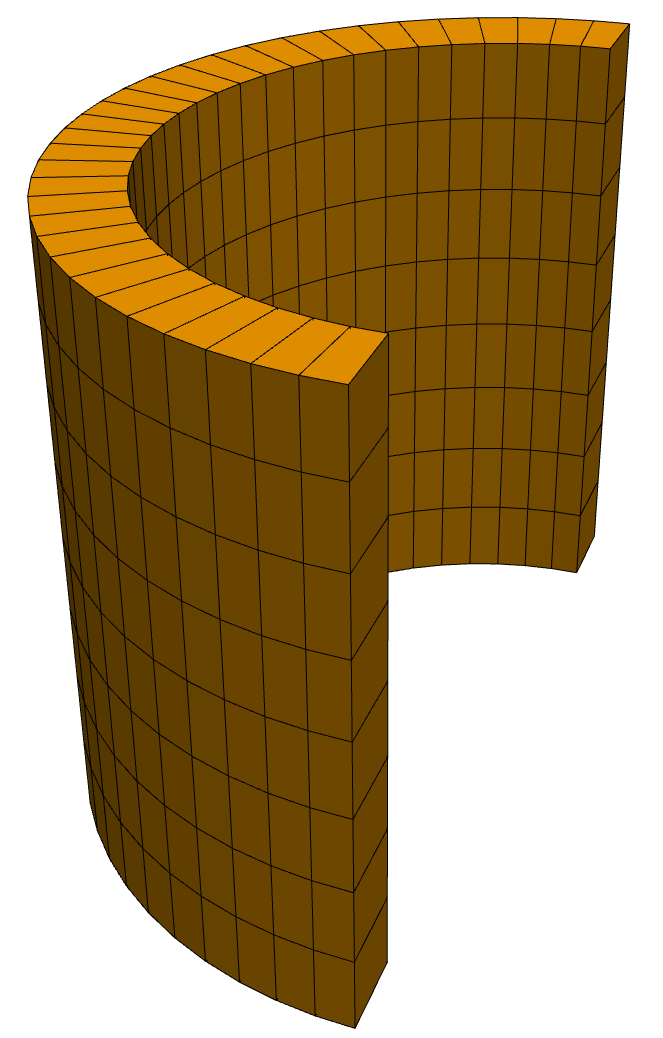}
		{\centering \scriptsize Galerkin 1}\\
		{\centering \scriptsize $32\times 1 \times 8$ elements \par}
	\end{minipage}
	\begin{minipage}{0.18\textwidth} 
		\centering
		\includegraphics[width=\linewidth]{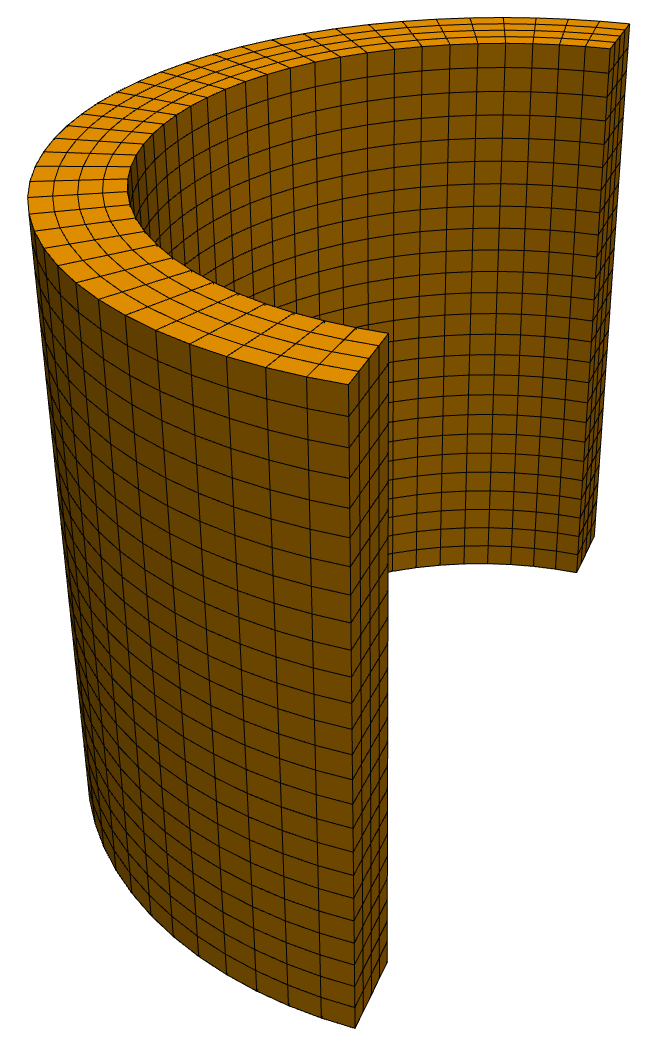}
		{\centering \scriptsize Galerkin 2}\\
		{\centering \scriptsize $38\times 4 \times 25$ elements \par}
	\end{minipage}
	\caption{Meshes of the half-open cylindrical geometry illustrating the interpolation and solution spaces in gray and orange, respectively. The gray meshes from left to right correspond to cases one through three in Tables \ref{tab:ex1-1_evtable} and \ref{tab:ex1-2_evtable}, which employ the first orange mesh in the corresponding solution space. Except for the first interpolation mesh, all meshes are nearly uniform in each parametric direction $(1,2,3)$.}%
    \label{fig:ex1_meshes}%
\end{figure}

\begin{figure}
    \centering
    \subfloat[1st eigenfunction]{{\includegraphics[width=0.32\textwidth]{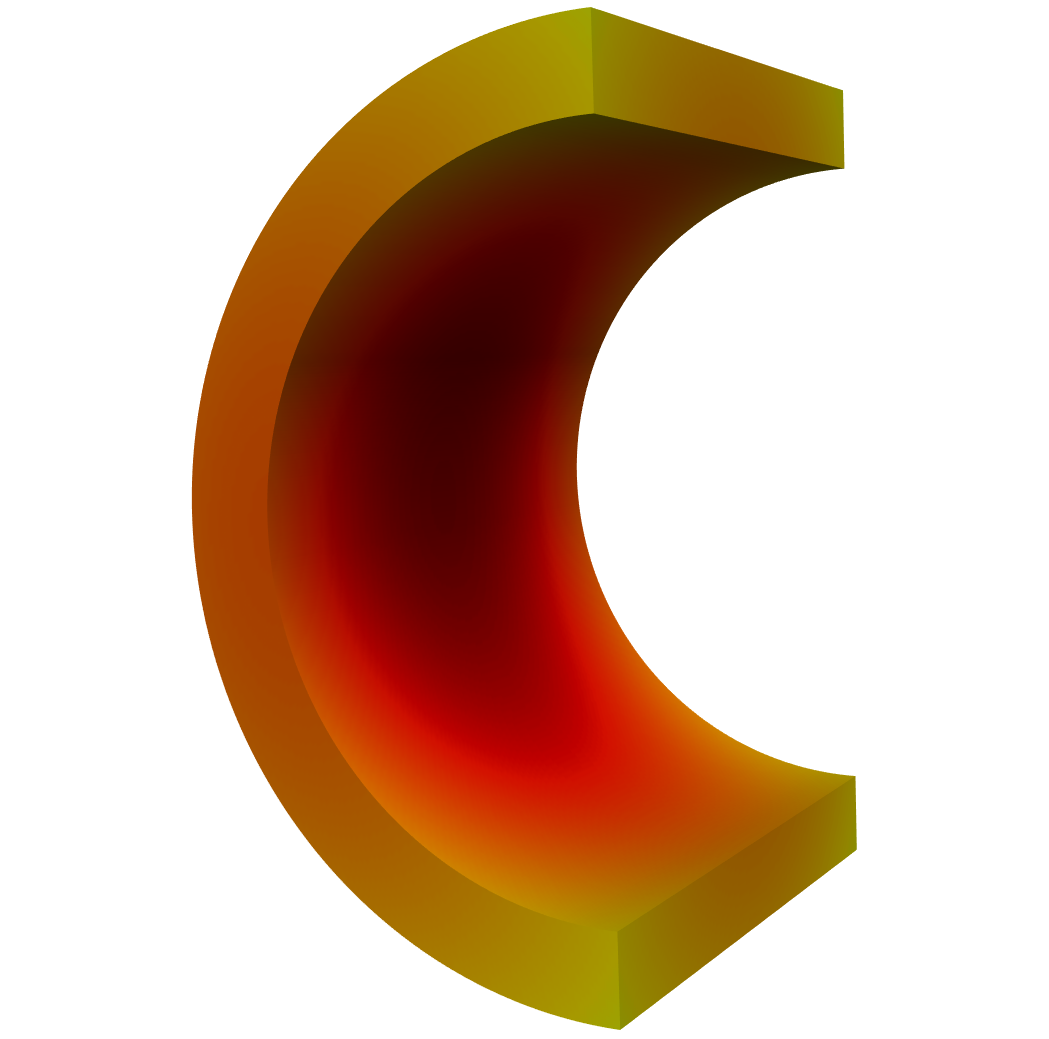} }}%
    \subfloat[2nd eigenfunction]{{\includegraphics[width=0.32\textwidth]{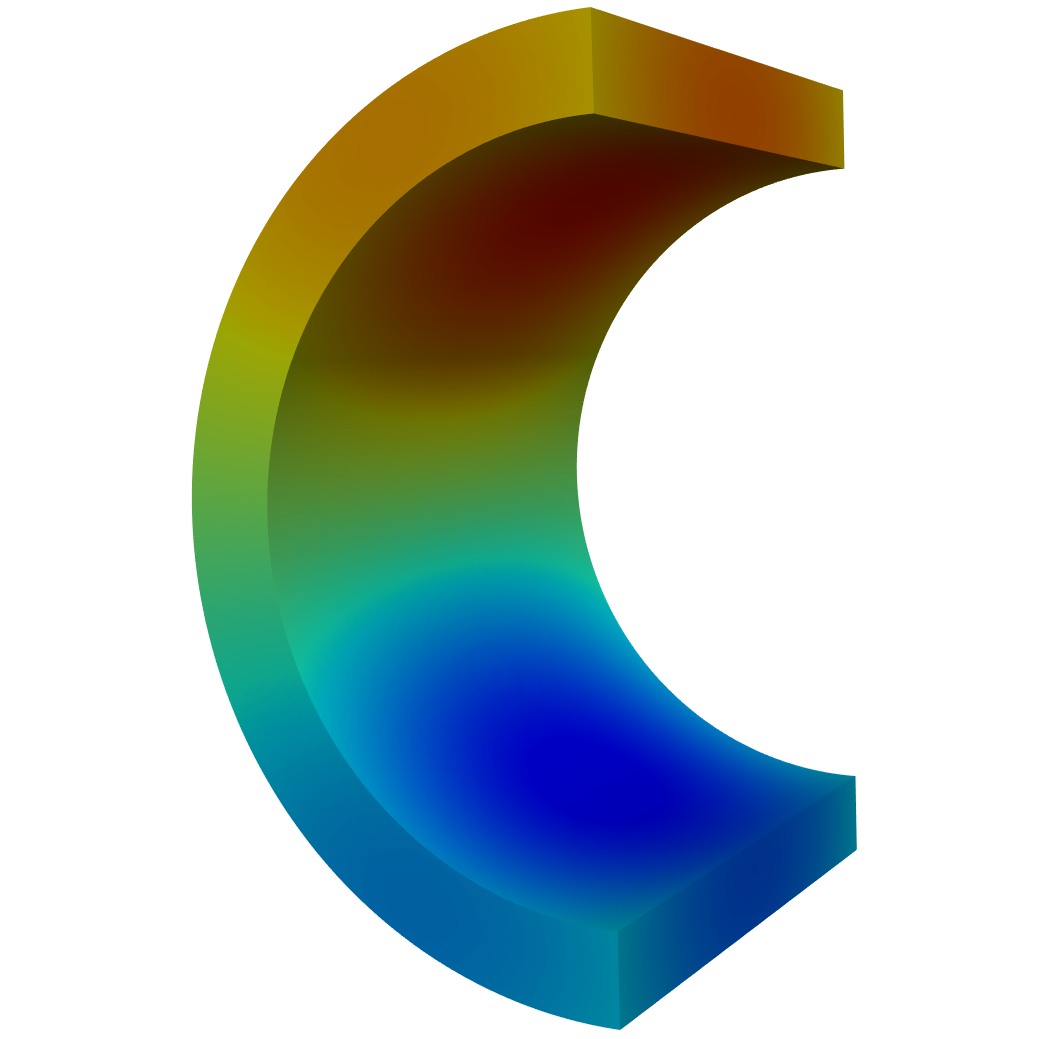} }}%
    \subfloat[3rd eigenfunction]{{\includegraphics[width=0.32\textwidth]{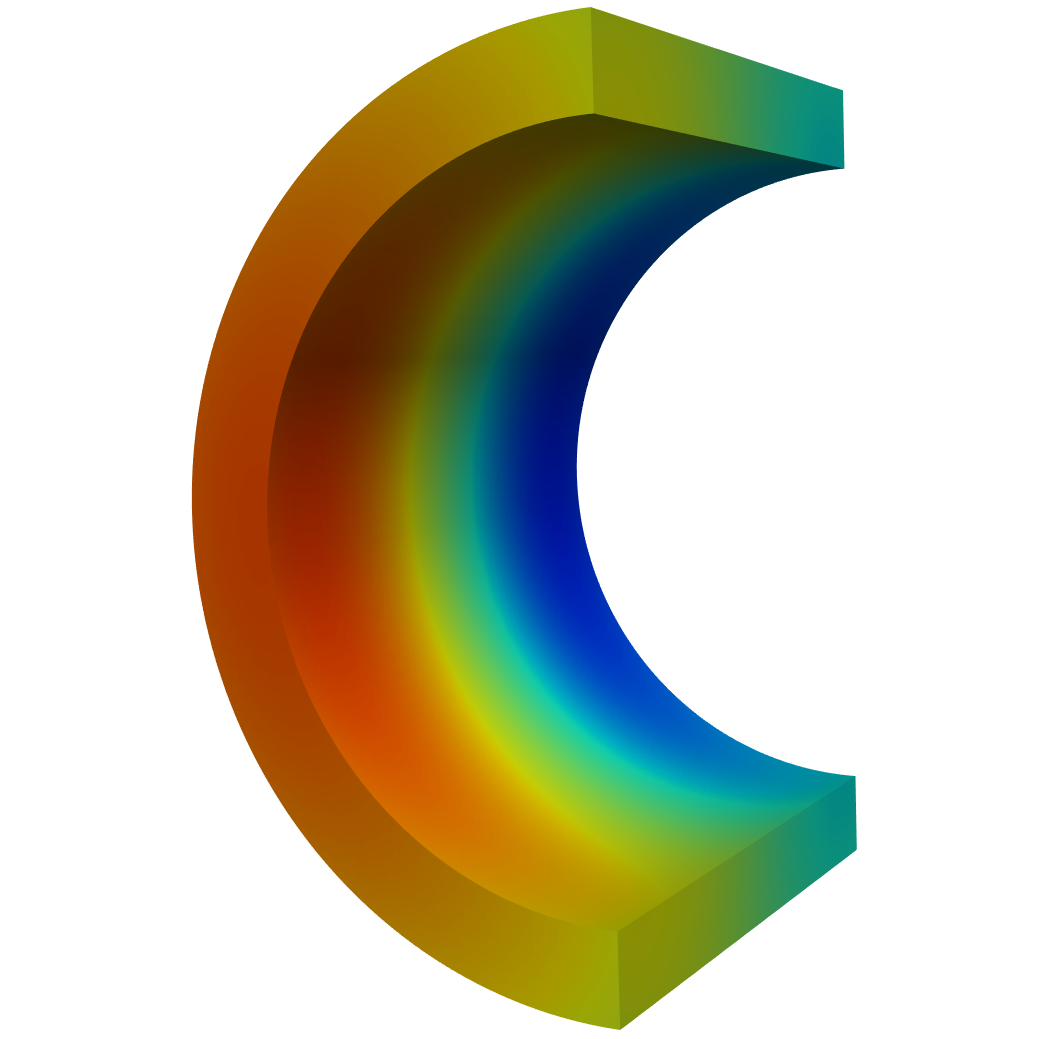} }}\\
    \subfloat[4th eigenfunction]{{\includegraphics[width=0.32\textwidth]{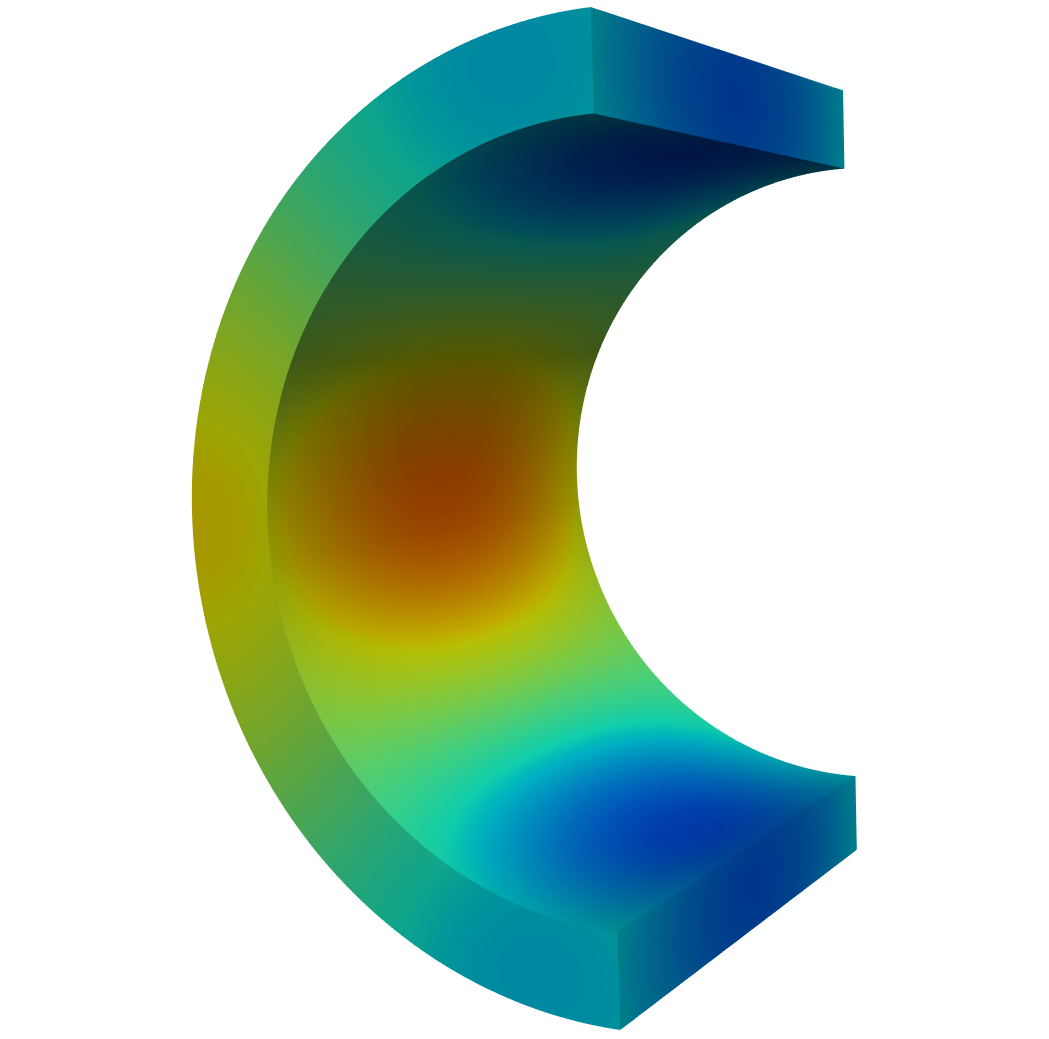} }}%
    \subfloat[5th eigenfunction]{{\includegraphics[width=0.32\textwidth]{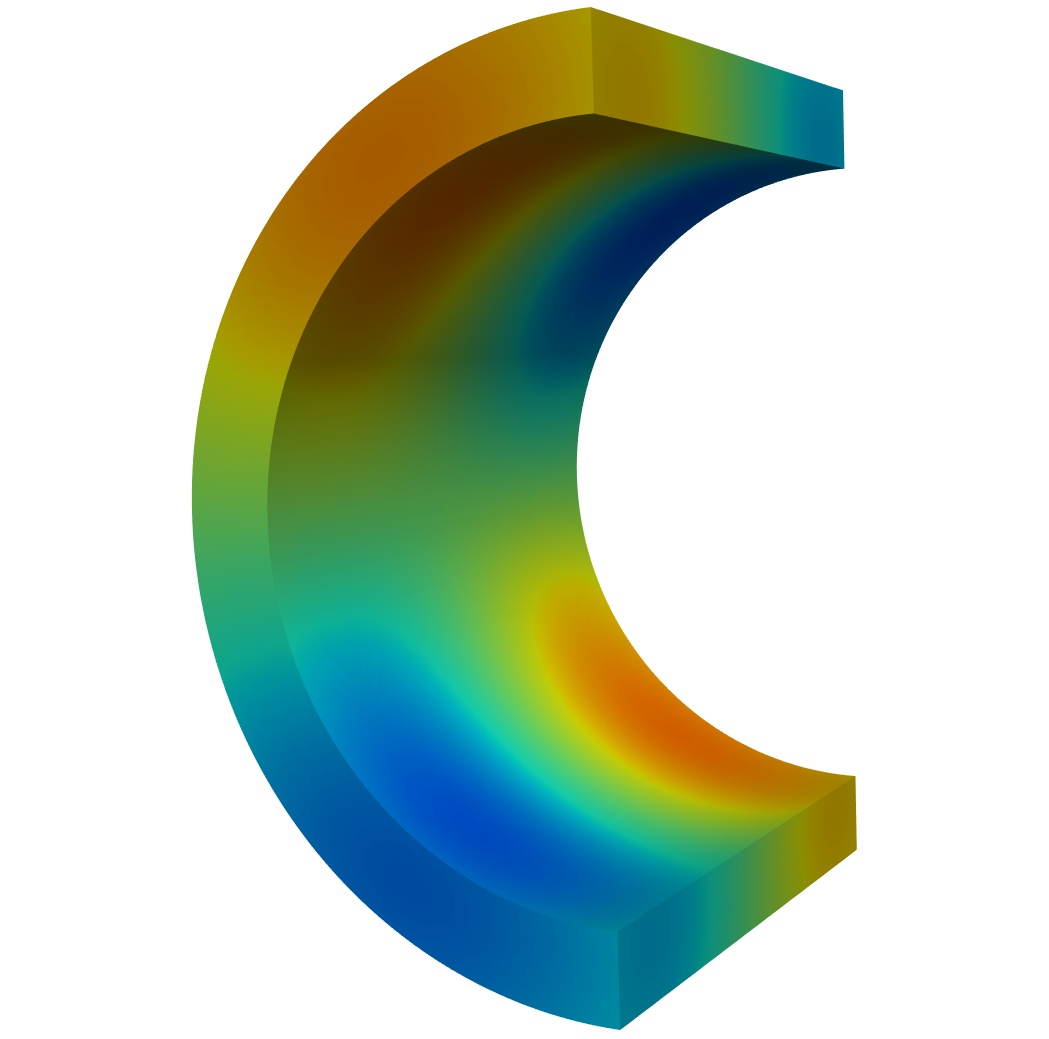} }}%
    \subfloat[6th eigenfunction]{{\includegraphics[width=0.32\textwidth]{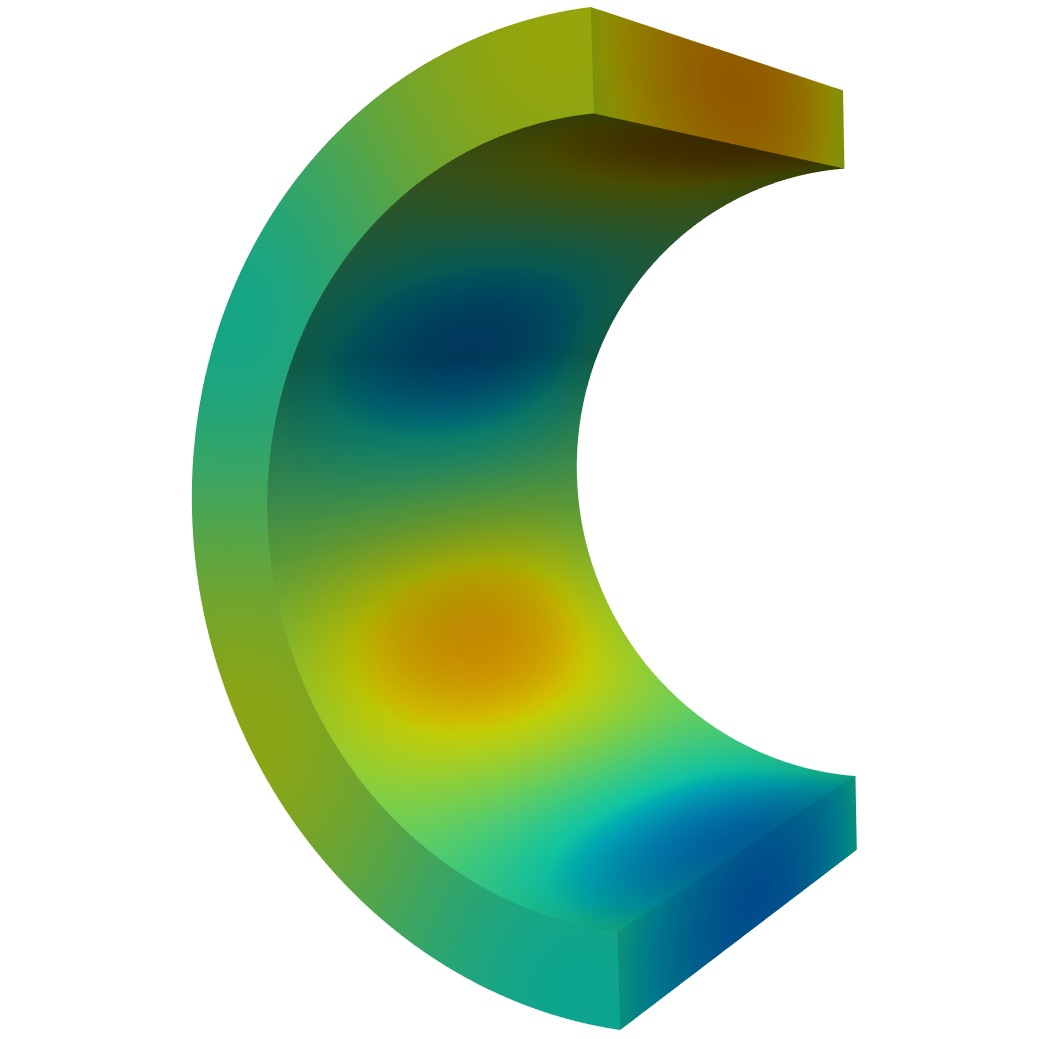} }}\\
    \subfloat[7th eigenfunction]{{\includegraphics[width=0.32\textwidth]{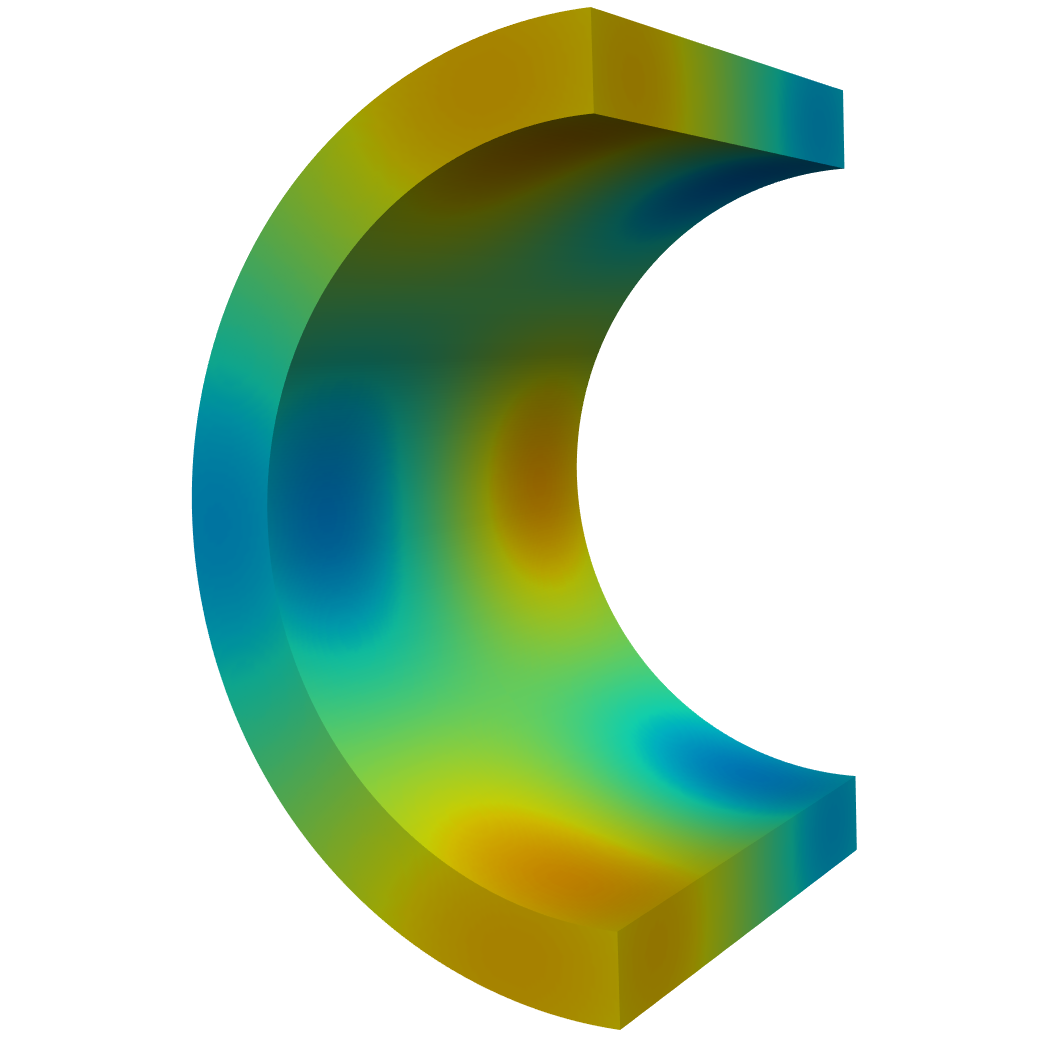} }}%
    \subfloat[8th eigenfunction]{{\includegraphics[width=0.32\textwidth]{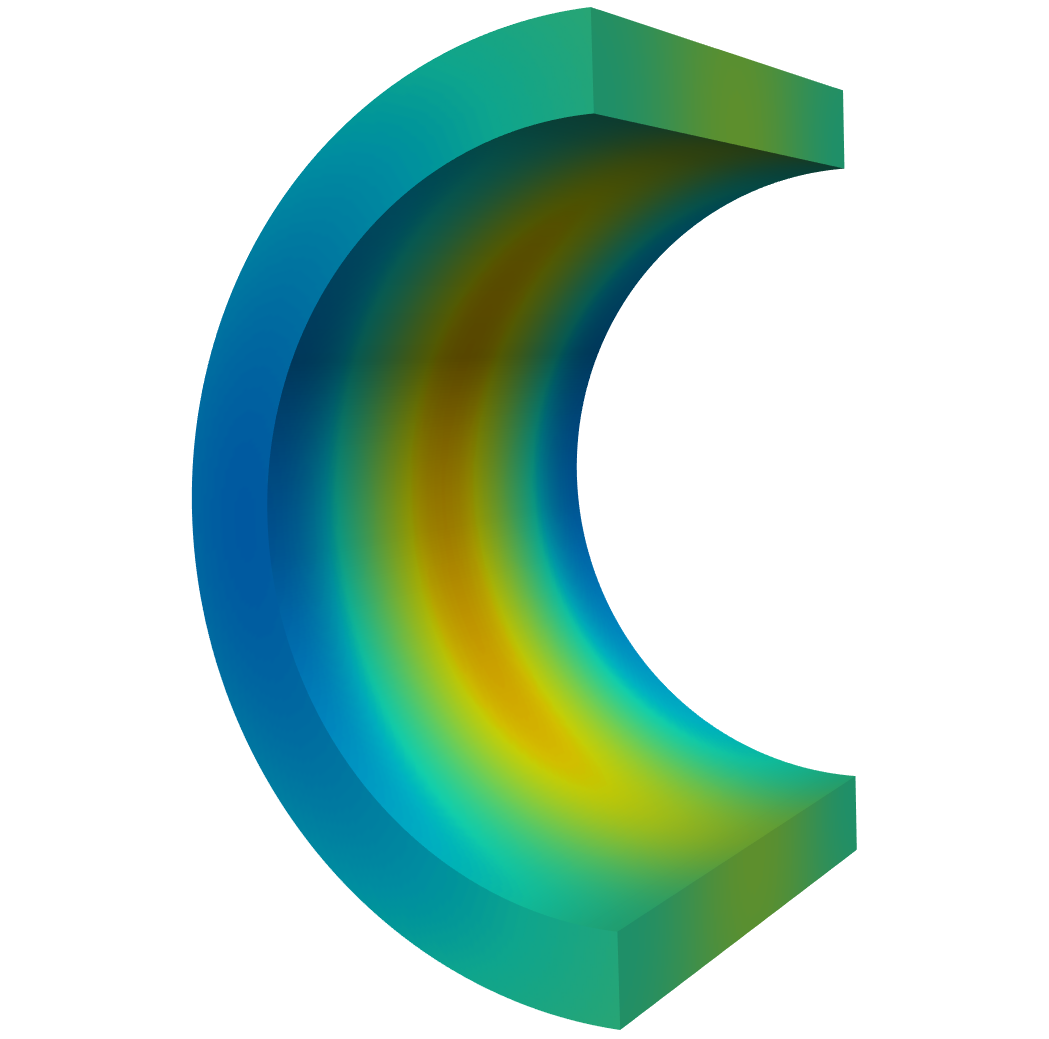} }}%
    \subfloat[9th eigenfunction]{{\includegraphics[width=0.32\textwidth]{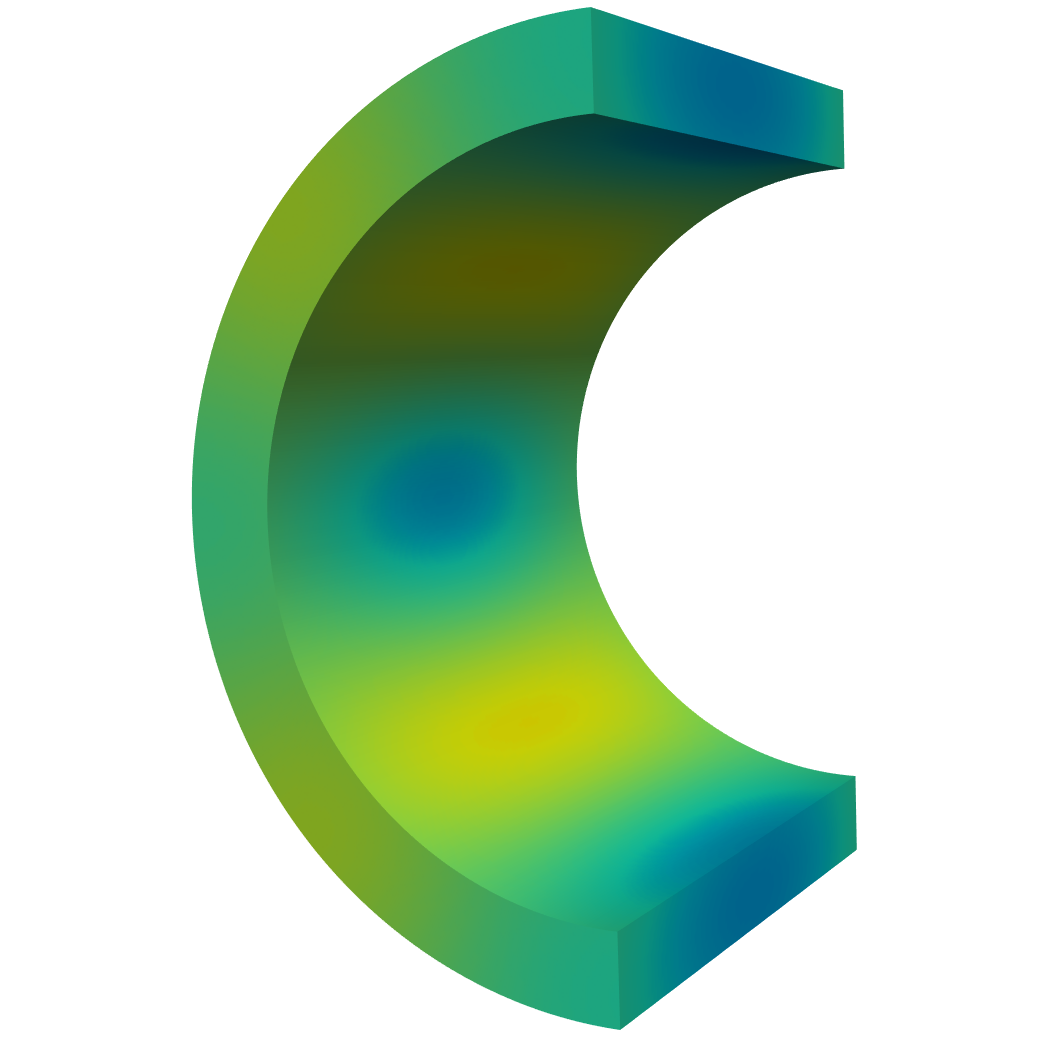} }}\\
		\vspace{1em}
    \includegraphics[width=0.45\textwidth]{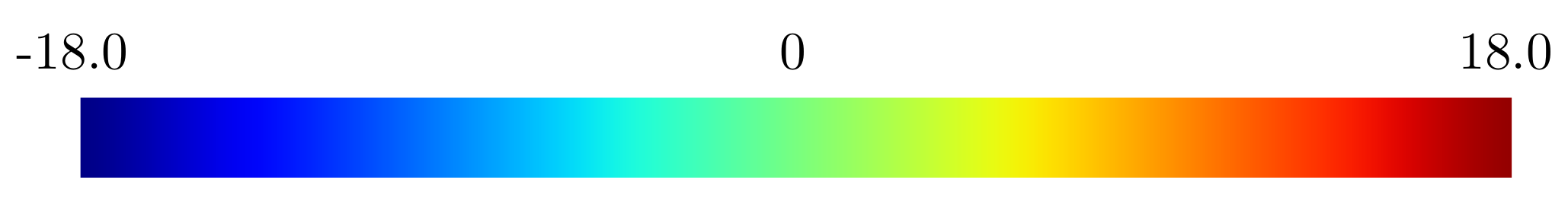}
    \caption{First nine normalized eigenfunctions weighted by the square root of the corresponding eigenvalues in Example 1--1. Results from the third benchmark case.}%
    \label{fig:ex1_ef}%
\end{figure}

In this example we consider the exponential kernel and since this kernel is $C^0$ along $\x = \xp$, there is no use in enforcing higher smoothness on the element boundaries of the interpolation space. Moreover, the coarse geometry is modeled by two patches with $C^0$ continuity between both patches, therefore, the Jacobian determinants will be discontinuous at the interface. Recalling Remark~\ref{remark_on_smoothness}, in order to attain optimal accuracy of our interpolation based quadrature, we enforce a discontinuous interpolation space in the circumferential direction at that interface. Under given considerations, the chosen interpolation space employs quadratic B-splines and is $C^0$ on most element boundaries, except at the discontinuity, where it is $C^{-1}$ in the circumferential direction. Since the system equations are not affected by the discontinuity, see \eqref{eq:tensor_structure_B} and \eqref{eq:ApproxGalA_mat}, we choose the continuity of this space analogously to the example presented in \cite{rahman_galerkin_2018} and use quadratic B-splines in each direction with $C^1$ continuity on the element boundaries. The solution space and the interpolation space meshes for each case are shown in Figure \ref{fig:ex1_meshes}. The first twenty largest eigenvalues together with information about the mesh and computational cost are tabulated in Table~\ref{tab:ex1-1_evtable}. The first nine eigenfunctions corresponding to the nine largest eigenvalues are visualized in Figure~\ref{fig:ex1_ef} by weighting each eigenfunction by the square root of the corresponding eigenvalue.

The presolution formation and assembly time includes the formation and assembly of the univariate mass matrices, the interpolation matrices and their factorizations as well as the computation of the Jacobian determinants at the Greville abscissa. The results in Table \ref{tab:ex1-1_evtable} indicate that these setup costs are negligibly small compared to the total solution time of the Lanczos eigenvalue solver, which is dominated by the matrix-free evaluation of the matrix-vector product (Step 5 in Algorithm \ref{alg:mf_eval}). The number of iterations of the Lanczos algorithm is equal in each case. The maximum resident memory increases in each case, but is negligibly small, when compared to standard methods.
\begin{table}[H]
\centering
	\caption{Enumeration of twenty largest eigenvalues corresponding to the half-open cylinder problem with the exponential kernel in Example 1--1. The numerical eigenvalues have been computed by the proposed isogeometric Galerkin method employing interpolation based quadrature for the three different cases of solution and interpolation spaces depicted in Figure \ref{fig:ex1_meshes} as well as the standard isogeometric Galerkin reference solutions computed with two different meshes and polynomial order $p=(2,2,2)$. Computations executed on a \textit{single} core.}
\label{tab:ex1-1_evtable}
\resizebox{0.9\textwidth}{!}{%
\begin{threeparttable}
\begin{tabular}{llllll}
\hline
Mode                                                                                                             & \multicolumn{5}{l}{Eigenvalue}                                                                                       		  					\\ \cline{2-6}
														 & Case 1           		& Case 2           		 & Case 3           		 & Galerkin 1\tnote{\**}		 & Galerkin 2\tnote{\**}                \\ \hline
1                                                                                                                & 162.9468999           		& 162.8100625           		 & 162.8071703           		 & 162.7991539          		 & 162.7965791                          \\
2                                                                                                                & 91.57710310           		& 91.44102567           		 & 91.43808750           		 & 91.43063070          		 & 91.42804062                          \\
3                                                                                                                & 57.68111166           		& 57.57587807           		 & 57.57369989           		 & 57.56702901          		 & 57.56447741                          \\
4                                                                                                                & 51.23142243           		& 51.09932034           		 & 51.09664525           		 & 51.09017918          		 & 51.08762278                          \\
5                                                                                                                & 38.91593553           		& 38.80608047           		 & 38.80371507           		 & 38.79740931          		 & 38.79483423                          \\
6                                                                                                                & 28.04104735           		& 27.91172258           		 & 27.90928395           		 & 27.90386143          		 & 27.90128438                          \\
7                                                                                                                & 25.17627947           		& 25.06401793           		 & 25.06174037           		 & 25.05611145          		 & 25.05356161                          \\
8                                                                                                                & 19.44500190           		& 19.37694239           		 & 19.37398575           		 & 19.36893419          		 & 19.36659412                          \\
9                                                                                                                & 16.28927839           		& 16.16360095           		 & 16.16148139           		 & 16.15700369          		 & 16.15443088                          \\
10                                                                                                               & 15.91731898          		& 15.80248489          		 & 15.80018997          		 & 15.79530798          		 & 15.79273209                          \\
11                                                                                                               & 15.22927329          		& 15.15403705          		 & 15.15115655          		 & 15.14622914          		 & 15.14385016                          \\
12                                                                                                               & 11.30279820          		& 11.22038002          		 & 11.21784016          		 & 11.21328690          		 & 11.21090778                          \\
13                                                                                                               & 10.30082590          		& 10.18526145          		 & 10.18310045          		 & 10.17896939          		 & 10.17639037                          \\
14                                                                                                               & 9.821983940          		& 9.698825760          		 & 9.697115542          		 & 9.693564320          		 & 9.690982310                          \\
15                                                                                                               & 8.151359815          		& 8.061279562          		 & 8.058884448          		 & 8.054783080          		 & 8.052352020                          \\
16                                                                                                               & 7.618618090          		& 7.582986789          		 & 7.580140848          		 & 7.578085990          		 & 7.576621410                          \\
17                                                                                                               & 6.835745032          		& 6.727654352          		 & 6.725577132          		 & 6.722378350          		 & 6.719925970                          \\
18                                                                                                               & 6.501550722          		& 6.450711848          		 & 6.447892827          		 & 6.445576690          		 & 6.443915210                          \\
19                                                                                                               & 6.300773813          		& 6.181552964          		 & 6.180140556          		 & 6.177345410          		 & 6.174771170                          \\
20                                                                                                               & 5.866056812          		& 5.769454716          		 & 5.767330624          		 & 5.763786920          		 & 5.761319370                          \\
\multicolumn{6}{c}{}                                                                                                                		                   		                    		                        		                                        \\
\multicolumn{6}{c}{\textit{Interpolation space} \hrulefill}                                                                         		                   		                    		                        		                                        \\
\begin{tabular}[c]{@{}l@{}}\hspace{0.8em} Number of elements \end{tabular}                                       & 256        		& 840        		 & 1596        		 & --                   		 & --                                   \\
\begin{tabular}[c]{@{}l@{}}\hspace{0.8em} Number of degrees of freedom \end{tabular}                             & 1980         		& 8990         		 & 16770         		 & --                   		 & --                                   \\
\begin{tabular}[c]{@{}l@{}}\hspace{0.8em} Mesh size \end{tabular}                                                & 2.857         		& 1.719         		 & 1.423         		 & --                   		 & --                                   \\
\begin{tabular}[c]{@{}l@{}}\hspace{0.8em} Mesh size/correlation length \end{tabular}                             & 0.571       		& 0.344       		 & 0.284       		 & --                   		 & --                                   \\
\begin{tabular}[c]{@{}l@{}}\hspace{0.8em} Formation and assembly of univariate matrices\end{tabular}             & $ 0.314 \,\mathrm{s} $ & $ 0.308 \,\mathrm{s} $	 & $ 0.301 \,\mathrm{s} $  & --                   		 & --                                   \\
\multicolumn{6}{c}{\textit{Solution space} \hrulefill}                                                                                                             		                    		                        		 		                        \\
\begin{tabular}[c]{@{}l@{}}\hspace{0.8em} Number of elements \end{tabular}                                       & \multicolumn{4}{c}{\ldots\, 256 \,\ldots}       		                    		            				 & 3800                                 \\
\begin{tabular}[c]{@{}l@{}}\hspace{0.8em} Number of degrees of freedom \end{tabular}                             & \multicolumn{4}{c}{\ldots\, 1050 \,\ldots}      		                    		             				 & 6642                                 \\
\begin{tabular}[c]{@{}l@{}}\hspace{0.8em} Mesh size \end{tabular}                                                & \multicolumn{4}{c}{\ldots\, 2.857 \,\ldots}     		                    		             				 & 1.073                                \\
\begin{tabular}[c]{@{}l@{}}\hspace{0.8em} Mesh size/correlation length \end{tabular}                             & \multicolumn{4}{c}{\ldots\, 0.571 \,\ldots}     		                    		             				 & 0.215                                \\
\begin{tabular}[c]{@{}l@{}}\hspace{0.8em} Formation and assembly of system matrices\end{tabular}                 & --               		& --               		 & --               		 & $4.86\,\mathrm{min}$ 		 & $17.3\,\mathrm{h}$                 \\
\multicolumn{6}{c}{\textit{Summary} \hrulefill}                                                                                     		                   		                    		                                                                		\\
	\begin{tabular}[c]{@{}l@{}}\hspace{0.8em} Number of iterations\end{tabular}                              & 63         		& 63         		 & 63         		 & 63                     		 & 63                                   \\
\begin{tabular}[c]{@{}l@{}}\hspace{0.8em} Maximum resident memory [GB]\end{tabular}                              & 0.438           		& 0.441           		 & 0.441           		 & 0.464                  		 & 1.566                                \\
\begin{tabular}[c]{@{}l@{}}\hspace{0.8em} Solution time\end{tabular}                                             & $ 5.031 \,\mathrm{s}$	& $ 64.13 \,\mathrm{s}$	 & $ 3.367 \,\mathrm{min}$      & $0.09\,\mathrm{s}$   		 & $6.21\,\mathrm s$                  \\
\begin{tabular}[c]{@{}l@{}}\hspace{0.8em} Total time\end{tabular}                                                & $ 5.345 \,\mathrm{s}$	& $ 64.44 \,\mathrm{s}$  	 & $ 3.372 \,\mathrm{min}$      & $4.86\,\mathrm{min}$ 		 & $17.3\,\mathrm h$                  \\
\multicolumn{6}{c}{}                                                                                                                                                                                                                      						        \\ \hline
\end{tabular}
\begin{tablenotes}
  \item[\**] exact kernel, NURBS trial and test space, elementwise assembly
\end{tablenotes}
\end{threeparttable}%
}
\end{table}

\begin{table}[H]
\centering
	\caption{Relative operator error with respect to the $2$- and Frobenius-norm in Example 1--1 (exponential kernel). For the comparison the exact operator $\mat{A}$ in \eqref{eq:GalA_mat} was estimated using a Gaussian quadrature rule with $(p+1)^3$ points.}
\label{tab:ex1-1-opnorms}
\renewcommand*{\arraystretch}{1.22}
\begin{tabular}{l|ccc}
Rel. matrix norm                                                                                & Case 1              & Case 2              & Case 3              \\ \hline
$\lVert \mat A - \tilde{\mat A} \rVert_2 \lVert \mat A \rVert_2^{-1}$                           & $9.95\cdot 10^{-4}$ & $7.99\cdot 10^{-5}$ & $6.12\cdot 10^{-5}$ \\
$\lVert \mat A - \tilde{\mat A} \rVert_{\mathrm{F}} \lVert \mat A \rVert_{\mathrm{F}}^{-1}$     & $3.49\cdot 10^{-3}$ & $1.98\cdot 10^{-4}$ & $1.56\cdot 10^{-4}$ \\ 
\end{tabular}
\end{table}

The two rightmost columns summarize the results obtained by the isogeometric Galerkin method proposed in \cite{rahman_galerkin_2018}. On the same mesh (Case 1) we observe a speed-up of roughly 2 orders in magnitude. This comparison might not be completely fair because a full Gaussian quadrature performed in \cite{rahman_galerkin_2018} is much more accurate than our interpolation based quadrature technique on the same mesh.  Nonetheless, the obtained accuracy in the eigenvalues is convincing, as evidenced also in Table \ref{tab:ex1-1-opnorms}. Therein we compare the relative error of the approximated operator $\mat{\tilde A}$ in \eqref{eq:ApproxGalA_mat} in terms of the 2- and Frobenius-norm with respect to the operator $\mat{A}$ in~\eqref{eq:GalA_mat} obtained by standard Gaussian quadrature  with $(p+1)^3$ quadrature points. The suboptimal convergence rate for the rough exponential kernel manifests again in the operator errors.

\subsubsection*{Example 1--2}
The second example employs the Gaussian covariance kernel with the same correlation length and variance as in Example 1--1. With that, we take advantage of higher smoothness across the element boundaries and rather than performing $h$-refinement as in Example 1--1, we set the interpolation mesh fixed (compare Case 1 in Example 1--1) and perform $k$-refinement for $p=2,4,8$. Nonetheless, at the discontinuity in the coarse geometry model, we enforce $C^{-1}$ in the circumferential direction.
\begin{table}[H]
\centering
	\caption{Enumeration of twenty largest eigenvalues corresponding to the half-open cylinder problem with the Gaussian kernel in Example 1--2. The numerical eigenvalues have been computed by the proposed isogeometric Galerkin method for polynomial degrees $p=2,4,8$ in each parametric direction on the coarsest solution and interpolation meshes depicted in Figure \ref{fig:ex1_meshes} as well as the standard isogeometric Galerkin reference solutions computed with two different meshes and polynomial order $p=(2,2,2)$. Computations executed on a \textit{single} core.}
\label{tab:ex1-2_evtable}
\resizebox{0.9\textwidth}{!}{%
\begin{threeparttable}
\begin{tabular}{llllll}
\hline
Mode                                                                                                             & \multicolumn{5}{l}{Eigenvalue}                                                                                        \\ \cline{2-6}
														 &$p=2$           &$p=4$            &$p=8$           & Galerkin 1\tnote{\**}& Galerkin 2\tnote{\**}                \\ \hline
1                                                                                                                & 124.0032406           & 123.9913749            & 123.9916387           & 123.9916388          & 123.99141826                         \\
2                                                                                                                & 102.6956247           & 102.6855546            & 102.6857821           & 102.6857823          & 102.68558059                         \\
3                                                                                                                & 75.56585684           & 75.56078583            & 75.56096426           & 75.56096463          & 75.560806410                         \\
4                                                                                                                & 75.34465933           & 75.39112714            & 75.39245804           & 75.39245720          & 75.392381700                         \\
5                                                                                                                & 62.39810201           & 62.43643727            & 62.43754500           & 62.43754437          & 62.437470230                         \\
6                                                                                                                & 49.86332482           & 49.86567375            & 49.86580222           & 49.86580308          & 49.865713360                         \\
7                                                                                                                & 45.91399154           & 45.94362160            & 45.94444340           & 45.94444308          & 45.944382590                         \\
8                                                                                                                & 33.55923787           & 33.66041670            & 33.66423475           & 33.66423503          & 33.664535290                         \\
9                                                                                                                & 30.29707291           & 30.32008760            & 30.32063646           & 30.32063662          & 30.320605640                         \\
10                                                                                                               & 29.81774644          & 29.82724920           & 29.82734187          & 29.82734291          & 29.827325750                         \\
11                                                                                                               & 27.79271639          & 27.87644349           & 27.87960795          & 27.87960820          & 27.879851690                         \\
12                                                                                                               & 20.45053462          & 20.51277792           & 20.51510945          & 20.51510970          & 20.515286220                         \\
13                                                                                                               & 18.11733255          & 18.13601905           & 18.13635697          & 18.13635738          & 18.136361050                         \\
14                                                                                                               & 16.33318523          & 16.34652184           & 16.34660003          & 16.3466016           & 16.346634880                         \\
15                                                                                                               & 13.49460845          & 13.53722675           & 13.53876834          & 13.53876867          & 13.538889150                         \\
16                                                                                                               & 11.29075514          & 11.39146819           & 11.39829646          & 11.39830777          & 11.399235970                         \\
17                                                                                                               & 9.924081589          & 9.939261559           & 9.939463418          & 9.939464250          & 9.9394922100                         \\
18                                                                                                               & 9.350652024          & 9.434037081           & 9.439692866          & 9.439702240          & 9.4404691900                         \\
19                                                                                                               & 8.282851361          & 8.296245384           & 8.296321088          & 8.296322560          & 8.2963738600                         \\
20                                                                                                               & 8.069634638          & 8.097318365           & 8.098244763          & 8.098245110          & 8.0983270800                         \\
\multicolumn{6}{c}{}                                                                                                                                                                                                                     \\
\multicolumn{6}{c}{\textit{Interpolation space} \hrulefill}                                                                                                                                                                              \\
\begin{tabular}[c]{@{}l@{}}\hspace{0.8em} Number of elements \end{tabular}                                       & 256        & 256         & 256        & --                   & --                                   \\
\begin{tabular}[c]{@{}l@{}}\hspace{0.8em} Number of degrees of freedom \end{tabular}                             & 1080         & 2400          & 6912         & --                   & --                                   \\
\begin{tabular}[c]{@{}l@{}}\hspace{0.8em} Mesh size \end{tabular}                                                & 2.857         & 2.857          & 2.857         & --                   & --                                   \\
\begin{tabular}[c]{@{}l@{}}\hspace{0.8em} Mesh size/correlation length \end{tabular}                             & 0.571       & 0.571        & 0.571       & --                   & --                                   \\
\begin{tabular}[c]{@{}l@{}}\hspace{0.8em} Formation and assembly of univariate matrices\end{tabular}             & $ 0.299 \,\mathrm{s} $ & $ 0.299 \,\mathrm{s} $	 & $ 0.301 \,\mathrm{s} $ & --                   & --                                   \\
\multicolumn{6}{c}{\textit{Solution space} \hrulefill}                                                                                                                                                                                   \\
\begin{tabular}[c]{@{}l@{}}\hspace{0.8em} Number of elements \end{tabular}                                       & \multicolumn{4}{c}{\ldots\, 256 \,\ldots}                                      & 3800                                 \\
\begin{tabular}[c]{@{}l@{}}\hspace{0.8em} Number of degrees of freedom \end{tabular}                             & \multicolumn{4}{c}{\ldots\, 1050 \,\ldots}                                     & 6642                                 \\
\begin{tabular}[c]{@{}l@{}}\hspace{0.8em} Mesh size \end{tabular}                                                & \multicolumn{4}{c}{\ldots\, 2.857 \,\ldots}                                    & 1.073                                \\
\begin{tabular}[c]{@{}l@{}}\hspace{0.8em} Mesh size/correlation length \end{tabular}                             & \multicolumn{4}{c}{\ldots\, 0.571 \,\ldots}                                    & 0.215                                \\
\begin{tabular}[c]{@{}l@{}}\hspace{0.8em} Formation and assembly of system matrices\end{tabular}                 & --               & --                & --               &$5.01\,\mathrm{min}$&$16.97\,\mathrm{h}$                 \\
\multicolumn{6}{c}{\textit{Summary} \hrulefill}                                                                                                                                                                                          \\
	\begin{tabular}[c]{@{}l@{}}\hspace{0.8em} Number of iterations\end{tabular}                              & 52         & 52          & 52         & 52                   & 52                                   \\
\begin{tabular}[c]{@{}l@{}}\hspace{0.8em} Maximum resident memory [GB]\end{tabular}                              & 0.437           & 0.436            & 0.437           & 0.464                & 1.615                                \\
\begin{tabular}[c]{@{}l@{}}\hspace{0.8em} Solution time\end{tabular}                                             & $ 0.817 \,\mathrm{s}$	& $ 3.412 \,\mathrm{s}$	 & $ 27.95 \,\mathrm{s}$      &$0.10\,\mathrm{s}$  &$5.61\,\mathrm s$                   \\
\begin{tabular}[c]{@{}l@{}}\hspace{0.8em} Total time\end{tabular}                                                & $ 1.116 \,\mathrm{s}$	& $ 3.711 \,\mathrm{s}$  	 & $ 28.25 \,\mathrm{s}$      &$5.02\,\mathrm{min}$&$16.97\,\mathrm h$                  \\
\multicolumn{6}{c}{}                                                                                                                                                                                                                                 \\ \hline
\end{tabular}
\begin{tablenotes}
  \item[\**] exact kernel, NURBS trial and test space, elementwise assembly
\end{tablenotes}
\end{threeparttable}%
}
\end{table}

As already discussed, it is evident, that by taking advantage of higher convergence rates the proposed method performs better for smooth kernels, which reflects in the error of the operator in Table \ref{tab:ex1-2-opnorms}. It is worth noting, that the timings versus accuracy are in favour of $k$-refinement. 
\begin{table}[H]
\centering
	\caption{Relative operator error with respect to the $2$- and Frobenius-norm in Example 1--2 (Gaussian kernel). For the comparison the exact operator $\mat{A}$ in \eqref{eq:GalA_mat} was estimated using a Gaussian quadrature rule with $(p+1)^3$ points.}
\label{tab:ex1-2-opnorms}
\renewcommand*{\arraystretch}{1.22}
\begin{tabular}{l|ccc}
Rel. matrix norm                                                                                     & Case 1  & Case 2  & Case 3  \\ \hline
$\lVert \mat A - \tilde{\mat A} \rVert_2 \lVert \mat A \rVert_2^{-1}$                                & $9.05\cdot 10^{-4}$ & $5.30\cdot 10^{-5}$ & $3.67\cdot 10^{-7}$ \\
$\lVert \mat A - \tilde{\mat A} \rVert_{\mathrm{F}} \lVert \mat A \rVert_{\mathrm{F}}^{-1}$          & $1.27\cdot 10^{-3}$ & $6.71\cdot 10^{-5}$ & $3.75\cdot 10^{-7}$ \\ 
\end{tabular}
\end{table}

\subsection{Random field with Gaussian kernel in a three-dimensional hemispherical shell}
\begin{figure}
    \centering
    \includegraphics[width=\textwidth]{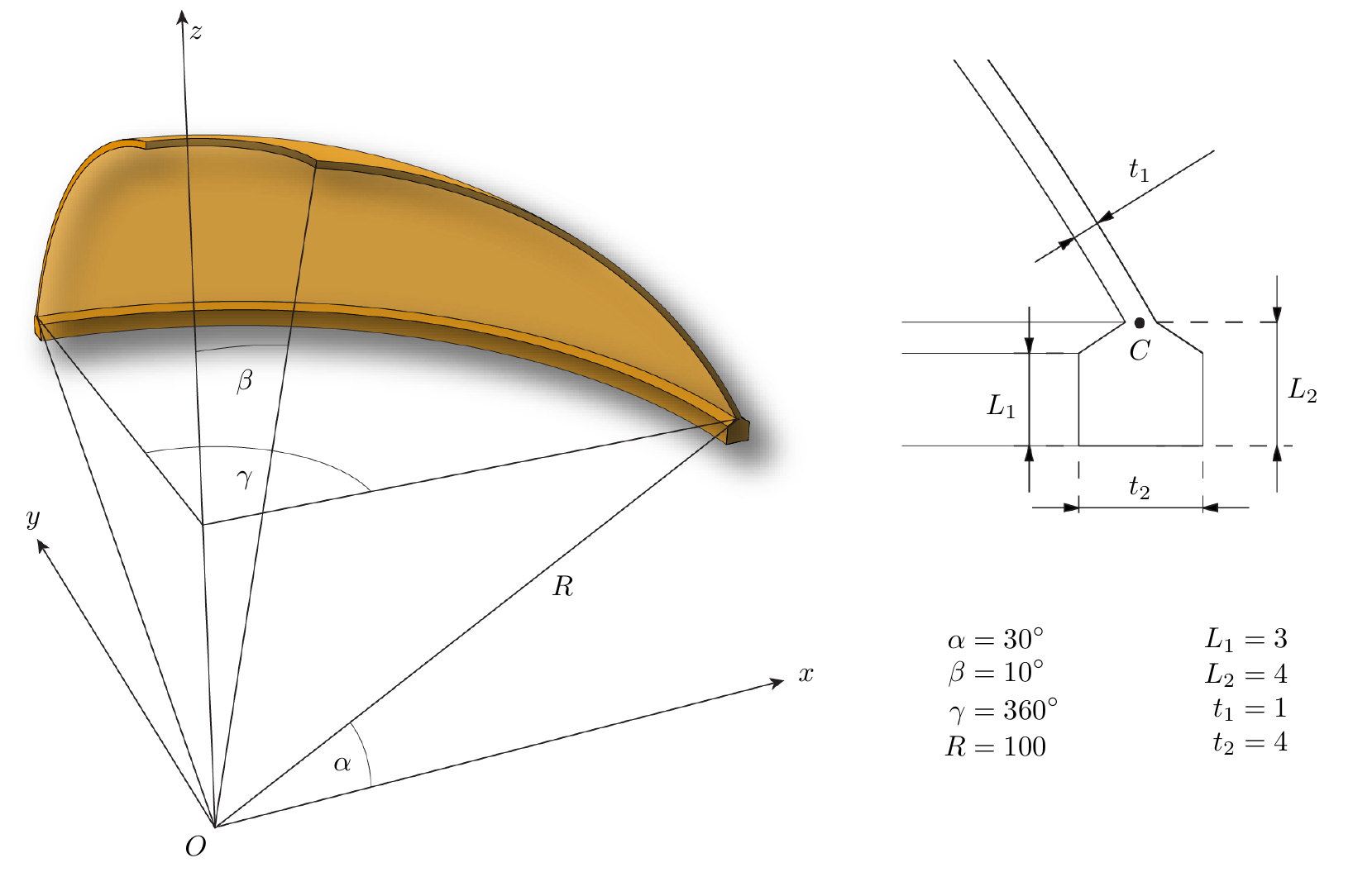}
    \caption{Hemispherical shell with a stiffener. The geometry is modeled as a single NURBS patch using polynomial degrees $p = \{2, 2, 2\}$ and knot vectors $\Xi_1 = (0,0,0,1,1,2,2,3,3,3), \; \Xi_2 = (0,0,0,1,1,2,2,3,3,4,4,4), \; \Xi_3 = (0,0,0,1,1,1)$.}%
    \label{fig:ex2_geo}%
\end{figure}

We consider the hemispherical shell with stiffener depicted in Figure~\ref{fig:ex2_geo}. This three dimensional multipatch shell structure is similar to the model published in \cite{rank_high_2005} but has a slightly different stiffener profile. We use the Gaussian covariance function with a correlation length $bL=0.5L$, where the characteristic domain length, $L\approx176$, is the diameter of the stiffener ring.

We study two examples across a range of polynomial degrees  $p= \{ 2,6,16 \}$. The two examples differ in the following way:
\begin{table}[H]
	\begin{tabular}{ll}
		Example 2--1    & The solution mesh is the same as the interpolation mesh; \\
		Example 2--2	& The solution mesh is twice as fine as the interpolation mesh in every component direction. \\
	\end{tabular}
\end{table}
In both studies we use interpolation and solution meshes obtained by \emph{p}-refinement of the geometrical model, followed by uniform \emph{h}-refinement. The continuity of these spaces is thus $C^0$ at knots that are present in the initial coarse geometrical model and $C^{p-1}$ at new knots introduced by \emph{h}-refinement. Again, where the solution space is $C^0$, we enforce $C^{-1}$ in the interpolation space in order to achieve optimal accuracy per degree of freedom.

\subsubsection*{Example 2--1}
\begin{figure}[H]
    \centering
    \subfloat[Interpolation space mesh in Example 2--1 and Example 2--2 with $23\times 84\times 2$ elements in parametric directions $(1,2,3)$]{{
	    \includegraphics[width=0.38\textwidth,valign=t]{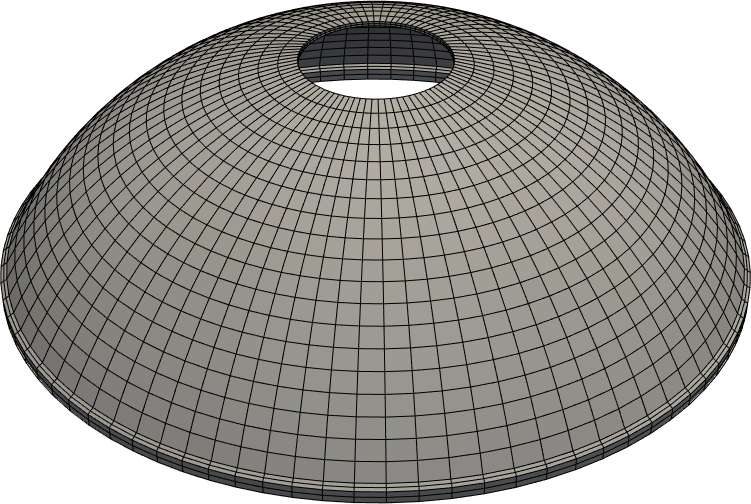}
	    \hspace{2.5em}
	    \framebox{\includegraphics[width=0.17\textwidth,valign=t]{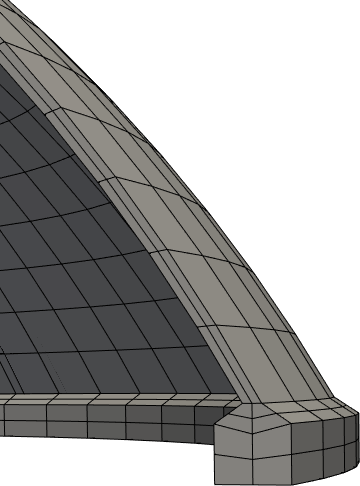}}
    	    \label{fig:ex2a_interpmesh}%
    }} 	\\
	\vspace{1em}
    \subfloat[Solution space mesh in Example 2--1 with $23\times 84\times 2$ elements in parametric directions $(1,2,3)$]{{
	    \includegraphics[width=0.38\textwidth, valign=t]{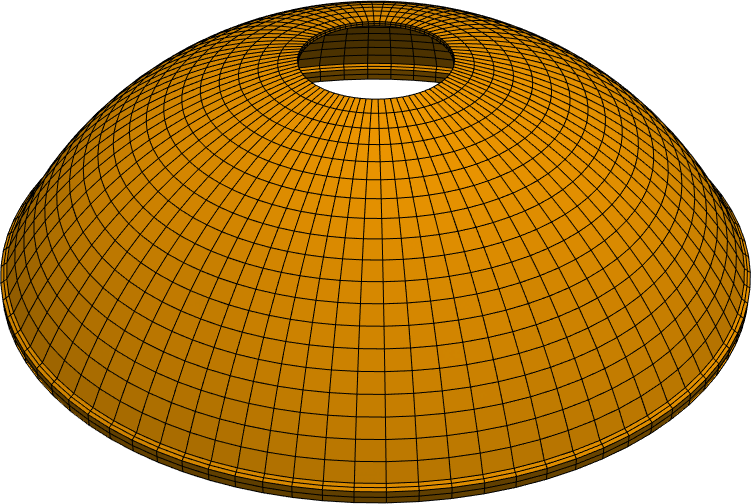}
	    \hspace{2.5em}
	    \framebox{\includegraphics[width=0.17\textwidth, valign=t]{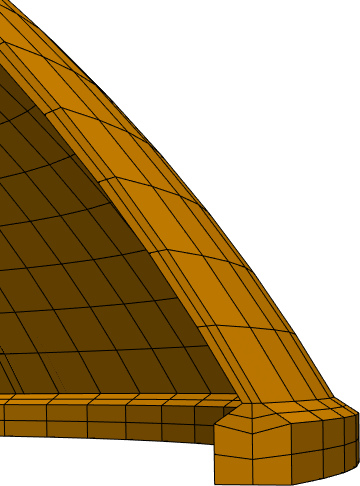}}
         \label{fig:ex2a_solmesh}%
    }}	\\
	\vspace{1em}
    \subfloat[Solution space mesh in Example 2--2 with $46\times 168\times 4$ elements in parametric directions $(1,2,3)$]{{
	    \includegraphics[width=0.38\textwidth,valign=t]{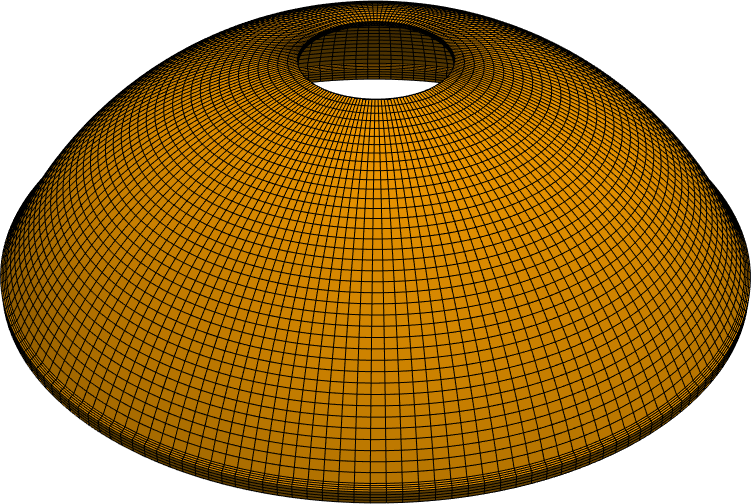}
	    \hspace{2.5em}
	    \framebox{\includegraphics[width=0.17\textwidth,valign=t]{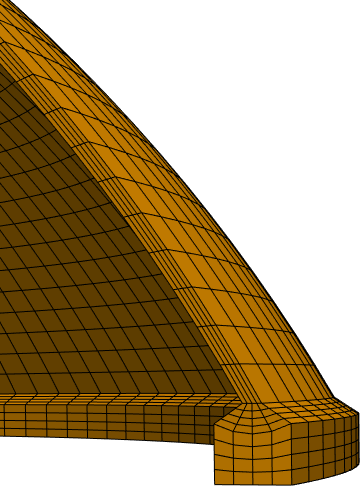}}
    	    \label{fig:ex2b_solmesh}%
    }}%
    \caption{The solution space and interpolation space meshes and the cross-sections used in Example 2--1 and Example 2--2.}%
	\label{fig:ex2_meshes}
\end{figure}

Figures \ref{fig:ex2a_interpmesh} and \ref{fig:ex2a_solmesh} depict the interpolation and solution meshes, respectively, that are used in this benchmark case. As described in this benchmark case the solution and interpolation space are identical.
The numerical results are summarized in Table \ref{tab:ex2a_ev}. As in the previous three-dimensional benchmarks the presolution setup costs hardly contribute to the total cost of the method. Again, the main computational cost lies in the matrix-free matrix-vector product that is evaluated in each iteration of the Lanczos eigenvalue solver. Due to the increased number of degrees of freedom for higher polynomial degrees the associated computational cost increases. Interestingly, the number of iterations is reduced in the case $p=16$ as compared to the lower polynomial degrees.

\begin{table}[H]
\centering
	\caption{Enumeration of twenty largest eigenvalues corresponding to the hemispherical shell with stiffener problem with Gaussian kernel in Example 2--1. The numerical eigenvalues have been computed by the proposed isogeometric Galerkin method employing interpolation based quadrature for the three different cases using solution and interpolation spaces depicted in Figure \ref{fig:ex2a_interpmesh} and \ref{fig:ex2a_solmesh} as well as standard isogeometric Galerkin reference solution computed using mesh depicted in Figure \ref{fig:ex2a_solmesh} and polynomial order $p=(2,2,2)$. Computations executed on a \textit{single} core.}
\label{tab:ex2a_ev}
\resizebox{0.85\textwidth}{!}{%
\begin{threeparttable}
\begin{tabular}{lllll}
\hline
Mode                                                                                                             & \multicolumn{4}{l}{Eigenvalue}                                                                  \\ \cline{2-5}
														 & $p=2$          &$p=6$            &$p=16$            & Galerkin 1\tnote{\**}     \\ \hline
1                                                                                                                & 14474.18249           & 14475.59075            & 14476.12924             & 14476.26164               \\
2                                                                                                                & 6530.062180           & 6531.224499            & 6531.666465             & 6531.775099               \\
3                                                                                                                & 6530.062180           & 6531.224499            & 6531.666444             & 6531.775099               \\
4                                                                                                                & 2091.112735           & 2091.638679            & 2091.836166             & 2091.884692               \\
5                                                                                                                & 2091.111050           & 2091.638679            & 2091.836147             & 2091.884692               \\
6                                                                                                                & 1971.193088           & 1971.617475            & 1971.784884             & 1971.826015               \\
7                                                                                                                & 552.5819743           & 552.6961143            & 552.7418006             & 552.7530118               \\
8                                                                                                                & 552.5819743           & 552.6961143            & 552.7417949             & 552.7530118               \\
9                                                                                                                & 522.0061723           & 522.1758813            & 522.2364134             & 522.2512911               \\
10                                                                                                               & 522.0061723          & 522.1758813           & 522.2364108            & 522.2512911               \\
11                                                                                                               & 113.1117043          & 113.1328509           & 113.1414885            & 113.1435938               \\
12                                                                                                               & 113.1116131          & 113.1328509           & 113.1414662            & 113.1435938               \\
13                                                                                                               & 106.0162787          & 106.0601883           & 106.0743281            & 106.0778059               \\
14                                                                                                               & 106.0161093          & 106.0601883           & 106.0743158            & 106.0778057               \\
15                                                                                                               & 101.1533173          & 101.1674081           & 101.1733615            & 101.1748288               \\
16                                                                                                               & 21.40721769          & 21.40924526           & 21.41014572            & 21.41036678               \\
17                                                                                                               & 21.40721769          & 21.40924526           & 21.41014233            & 21.41036678               \\
18                                                                                                               & 19.11433927          & 19.11772882           & 19.11904088            & 19.11936197               \\
19                                                                                                               & 19.11433927          & 19.11772882           & 19.11903300            & 19.11936197               \\
20                                                                                                               & 18.06993704          & 18.07977268           & 18.08244326            & 18.08309832               \\
\multicolumn{5}{c}{}                                                                                                                                                                                     \\
\multicolumn{5}{c}{\textit{Interpolation space} \hrulefill}                                                                                                                                              \\
\begin{tabular}[c]{@{}l@{}}\hspace{0.8em} Number of elements \end{tabular}                                       & 3864        & 3864         & 3864          & --                        \\
\begin{tabular}[c]{@{}l@{}}\hspace{0.8em} Number of degrees of freedom \end{tabular}                             & 10672         & 35424          & 189144           & --                        \\
\begin{tabular}[c]{@{}l@{}}\hspace{0.8em} Mesh size \end{tabular}                                                & 7.992         & 7.992          & 7.992           & --                        \\
\begin{tabular}[c]{@{}l@{}}\hspace{0.8em} Mesh size/correlation length \end{tabular}                             & 0.091       & 0.091        & 0.091         & --                        \\
\begin{tabular}[c]{@{}l@{}}\hspace{0.8em} Formation and assembly of univariate matrices\end{tabular}             & $ 0.354 \,\mathrm{s} $ & $ 0.427 \,\mathrm{s} $	 & $ 3.530 \,\mathrm{s} $  & --                        \\
\multicolumn{5}{c}{\textit{Solution space} \hrulefill}                                                                                                                                                   \\
\begin{tabular}[c]{@{}l@{}}\hspace{0.8em} Number of elements \end{tabular}                                       & 3864             & 3864              & 3864               & 3864                      \\
\begin{tabular}[c]{@{}l@{}}\hspace{0.8em} Number of degrees of freedom \end{tabular}                             & 9612             & 32760             & 180090             & 9612                      \\
\begin{tabular}[c]{@{}l@{}}\hspace{0.8em} Mesh size \end{tabular}                                                & 7.992            & 7.992             & 7.992              & 7.992                     \\
\begin{tabular}[c]{@{}l@{}}\hspace{0.8em} Mesh size/correlation length \end{tabular}                             & 0.091            & 0.091             & 0.091              & 0.091                     \\
\begin{tabular}[c]{@{}l@{}}\hspace{0.8em} Formation and assembly of system matrices\end{tabular}                 & --               & --                & --                 &$17.29\,\mathrm{h}$      \\
\multicolumn{5}{c}{\textit{Summary} \hrulefill}                                                                                                                                                          \\
	\begin{tabular}[c]{@{}l@{}}\hspace{0.8em} Number of iterations\end{tabular}                              & 52         & 52          & 41           & 52                        \\
\begin{tabular}[c]{@{}l@{}}\hspace{0.8em} Maximum resident memory [GB]\end{tabular}                              & 0.209           & 0.275            & 1.282             & 2.709                     \\
\begin{tabular}[c]{@{}l@{}}\hspace{0.8em} Solution time\end{tabular}                                             & $ 70.28 \,\mathrm{s}$	& $ 12.33 \,\mathrm{min}$	 & $ 5.00 \,\mathrm{h}$      &$13.16\,\mathrm{s}$      \\
\begin{tabular}[c]{@{}l@{}}\hspace{0.8em} Total time\end{tabular}                                                & $ 70.63 \,\mathrm{s}$	& $ 12.33 \,\mathrm{min}$  	 & $ 5.00 \,\mathrm{h}$      &$17.29\,\mathrm{h}$      \\
\multicolumn{5}{c}{}                                                                                                                                                                                     \\ \hline
\end{tabular}
\begin{tablenotes}
  \item[\**] exact kernel, NURBS trial and test space, elementwise assembly
\end{tablenotes}
\end{threeparttable}%
}
\end{table}

\subsubsection*{Example 2--2}
In the second benchmark the element size of the solution mesh is halved, see Figure \ref{fig:ex2b_solmesh}. The interpolation mesh is kept the same as in the first benchmark. The obtained results are presented in Table \ref{tab:ex2b_ev}. By comparing these results with those in Table \ref{tab:ex2a_ev} it may be observed that the dimension of the solution space does not significantly affect the total solver costs. Indeed, the number of degrees of freedom are more than doubled, yet the timings stay more or less the same, independent of polynomial degree. The increased dimension of the solution space mesh is reflected in the maximum resident memory, which has increased as compared to to the results in Table \ref{tab:ex2a_ev}. As witnessed in the previous benchmark, the higher order computations required fewer iterations than the lower order ones. 

\begin{remark} Note that the flexibility in mesh size of the interpolation versus the trial space mesh provides a mechanism by which the error due to quadrature versus the error due to discretization can be effectively controlled.
\end{remark}

\begin{table}[H]
\centering
	\caption{Enumeration of twenty largest eigenvalues corresponding to the hemispherical shell with stiffener problem with Gaussian kernel in Example 2--2. The numerical eigenvalues have been computed by the proposed isogeometric Galerkin method employing interpolation based quadrature for the three different cases using solution and interpolation spaces depicted in Figure \ref{fig:ex2a_interpmesh} and \ref{fig:ex2b_solmesh} as well as standard isogeometric Galerkin reference solution computed using mesh depicted in Figure \ref{fig:ex2a_solmesh} and polynomial order $p=(2,2,2)$. Computations executed on a \textit{single} core.}
\label{tab:ex2b_ev}
\resizebox{0.85\textwidth}{!}{%
\begin{threeparttable}
\begin{tabular}{lllll}
\hline
Mode                                                                                                             & \multicolumn{4}{l}{Eigenvalue}                                                                  \\ \cline{2-5}
														 &$p=2$           &$p=6$            &$p=16$            & Galerkin 1\tnote{\**}     \\ \hline
1                                                                                                                & 14475.49870           & 14475.83951            & 14476.16143             & 14476.26164               \\
2                                                                                                                & 6531.142590           & 6531.428669            & 6531.692867             & 6531.775099               \\
3                                                                                                                & 6531.142590           & 6531.428669            & 6531.692860             & 6531.775099               \\
4                                                                                                                & 2091.595539           & 2091.729909            & 2091.847949             & 2091.884692               \\
5                                                                                                                & 2091.593853           & 2091.729909            & 2091.847938             & 2091.884692               \\
6                                                                                                                & 1971.603041           & 1971.694851            & 1971.794886             & 1971.826015               \\
7                                                                                                                & 552.6940460           & 552.7172396            & 552.7445251             & 552.7530118               \\
8                                                                                                                & 552.6940460           & 552.7172396            & 552.7445226             & 552.7530118               \\
9                                                                                                                & 522.1541843           & 522.2038475            & 522.2400300             & 522.2512911               \\
10                                                                                                               & 522.1541843          & 522.2038475           & 522.2400296            & 522.2512911               \\
11                                                                                                               & 113.1329195          & 113.1368433           & 113.1419923            & 113.1435938               \\
12                                                                                                               & 113.1328283          & 113.1368433           & 113.1419915            & 113.1435938               \\
13                                                                                                               & 106.0508639          & 106.0667227           & 106.0751804            & 106.0778059               \\
14                                                                                                               & 106.0506944          & 106.0667227           & 106.0751771            & 106.0778057               \\
15                                                                                                               & 101.1680420          & 101.1701707           & 101.1737246            & 101.1748288               \\
16                                                                                                               & 21.40946576          & 21.40966457           & 21.41020520            & 21.41036678               \\
17                                                                                                               & 21.40946576          & 21.40966457           & 21.41020194            & 21.41036678               \\
18                                                                                                               & 19.11757482          & 19.11833648           & 19.11912297            & 19.11936197               \\
19                                                                                                               & 19.11757482          & 19.11833648           & 19.11911862            & 19.11936197               \\
20                                                                                                               & 18.07646765          & 18.08100654           & 18.08261032            & 18.08309832               \\
\multicolumn{5}{c}{}                                                                                                                                                                                     \\
\multicolumn{5}{c}{\textit{Interpolation space} \hrulefill}                                                                                                                                              \\
\begin{tabular}[c]{@{}l@{}}\hspace{0.8em} Number of elements \end{tabular}                                       & 3864        & 3864         & 3864          & --                        \\
\begin{tabular}[c]{@{}l@{}}\hspace{0.8em} Number of degrees of freedom \end{tabular}                             & 10672         & 35424          & 189144           & --                        \\
\begin{tabular}[c]{@{}l@{}}\hspace{0.8em} Mesh size \end{tabular}                                                & 7.992         & 7.992          & 7.992           & --                        \\
\begin{tabular}[c]{@{}l@{}}\hspace{0.8em} Mesh size/correlation length \end{tabular}                             & 0.091       & 0.091        & 0.091         & --                        \\
\begin{tabular}[c]{@{}l@{}}\hspace{0.8em} Formation and assembly of univariate matrices\end{tabular}             & $ 0.367 \,\mathrm{s} $ & $ 1.229 \,\mathrm{s} $	 & $ 23.21 \,\mathrm{s} $ & --                        \\
\multicolumn{5}{c}{\textit{Solution space} \hrulefill}                                                                                                                                                   \\
\begin{tabular}[c]{@{}l@{}}\hspace{0.8em} Number of elements \end{tabular}                                       & 30912            & 30912             & 30912              & 3864                     \\
\begin{tabular}[c]{@{}l@{}}\hspace{0.8em} Number of degrees of freedom \end{tabular}                             & 51900            & 117180            & 421360             & 9612                      \\
\begin{tabular}[c]{@{}l@{}}\hspace{0.8em} Mesh size \end{tabular}                                                & 4.019            & 4.019             & 4.019              & 7.992                     \\
\begin{tabular}[c]{@{}l@{}}\hspace{0.8em} Mesh size/correlation length \end{tabular}                             & 0.046            & 0.046             & 0.046              & 0.091                     \\
\begin{tabular}[c]{@{}l@{}}\hspace{0.8em} Formation and assembly of system matrices\end{tabular}                 & --               & --                & --                 &$17.29\,\mathrm{h}$      \\
\multicolumn{5}{c}{\textit{Summary} \hrulefill}                                                                                                                                                          \\
	\begin{tabular}[c]{@{}l@{}}\hspace{0.8em} Number of iterations\end{tabular}                              & 52         & 52          & 41           & 52                        \\
\begin{tabular}[c]{@{}l@{}}\hspace{0.8em} Maximum resident memory [GB]\end{tabular}                              & 0.261           & 0.741            & 7.346             & 2.709                     \\
\begin{tabular}[c]{@{}l@{}}\hspace{0.8em} Solution time\end{tabular}                                             & $ 70.05 \,\mathrm{s}$	& $ 12.433 \,\mathrm{min}$	 & $ 4.850 \,\mathrm{h}$     &$13.16\,\mathrm{s}$      \\
\begin{tabular}[c]{@{}l@{}}\hspace{0.8em} Total time\end{tabular}                                                & $ 70.42 \,\mathrm{s}$	& $ 12.433 \,\mathrm{min}$  	 & $ 4.858 \,\mathrm{h}$     &$17.29\,\mathrm{h}$      \\
\multicolumn{5}{c}{}                                                                                                                                                                                               \\ \hline
\end{tabular}
\begin{tablenotes}
  \item[\**] exact kernel, NURBS trial and test space, elementwise assembly
\end{tablenotes}
\end{threeparttable}%
}
\end{table}

\begin{figure}%
    \centering
    \subfloat[1st eigenfunction]{{\includegraphics[width=0.27\textwidth]{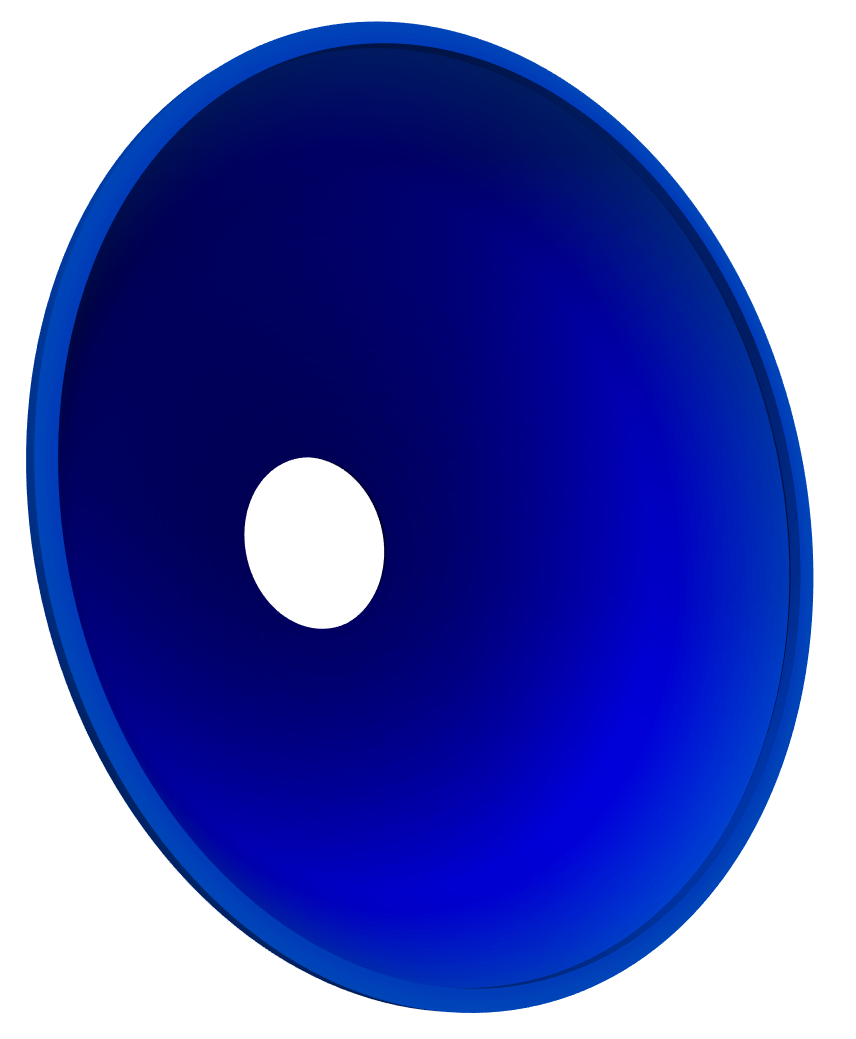} }}%
    \hfill
    \subfloat[2nd eigenfunction]{{\includegraphics[width=0.27\textwidth]{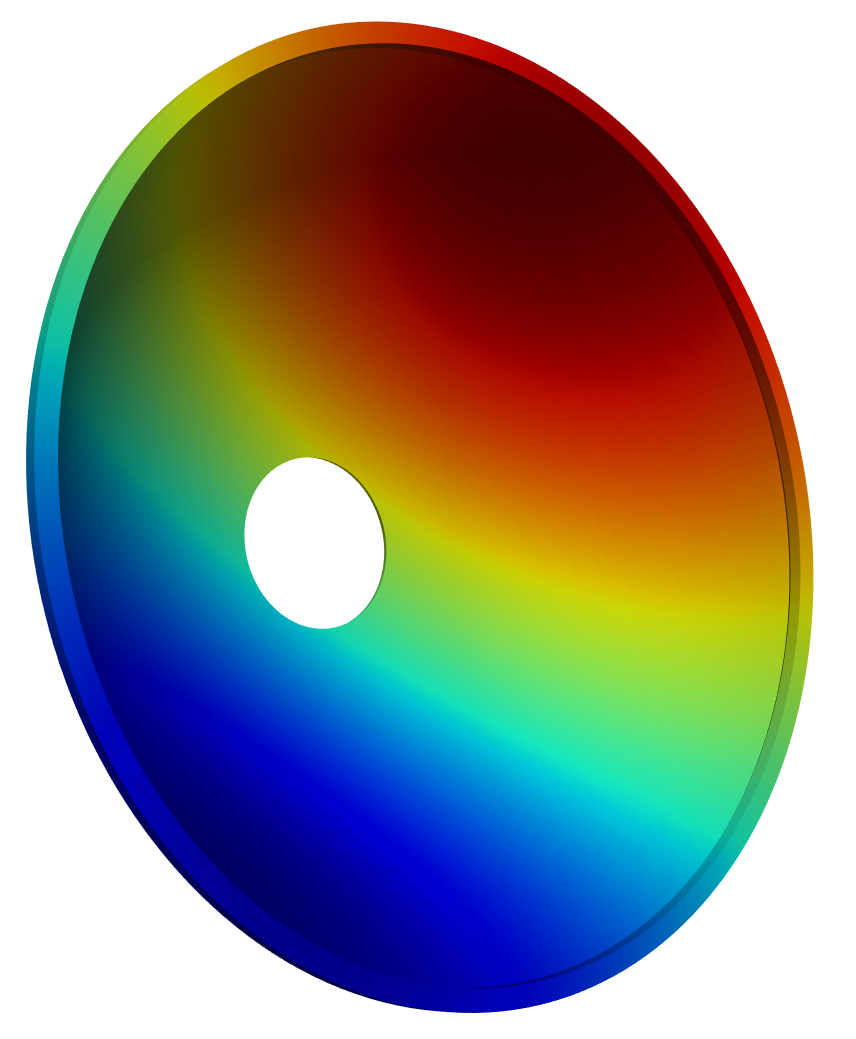} }}%
    \hfill
    \subfloat[3rd eigenfunction]{{\includegraphics[width=0.27\textwidth]{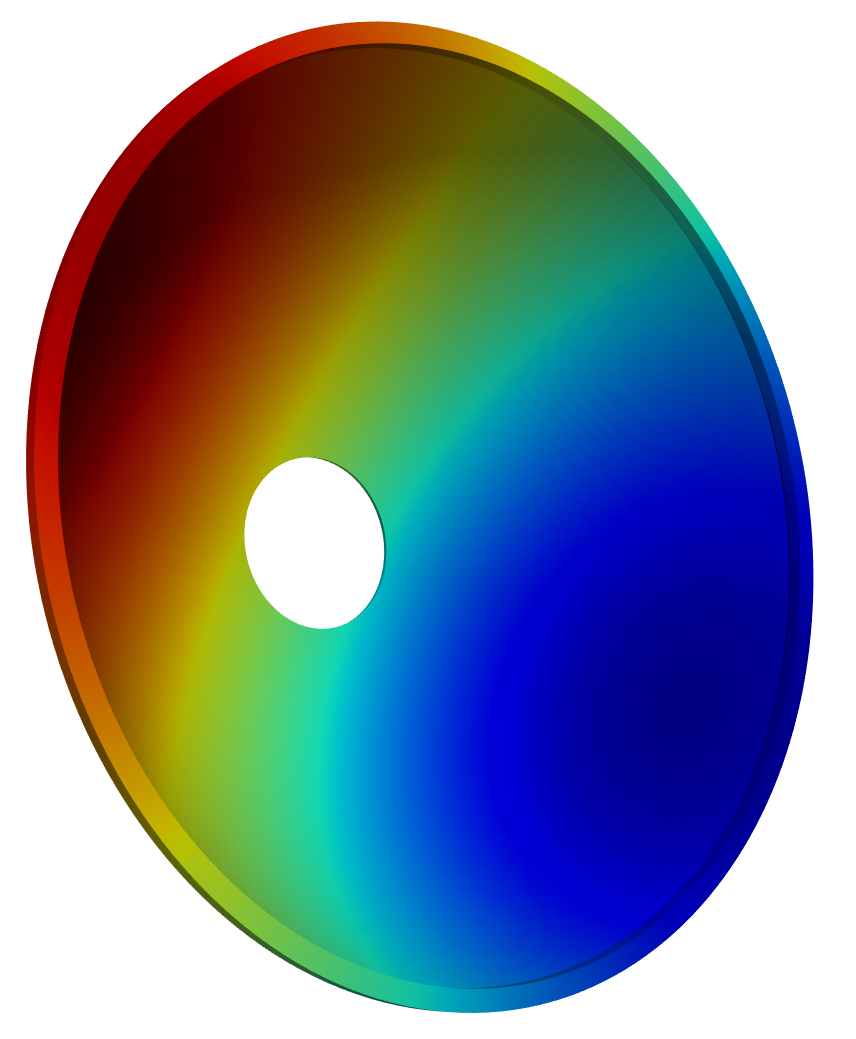} }}\\
    \subfloat[4th eigenfunction]{{\includegraphics[width=0.27\textwidth]{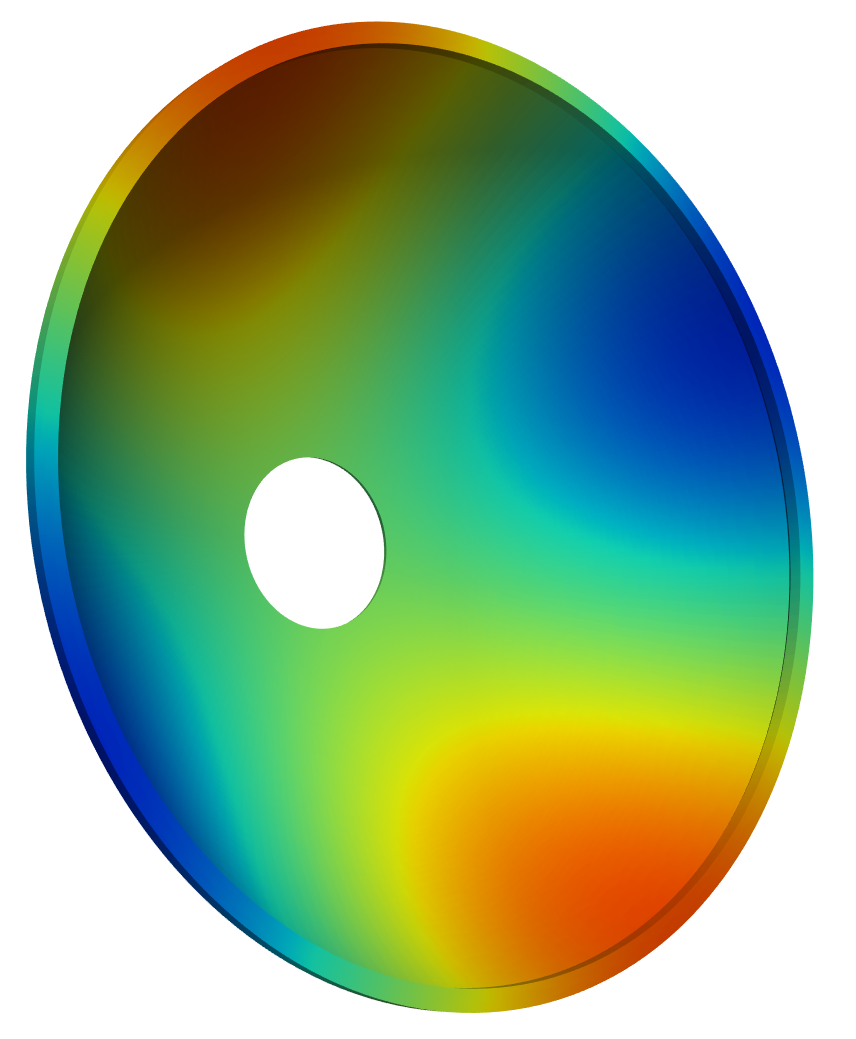} }}%
    \hfill
    \subfloat[5th eigenfunction]{{\includegraphics[width=0.27\textwidth]{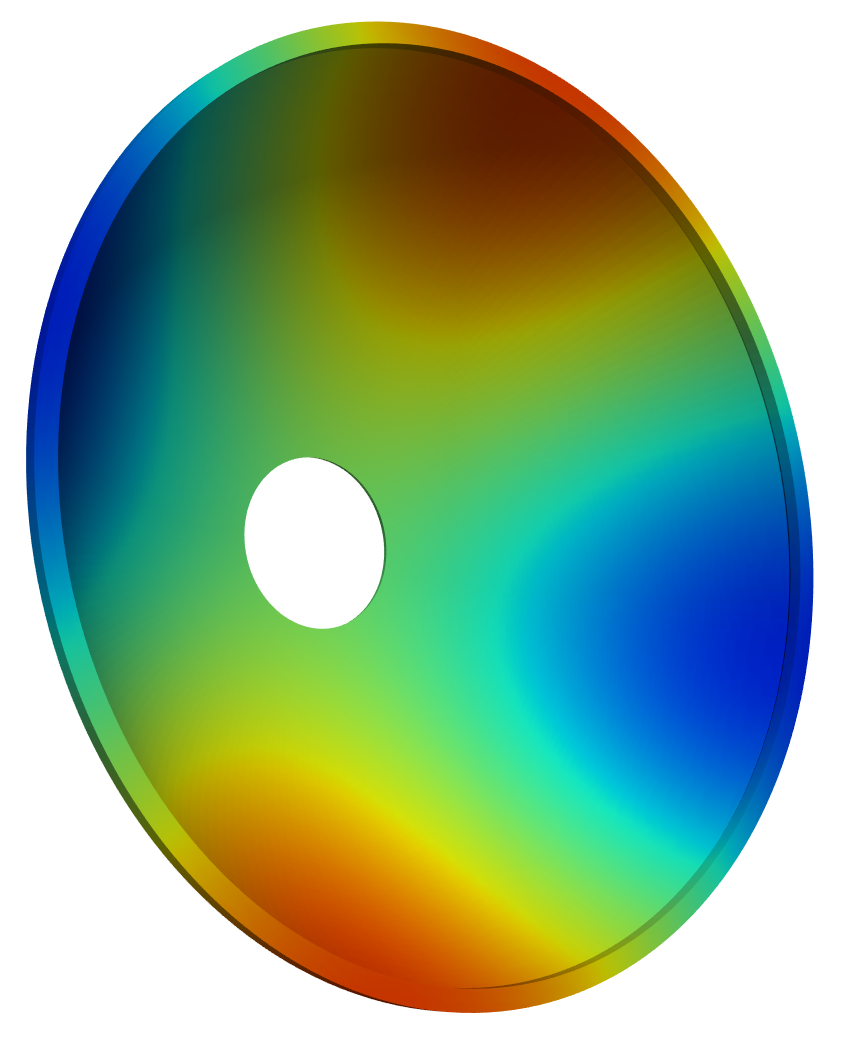} }}%
    \hfill
    \subfloat[6th eigenfunction]{{\includegraphics[width=0.27\textwidth]{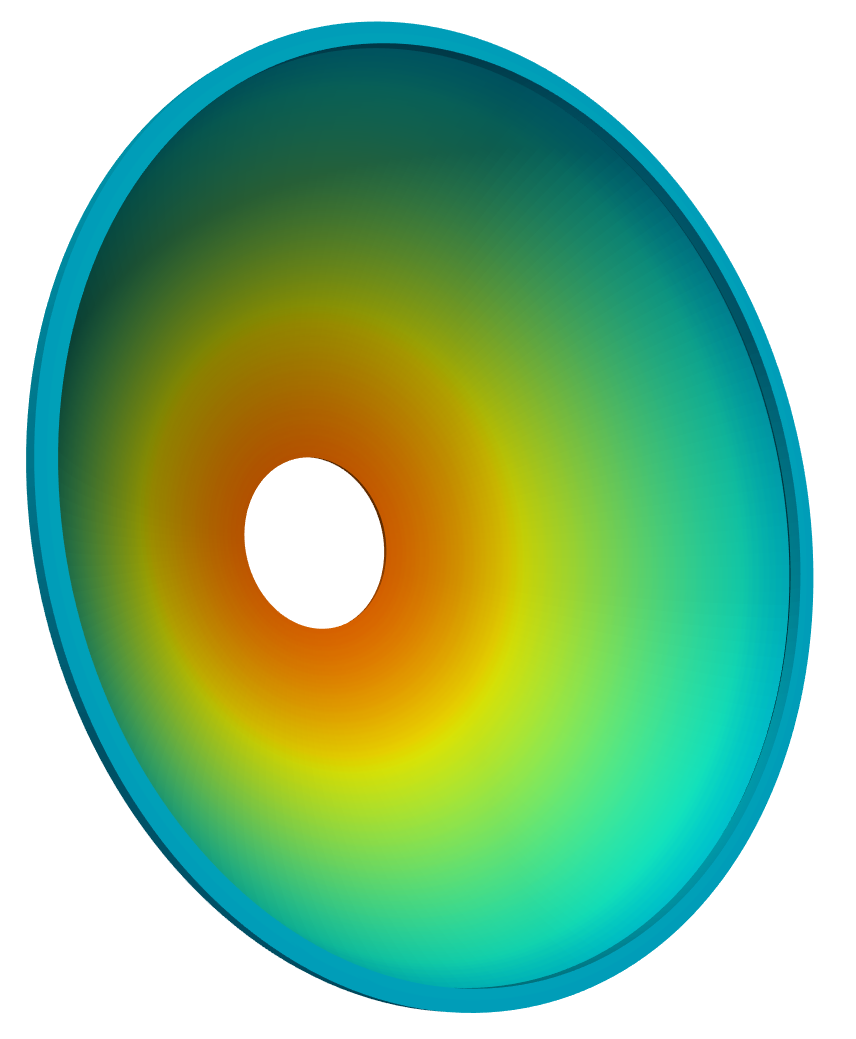} }}\\
    \subfloat[7th eigenfunction]{{\includegraphics[width=0.27\textwidth]{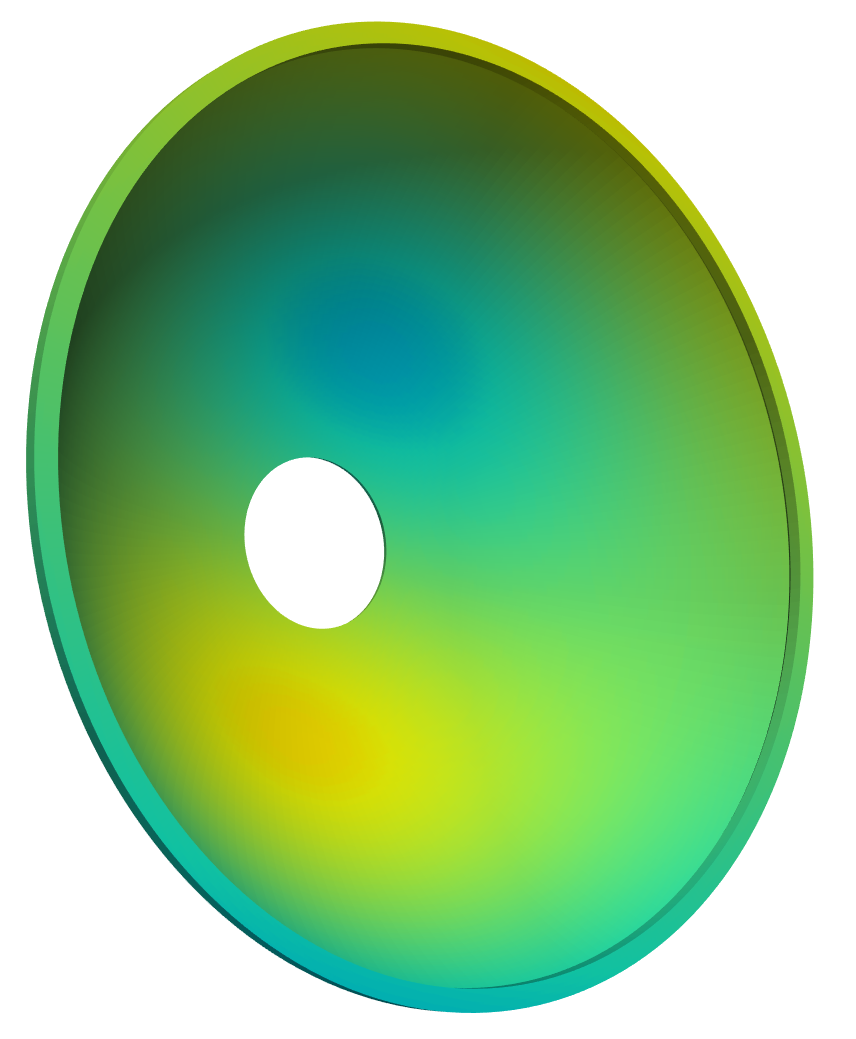} }}%
    \hfill
    \subfloat[8th eigenfunction]{{\includegraphics[width=0.27\textwidth]{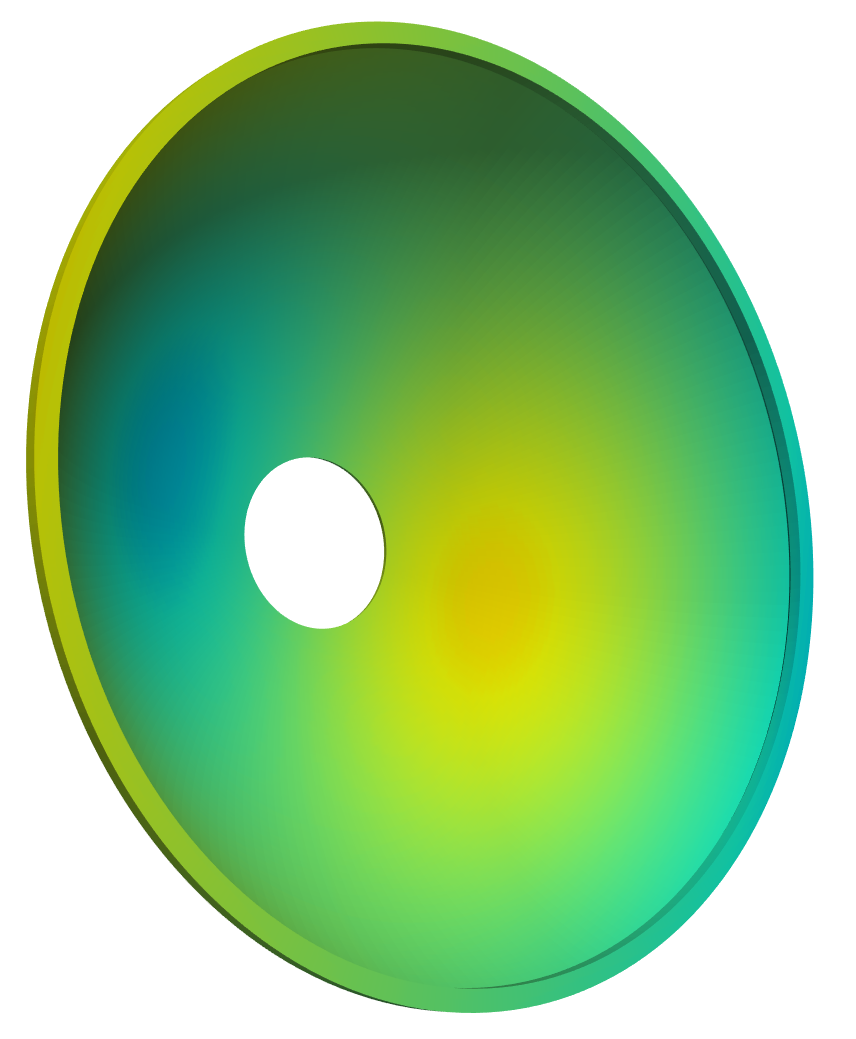} }}%
    \hfill
    \subfloat[9th eigenfunction]{{\includegraphics[width=0.27\textwidth]{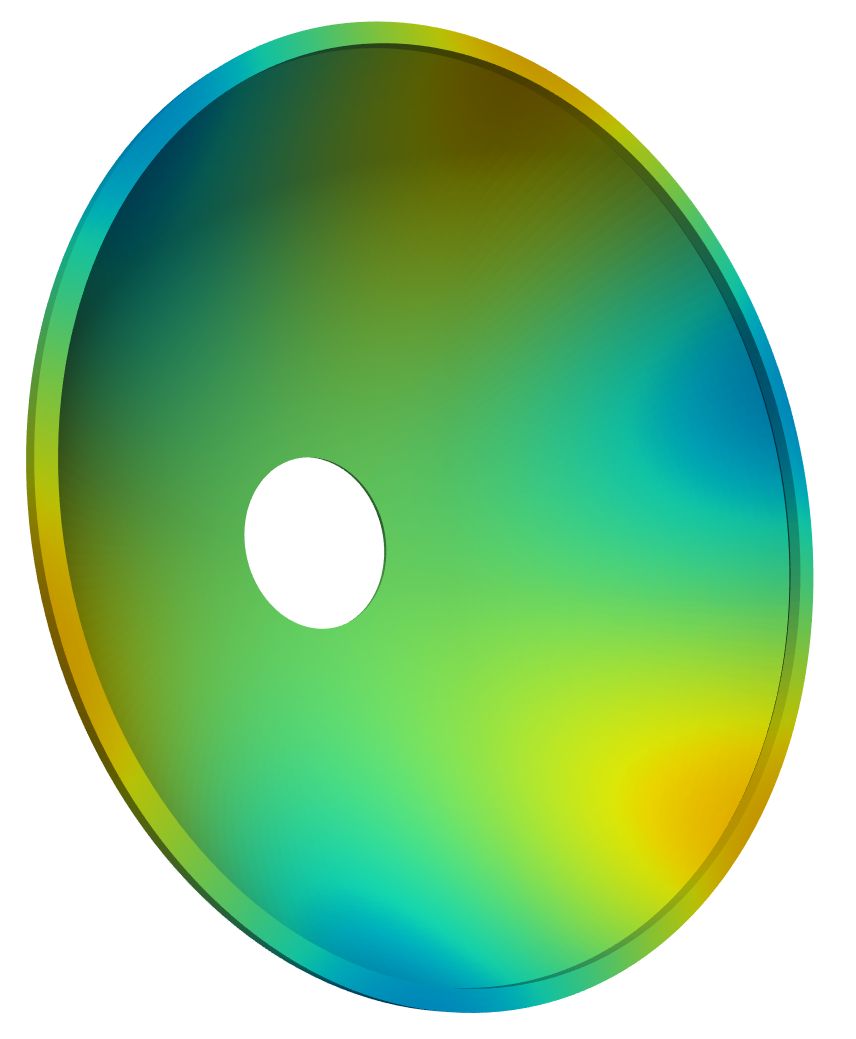} }}\\
		\vspace{1em}
    \includegraphics[width=0.45\textwidth]{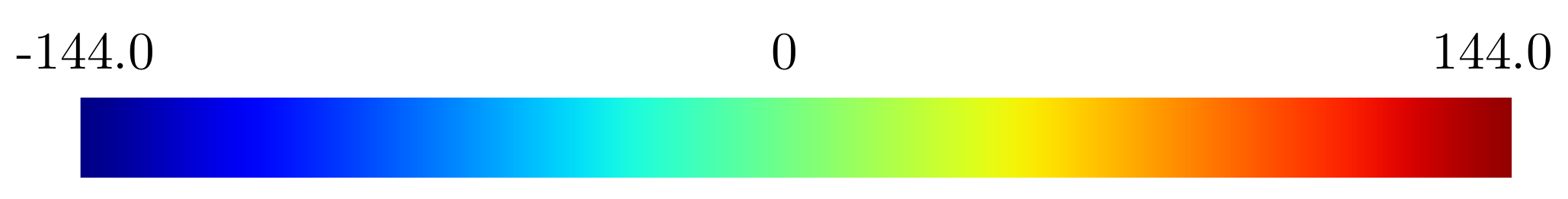}
    \caption{First nine normalized eigenfunctions weighted by the square root of the corresponding eigenvalues in Example 2--2.}%
    \label{fig:ex2b_ef}%
\end{figure}

\begin{remark}
A relevant question in random field discretization is what mesh size is necessary to attain
acceptable approximations. The mesh size should clearly depend on the correlation length $bL$.
A~rule of thumb, proposed in \cite{der_kiureghian_stochastic_1988}, is that the element size is approximately in the range from a half to a quarter of the given correlation length. Similar rules
have been established by other authors, see \cite{stefanou_stochastic_2009} and references therein. Especially, in a three-dimensional
problem this may lead to a large number of degrees of freedom. Engineering models of practical interest
are generally more complex than the models shown in this paper and may require millions of degrees of freedom.
\end{remark}

\subsubsection*{Parallel execution}

\begin{figure}[H]%
    \centering
	{\includegraphics[width=0.8\textwidth]{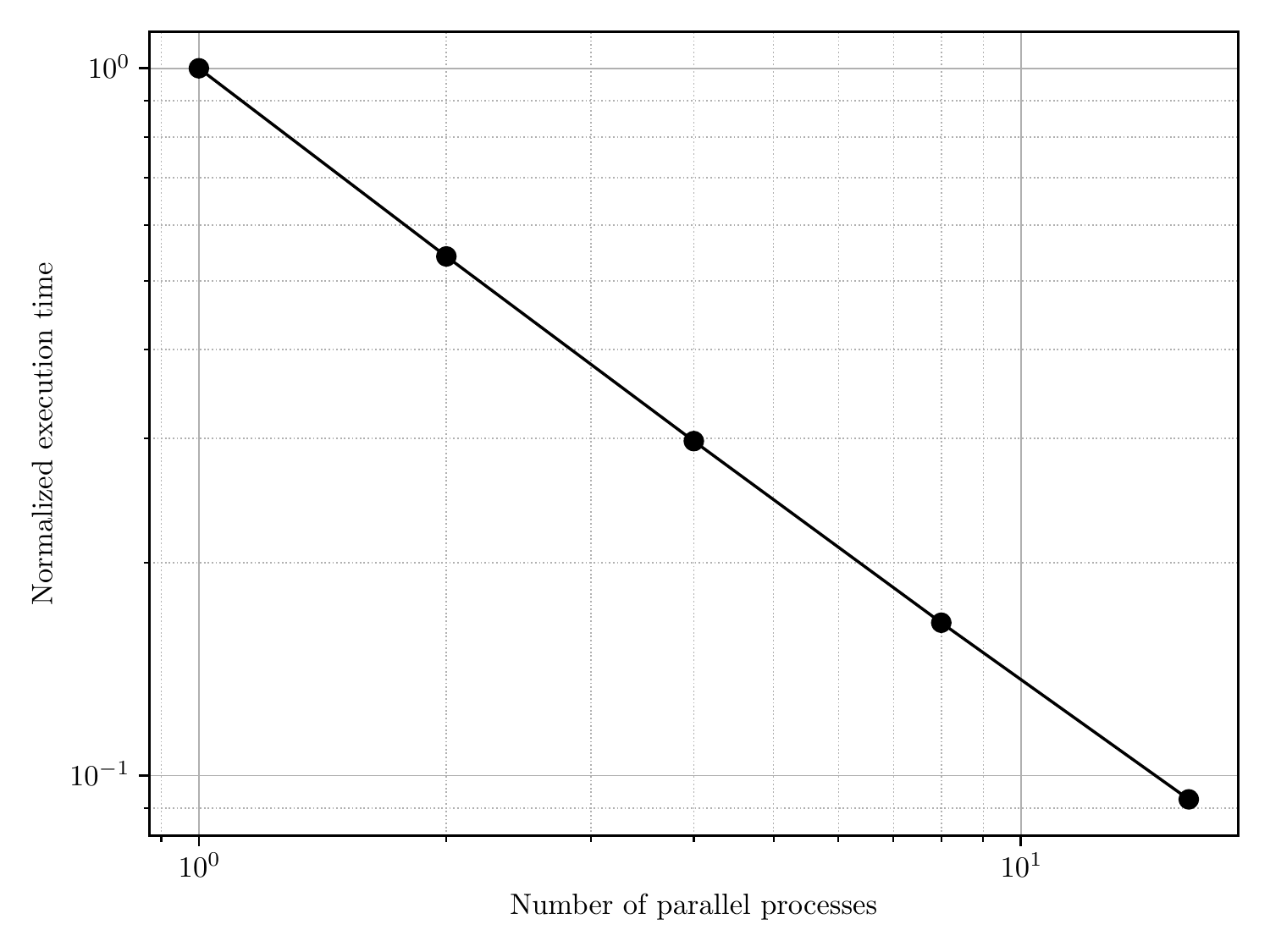}}
     \caption{Scaling of the execution time in Example 2--2 in the case of $p=6$ and $1,2,8$ and $16$ parallel processes. Timings normalized with respect to the execution time in a single process.} %
    \label{fig:scaling}
\end{figure}

In order to provide comparable results in the benchmarks, all the computations have been performed in a sequential manner in a single process. The proposed method is well suited for parallel execution, as discussed in section \ref{sec:ibq-complexity}. Figure \ref{fig:scaling} shows nearly optimal scaling with the number of cores for Example 2--2 in the case of $p=6$. Note, that \textit{only} the kernel evaluation and matrix-vector product in Step 5 of Algorithm \ref{alg:mf_eval} have been performed in parallel using shared-memory parallelism. All the other steps, as well as the Lanczos eigensolver are still run sequentially. In this particular case $99.84\%$ of the Algorithm \ref{alg:mf_eval} execution time was spent in the parallel execution mode.\B

\section{Conclusion} \label{sec:conclusion}
This paper presented an efficient matrix-free Galerkin method for the Karhunen-Loève series expansion (KLE) of random fields. The KLE requires the solution of a generalized eigenvalue problem corresponding to the homogeneous Fredholm integral eigenvalue problem of the second kind, and is computationally challenging for several reasons. Firstly, the Galerkin method requires numerical integration over a $2d$ dimensional domain, where $d$, in this work, denotes the spatial dimension. Consequently, classical formation and assembly procedures have a time complexity that scales $\mathcal{O}{\left( N_e^2\cdot (p+1)^{3d} \right)}$ with increasing polynomial degree~$p$ and number of elements $N_e$. Secondly, the main system matrix is dense and requires $\mathcal{O}\left( N^2 \right)$ bytes of storage, where $N$ is the global number of degrees of freedom. This means that a discretization involving a hundred thousand degrees of freedom requires at least 80GB of RAM to store the main system matrix in double precision. Hence, the computational complexity as well as memory requirements of standard solution techniques become quickly computationally intractable with increasing polynomial degree, problem size and spatial dimension.

We proposed an efficient solution methodology that significantly ameliorates the aforementioned computational challenges. Our approach is based on the following key ingredients:
\begin{enumerate}
	\item A trial space of rational spline functions, whose Gramian or mass matrix has a Kronecker structure independent of the geometric mapping;
	\item An inexpensive reformulation of the generalized algebraic eigenvalue problem into a standard algebraic eigenvalue problem;
	\item A degenerate kernel approximation of the covariance function using smooth tensor product splines;
	\item Formulation of an efficient matrix-free and parallel matrix-vector product for iterative eigenvalue solvers, which utilizes the Kronecker structure of the system matrices.
\end{enumerate}
In Step 2 the reformulation to a standard eigenvalue problem significantly reduces the computational cost while improving conditioning. This can be done efficiently due to the Kronecker structure of the mass matrix, which is a result of the particular choice of the trial space, see Step 1. In Step 3 the degenerate kernel approximation enables us to evaluate the resulting integrals exactly with a minimal number of evaluation points. Both steps involve matrices that are endowed with a Kronecker structure and can be performed matrix-free in $\mathcal{O} \left(N\cdot N^{1/d} \right)$ time. The leading cost of the method is due to the Lanczos eigenvalue algorithm, which involves dense matrix-vector multiplications. As noted in Step 4, we perform this step matrix-free, by computing the necessary components on the fly and in parallel in approximately $\mathcal O\left(N^2 N_{\text{iter}} / N_{\text{thread}}\right)$ time. Here $N_{\text{iter}}$ denotes the number of iterations of the eigensolver and $N_{\text{thread}}$ is the number of simultaneous processes. Several three dimensional benchmark problems involving non-trivial geometrical mappings have illustrated exceptional efficiency and effectiveness of the proposed solution methodology. In particular, we showed that the proposed methodology scales favorably with polynomial degree and works particularly well for smooth covariance functions, such as the Gaussian kernel. The Python implementation used to generate these results and the associated reference benchmarks has been provided as open-source software and is available for download at \url{https://github.com/m1ka05/tensiga}.

In a follow-up study we plan to extensively study the accuracy of the proposed solution methodology. There are two sources of error: (1) a quadrature error due to approximation of the covariance function; and (2) a discretization error due to the finite dimensional representation of the eigenmodes. We will perform a priori as well as a posteriori error analysis and formulate criteria for bounding the error due to quadrature by the discretization error. Within the same context of accuracy and robustness it is interesting to extend the spectral analysis results in \cite{hughes_finite_2014} to generalized eigenvalue problems corresponding to Fredholm integral equations for different covariance functions as well as polynomial order of the approximation.

We also plan to further improve the efficiency of the proposed method where possible. The proposed matrix-free algorithm lends itself for acceleration on graphics processing units (GPUs). Furthermore, exploiting particular structure (such as sparsity or symmetry) of the covariance function may lead to improved solver cost. For example, the hierarchical matrix method proposed in \cite{khoromskij_application_2009} performs the matrix-vector products in $\mathcal O\left(N \log{N}\right)$ time, by exploiting certain structure of the covariance function.

There are several other interesting avenues for future research. Some or all of the techniques proposed here could be applied to linear as well as non-linear Fredholm integral differential equations of the first as well as the second kind. While the proposed method is designed for smooth kernels it would be interesting to develop similar methods that are tailored towards continuous kernels, such as the exponential kernel, or even singular kernels, which are typical in boundary integral equations, see \cite{calabro_efficient_2018, falini_adaptive_2019, giannelli_study_2019} for similar ideas making use of quasi-interpolation.

Finally, we would like to mention that similar techniques can be applied in the context of the collocation method. The computational cost of such a method would be similar to that of the proposed Galerkin method.

\section*{Acknowledgments}
M.L. Mika, R.R. Hiemstra and D. Schillinger gratefully acknowledge funding from the German Research Foundation through the DFG Emmy Noether Award SCH 1249/2-1. T.J.R. Hughes and R.R. Hiemstra were partially supported by the National Science Foundation Industry/University Cooperative Research Center (IUCRC) for Efficient Vehicles and Sustainable Transportation Systems (EV-STS), and the United States Army CCDC Ground Vehicle Systems Center (TARDEC/NSF Project \# 1650483 AMD 2). Any opinions, findings, and conclusions or recommendations expressed in this material are those of the authors and do not necessarily reflect the views of the National Science Foundation. The authors thank Mona Dannert and Udo Nackenhorst for very helpful discussions and comments.






\bibliographystyle{acm}
\bibliography{zotero,references}

\end{document}